
%
%
\input harvmac
%
%
%
%
%
%
%
%
%
\newif\ifdraft

\noblackbox
\catcode`\@=11
\newif\iffrontpage
%
\ifx\answ\bigans
\def\titleft{\titsm}
\magnification=1200\baselineskip=15pt plus 2pt minus 1pt
%
\advance\hoffset by-0.075truein
\hsize=6.15truein\vsize=600.truept\hsbody=\hsize\hstitle=\hsize
\else\let\lr=L
\def\titleft{\titla}
\magnification=1000\baselineskip=14pt plus 2pt minus 1pt
%
\vsize=6.5truein
\hstitle=8truein\hsbody=4.75truein
\fullhsize=10truein\hsize=\hsbody
\fi
\parskip=4pt plus 10pt minus 4pt

\font\titla=cmr10 scaled\magstep3
\font\tenmss=cmss10
\font\absmss=cmss10 scaled\magstep1
\newfam\mssfam
\font\footrm=cmr8  \font\footrms=cmr5
\font\footrmss=cmr5   \font\footi=cmmi8
\font\footis=cmmi5   \font\footiss=cmmi5
\font\footsy=cmsy8   \font\footsys=cmsy5
\font\footsyss=cmsy5   \font\footbf=cmbx8
\font\footmss=cmss8
\def\footfont{\def\rm{\fam0\footrm}
\textfont0=\footrm \scriptfont0=\footrms
\scriptscriptfont0=\footrmss
\textfont1=\footi \scriptfont1=\footis
\scriptscriptfont1=\footiss
\textfont2=\footsy \scriptfont2=\footsys
\scriptscriptfont2=\footsyss
\textfont\itfam=\footi \def\it{\fam\itfam\footi}
\textfont\mssfam=\footmss \def\mss{\fam\mssfam\footmss}
\textfont\bffam=\footbf \def\bf{\fam\bffam\footbf} \rm}
\def\tenpoint{\def\rm{\fam0\tenrm}
\textfont0=\tenrm \scriptfont0=\sevenrm
\scriptscriptfont0=\fiverm
\textfont1=\teni  \scriptfont1=\seveni
\scriptscriptfont1=\fivei
\textfont2=\tensy \scriptfont2=\sevensy
\scriptscriptfont2=\fivesy
\textfont\itfam=\tenit \def\it{\fam\itfam\tenit}
\textfont\mssfam=\tenmss \def\mss{\fam\mssfam\tenmss}
\textfont\bffam=\tenbf \def\bf{\fam\bffam\tenbf} \rm}
\ifx\answ\bigans\def\abstractfont{\tenpoint}\else
\def\abstractfont{\def\rm{\fam0\absrm}
\textfont0=\absrm \scriptfont0=\absrms
\scriptscriptfont0=\absrmss
\textfont1=\absi \scriptfont1=\absis
\scriptscriptfont1=\absiss
\textfont2=\abssy \scriptfont2=\abssys
\scriptscriptfont2=\abssyss
\textfont\itfam=\bigit \def\it{\fam\itfam\bigit}
\textfont\mssfam=\absmss \def\mss{\fam\mssfam\absmss}
\textfont\bffam=\absbf \def\bf{\fam\bffam\absbf}\rm}\fi
%
\def\f@@t{\baselineskip10pt\lineskip0pt\lineskiplimit0pt
\bgroup\aftergroup\@foot\let\next}
\setbox\strutbox=\hbox{\vrule height 8.pt depth 3.5pt width\z@}
\def\vfootnote#1{\insert\footins\bgroup
\baselineskip10pt\footfont
\interlinepenalty=\interfootnotelinepenalty
\floatingpenalty=20000
\splittopskip=\ht\strutbox \boxmaxdepth=\dp\strutbox
\leftskip=24pt \rightskip=\z@skip
\parindent=12pt \parfillskip=0pt plus 1fil
\spaceskip=\z@skip \xspaceskip=\z@skip
\Textindent{$#1$}\footstrut\futurelet\next\fo@t}
\def\Textindent#1{\noindent\llap{#1\enspace}\ignorespaces}
\def\footnote#1{\attach{#1}\vfootnote{#1}}%

\def\foot{\attach\footsymbolgen\vfootnote{\footsymbol}}
\let\footsymbol=\star
\newcount\lastf@@t           \lastf@@t=-1
\newcount\footsymbolcount    \footsymbolcount=0
\def\footsymbolgen{\relax\footsym
\global\lastf@@t=\pageno\footsymbol}
\def\footsym{\ifnum\footsymbolcount<0
\global\footsymbolcount=0\fi
{\iffrontpage \else \advance\lastf@@t by 1 \fi
\ifnum\lastf@@t<\pageno \global\footsymbolcount=0
\else \global\advance\footsymbolcount by 1 \fi }
\ifcase\footsymbolcount \fd@f\star\or
\fd@f\dagger\or \fd@f\ast\or
\fd@f\ddagger\or \fd@f\natural\or
\fd@f\diamond\or \fd@f\bullet\or
\fd@f\nabla\else \fd@f\dagger
\global\footsymbolcount=0 \fi }
\def\fd@f#1{\xdef\footsymbol{#1}}
\def\space@ver#1{\let\@sf=\empty \ifmmode #1\else \ifhmode
\edef\@sf{\spacefactor=\the\spacefactor}
\unskip${}#1$\relax\fi\fi}
\def\attach#1{\space@ver{\strut^{\mkern 2mu #1} }\@sf\ }
%
\newif\ifnref
\def\rrr#1#2{\relax\ifnref\nref#1{#2}\else\ref#1{#2}\fi}

\def\nrf#1{\nreftrue{#1}\nreffalse}

\def\multref#1#2#3{\nrf{#1#2#3}\refs{#1{--}#3}}
\nreffalse
\def\refout{\listrefs}
%
\def\eqn#1{\xdef #1{(\secsym\the\meqno)}
\writedef{#1\leftbracket#1}%
\global\advance\meqno by1\eqno#1\eqlabeL#1}
\def\eqnalign#1{\xdef #1{(\secsym\the\meqno)}
\writedef{#1\leftbracket#1}%
\global\advance\meqno by1#1\eqlabeL{#1}}
%
\def\chap#1{\newsec{#1}}
\def\chapter#1{\chap{#1}}
\def\sect#1{\subsec{#1}}
\def\section#1{\sect{#1}}
\def\\{\ifnum\lastpenalty=-10000\relax
\else\hfil\penalty-10000\fi\ignorespaces}
\def\note#1{\leavevmode%
\edef\@@marginsf{\spacefactor=\the\spacefactor\relax}%
\ifdraft\strut\vadjust{%
\hbox to0pt{\hskip\hsize\hskip.05in%
\vbox to0pt{\vskip-\dp\strutbox%
\sevenrm\baselineskip=10pt plus 1pt minus 1pt%
\ifx\answ\bigans\hsize=.9in\else\hsize=.4in\fi%
\tolerance=5000 \hbadness=5000%
\leftskip=0pt \rightskip=0pt \everypar={}%
\raggedright\parskip=0pt \parindent=0pt%
\vskip-\ht\strutbox\noindent\strut#1\par%
\vss}\hss}}\fi\@@marginsf\kern-.01cm}
\def\titlepage{%
\frontpagetrue\nopagenumbers\abstractfont%
\hsize=\hstitle\rightline{\vbox{\baselineskip=10pt%
{\abstractfont\pubnum}}}\pageno=0}
\frontpagefalse
\def\pubnum{}
\def\pdate{\number\month/\number\yearltd}
\def\makefootline{\iffrontpage\vskip .27truein
\line{\the\footline}
\vskip -.1truein\line{\pdate\hfil}
\else\vskip.5cm\line{\hss \tenrm $-$ \folio\ $-$ \hss}\fi}
\def\title#1{\vskip .7truecm\titlestyle{\titleft #1}}
\def\titlestyle#1{\par\begingroup \interlinepenalty=9999
\leftskip=0.02\hsize plus 0.23\hsize minus 0.02\hsize
\rightskip=\leftskip \parfillskip=0pt
\hyphenpenalty=9000 \exhyphenpenalty=9000
\tolerance=9999 \pretolerance=9000
\spaceskip=0.333em \xspaceskip=0.5em
\noindent #1\par\endgroup }
\def\autskip{\ifx\answ\bigans\vskip.5truecm\else\vskip.1cm\fi}
\def\author#1{\vskip .7in \centerline{#1}}
\def\andauthor#1{\autskip
\centerline{\it and} \autskip\centerline{#1}}
\def\address#1{\ifx\answ\bigans\vskip.2truecm
\else\vskip.1cm\fi{\it \centerline{#1}}}
\def\abstract#1{\vskip .5in\vfil\centerline
{\bf Abstract}\penalty1000
{{\smallskip\ifx\answ\bigans\leftskip 2pc \rightskip 2pc
\else\leftskip 5pc \rightskip 5pc\fi
\noindent\abstractfont \baselineskip=12pt
{#1} \smallskip}}
\penalty-1000}
\def\endpage{\tenpoint\supereject\global\hsize=\hsbody%
\frontpagefalse\footline={\hss\tenrm\folio\hss}}
%
\def\CERN{\address{CERN, Geneva, Switzerland}}
\def\inbar{\vrule height1.5ex width.4pt depth0pt}
\def\IC{\relax\,\hbox{$\inbar\kern-.3em{\mss C}$}}
\def\IF{\relax{\rm I\kern-.18em F}}
\def\IH{\relax{\rm I\kern-.18em H}}
\def\II{\relax{\rm I\kern-.17em I}}
\def\IN{\relax{\rm I\kern-.18em N}}
\def\IP{\relax{\rm I\kern-.18em P}}
\def\IQ{\relax\,\hbox{$\inbar\kern-.3em{\rm Q}$}}
\def\IR{\relax{\rm I\kern-.18em R}}
\def\ZZ{\relax{\hbox{\mss Z\kern-.42em Z}}}
\def\nup#1({Nucl.\ Phys.\ $\us {B#1}$\ (}
\def\plt#1({Phys.\ Lett.\ $\us  {B#1}$\ (}
\def\plb#1({Phys.\ Lett.\ $\us  {#1B}$\ (}
\def\cmp#1({Comm.\ Math.\ Phys.\ $\us  {#1}$\ (}
\def\prp#1({Phys.\ Rep.\ $\us  {#1}$\ (}
\def\prl#1({Phys.\ Rev.\ Lett.\ $\us  {#1}$\ (}
\def\prv#1({Phys.\ Rev. $\us  {#1}$\ (}
\def\und#1({            $\us  {#1}$\ (}
\def\tit#1,{{\it #1},\ }
%

\def\tilde{\widetilde}
\def\bar{\overline}
\def\us#1{\bf{#1}}
\def\hat{\widehat}

\def\Coe#1.#2.{{#1\over #2}}

\def\coe#1.#2.{\relax{\textstyle {#1 \over #2}}\displaystyle}

\def\notin{\hbox{{$\in$}\kern-.51em\hbox{/}}}

\catcode`\@=12
%
\newbox\hdbox%
\newcount\hdrows%
\newcount\multispancount%
\newcount\ncase%
\newcount\ncols
\newcount\nrows%
\newcount\nspan%
\newcount\ntemp%
\newdimen\hdsize%
\newdimen\newhdsize%
\newdimen\parasize%
\newdimen\spreadwidth%
\newdimen\thicksize%
\newdimen\thinsize%
\newdimen\tablewidth%
\newif\ifcentertables%
\newif\ifendsize%
\newif\iffirstrow%
\newif\iftableinfo%
\newtoks\dbt%
\newtoks\hdtks%
\newtoks\savetks%
\newtoks\tableLETtokens%
\newtoks\tabletokens%
\newtoks\widthspec%
%
%
%
%
\tableinfotrue%
\catcode`\@=11
%
%
\def\tstrut{\vrule height3.1ex depth1.2ex width0pt}%
\def\and{\char`\&}
\def\tablerule{\noalign{\hrule height\thinsize depth0pt}}%
\thicksize=1.5pt
\thinsize=0.6pt
\def\thickrule{\noalign{\hrule height\thicksize depth0pt}}%
\def\ctr#1{\hfil\ #1\hfil}%
%
%
%
%
\tablewidth=-\maxdimen%
\spreadwidth=-\maxdimen%
\def\tabskipglue{0pt plus 1fil minus 1fil}%
%
%
\centertablestrue%
%
%
%
%
\parasize=4in%
\gdef\ARGS{########}
\gdef\headerARGS{####}
\def\@mpersand{&}
{\catcode`\|=13
\gdef\letbarzero{\let|0}
\gdef\letbartab{\def|{&&}}%
\gdef\letvbbar{\let\vb|}%
}
{\catcode`\&=4
\def\ampskip{&\omit\hfil&}
\catcode`\&=13
\let&0
\xdef\letampskip{\def&{\ampskip}}%
\gdef\letnovbamp{\let\novb&\let\tab&}
}
\def\begintable{
   \begingroup%
   \catcode`\|=13\letbartab\letvbbar%
   \catcode`\&=13\letampskip\letnovbamp%
   \def\multispan##1{
      \omit \mscount##1%
      \multiply\mscount\tw@\advance\mscount\m@ne%
      \loop\ifnum\mscount>\@ne \sp@n\repeat%
   }
   \def\|{%
      &\omit\widevline&%
   }%
   \ruledtable
}
\long\def\ruledtable#1\endtable{%
%
%
%
   \offinterlineskip
   \tabskip 0pt
   \def\widevline{\vrule width\thicksize}
   \def\endrow{\@mpersand\omit\hfil\crnorm\@mpersand}%
   \def\crthick{\@mpersand\crnorm\thickrule\@mpersand}%
   \def\crthickneg##1{\@mpersand\crnorm\thickrule
          \noalign{{\skip0=##1\vskip-\skip0}}\@mpersand}%
   \def\crnorule{\@mpersand\crnorm\@mpersand}%
   \def\crnoruleneg##1{\@mpersand\crnorm
          \noalign{{\skip0=##1\vskip-\skip0}}\@mpersand}%
   \let\nr=\crnorule
   \def\endtable{\@mpersand\crnorm\thickrule}%
   \let\crnorm=\cr
%
%
   \edef\cr{\@mpersand\crnorm\tablerule\@mpersand}%
   \def\crneg##1{\@mpersand\crnorm\tablerule
          \noalign{{\skip0=##1\vskip-\skip0}}\@mpersand}%
   \let\ctneg=\crthickneg
   \let\nrneg=\crnoruleneg
   \the\tableLETtokens
%
%
   \tabletokens={&#1}
%
%
   \countROWS\tabletokens\into\nrows%
   \countCOLS\tabletokens\into\ncols%
%
%
   \advance\ncols by -1%
   \divide\ncols by 2%
   \advance\nrows by 1%
%
%
   \iftableinfo %
      \immediate\write16{[Nrows=\the\nrows, Ncols=\the\ncols]}%
   \fi%
%
%
   \ifcentertables
      \ifhmode \par\fi
      \hbox to \hsize{
      \hss
   \else %
      \hbox{%
   \fi
      \vbox{%
         \makePREAMBLE{\the\ncols}
         \edef\next{\preamble}
         \let\preamble=\next
         \makeTABLE{\preamble}{\tabletokens}
      }
      \ifcentertables \hss}\else }\fi
   \endgroup
   \tablewidth=-\maxdimen
   \spreadwidth=-\maxdimen
}
\def\makeTABLE#1#2{
   {
   \let\ifmath0
   \let\header0
   \let\multispan0
%
%
   \ncase=0%
   \ifdim\tablewidth>-\maxdimen \ncase=1\fi%
   \ifdim\spreadwidth>-\maxdimen \ncase=2\fi%
   \relax
%
   \ifcase\ncase %
      \widthspec={}%
   \or %
      \widthspec=\expandafter{\expandafter t\expandafter o%
                 \the\tablewidth}%
   \else %
      \widthspec=\expandafter{\expandafter s\expandafter p\expandafter r%
                 \expandafter e\expandafter a\expandafter d%
                 \the\spreadwidth}%
   \fi %
   \xdef\next{
      \halign\the\widthspec{%
      #1
      \noalign{\hrule height\thicksize depth0pt}
      \the#2\endtable
%
      }
   }
   }
   \next
}
\def\makePREAMBLE#1{
   \ncols=#1
   \begingroup
   \let\ARGS=0
   \edef\xtp{\widevline\ARGS\tabskip\tabskipglue%
   &\ctr{\ARGS}\tstrut}
   \advance\ncols by -1
   \loop
      \ifnum\ncols>0 %
      \advance\ncols by -1%
      \edef\xtp{\xtp&\vrule width\thinsize\ARGS&\ctr{\ARGS}}%
   \repeat
   \xdef\preamble{\xtp&\widevline\ARGS\tabskip0pt%
   \crnorm}
   \endgroup
}
\def\countROWS#1\into#2{
   \let\countREGISTER=#2%
   \countREGISTER=0%
   \expandafter\ROWcount\the#1\endcount%
}%
\def\ROWcount{%
   \afterassignment\subROWcount\let\next= %
}%
\def\subROWcount{%
   \ifx\next\endcount %
      \let\next=\relax%
   \else%
      \ncase=0%
      \ifx\next\cr %
         \global\advance\countREGISTER by 1%
         \ncase=0%
      \fi%
      \ifx\next\endrow %
         \global\advance\countREGISTER by 1%
         \ncase=0%
      \fi%
      \ifx\next\crthick %
         \global\advance\countREGISTER by 1%
         \ncase=0%
      \fi%
      \ifx\next\crnorule %
         \global\advance\countREGISTER by 1%
         \ncase=0%
      \fi%
      \ifx\next\crthickneg %
         \global\advance\countREGISTER by 1%
         \ncase=0%
      \fi%
      \ifx\next\crnoruleneg %
         \global\advance\countREGISTER by 1%
         \ncase=0%
      \fi%
      \ifx\next\crneg %
         \global\advance\countREGISTER by 1%
         \ncase=0%
      \fi%
      \ifx\next\header %
         \ncase=1%
      \fi%
      \relax%
      \ifcase\ncase %
         \let\next\ROWcount%
      \or %
         \let\next\argROWskip%
      \else %
      \fi%
   \fi%
   \next%
}
\def\counthdROWS#1\into#2{%
\dvr{10}%
   \let\countREGISTER=#2%
   \countREGISTER=0%
\dvr{11}%
\dvr{13}%
   \expandafter\hdROWcount\the#1\endcount%
\dvr{12}%
}%
\def\hdROWcount{%
   \afterassignment\subhdROWcount\let\next= %
}%
\def\subhdROWcount{%
   \ifx\next\endcount %
      \let\next=\relax%
   \else%
      \ncase=0%
      \ifx\next\cr %
         \global\advance\countREGISTER by 1%
         \ncase=0%
      \fi%
      \ifx\next\endrow %
         \global\advance\countREGISTER by 1%
         \ncase=0%
      \fi%
      \ifx\next\crthick %
         \global\advance\countREGISTER by 1%
         \ncase=0%
      \fi%
      \ifx\next\crnorule %
         \global\advance\countREGISTER by 1%
         \ncase=0%
      \fi%
      \ifx\next\header %
         \ncase=1%
      \fi%
\relax%
      \ifcase\ncase %
         \let\next\hdROWcount%
      \or%
         \let\next\arghdROWskip%
      \else %
      \fi%
   \fi%
   \next%
}%
{\catcode`\|=13\letbartab
\gdef\countCOLS#1\into#2{%
   \let\countREGISTER=#2%
   \global\countREGISTER=0%
   \global\multispancount=0%
   \global\firstrowtrue
   \expandafter\COLcount\the#1\endcount%
   \global\advance\countREGISTER by 3%
   \global\advance\countREGISTER by -\multispancount
}%
\gdef\COLcount{%
   \afterassignment\subCOLcount\let\next= %
}%
{\catcode`\&=13%
\gdef\subCOLcount{%
   \ifx\next\endcount %
      \let\next=\relax%
   \else%
      \ncase=0%
      \iffirstrow
         \ifx\next& %
            \global\advance\countREGISTER by 2%
            \ncase=0%
         \fi%
         \ifx\next\span %
            \global\advance\countREGISTER by 1%
            \ncase=0%
         \fi%
         \ifx\next| %
            \global\advance\countREGISTER by 2%
            \ncase=0%
         \fi
         \ifx\next\|
            \global\advance\countREGISTER by 2%
            \ncase=0%
         \fi
         \ifx\next\multispan
            \ncase=1%
            \global\advance\multispancount by 1%
         \fi
         \ifx\next\header
            \ncase=2%
         \fi
         \ifx\next\cr       \global\firstrowfalse \fi
         \ifx\next\endrow   \global\firstrowfalse \fi
         \ifx\next\crthick  \global\firstrowfalse \fi
         \ifx\next\crnorule \global\firstrowfalse \fi
         \ifx\next\crnoruleneg \global\firstrowfalse \fi
         \ifx\next\crthickneg  \global\firstrowfalse \fi
         \ifx\next\crneg       \global\firstrowfalse \fi
      \fi
\relax
      \ifcase\ncase %
         \let\next\COLcount%
      \or %
         \let\next\spancount%
      \or %
         \let\next\argCOLskip%
      \else %
      \fi %
   \fi%
   \next%
}%
\gdef\argROWskip#1{%
   \let\next\ROWcount \next%
}
\gdef\arghdROWskip#1{%
   \let\next\ROWcount \next%
}
\gdef\argCOLskip#1{%
   \let\next\COLcount \next%
}
}
}
\def\spancount#1{
   \nspan=#1\multiply\nspan by 2\advance\nspan by -1%
   \global\advance \countREGISTER by \nspan
   \let\next\COLcount \next}%
\def\dvr#1{\relax}%
\def\header#1{%
\dvr{1}{\let\cr=\@mpersand%
\hdtks={#1}%
\counthdROWS\hdtks\into\hdrows%
\advance\hdrows by 1%
\ifnum\hdrows=0 \hdrows=1 \fi%
\dvr{5}\makehdPREAMBLE{\the\hdrows}%
\dvr{6}\getHDdimen{#1}%
{\parindent=0pt\hsize=\hdsize{\let\ifmath0%
\xdef\next{\valign{\headerpreamble #1\crnorm}}}\dvr{7}\next\dvr{8}%
}%
}\dvr{2}}
\def\makehdPREAMBLE#1{
\dvr{3}%
\hdrows=#1
{
\let\headerARGS=0%
\let\cr=\crnorm%
\edef\xtp{\vfil\hfil\hbox{\headerARGS}\hfil\vfil}%
\advance\hdrows by -1
\loop
\ifnum\hdrows>0%
\advance\hdrows by -1%
\edef\xtp{\xtp&\vfil\hfil\hbox{\headerARGS}\hfil\vfil}%
\repeat%
\xdef\headerpreamble{\xtp\crcr}%
}
\dvr{4}}
\def\getHDdimen#1{%
\hdsize=0pt%
\getsize#1\cr\end\cr%
}
\def\getsize#1\cr{%
\endsizefalse\savetks={#1}%
\expandafter\lookend\the\savetks\cr%
\relax \ifendsize \let\next\relax \else%
\setbox\hdbox=\hbox{#1}\newhdsize=1.0\wd\hdbox%
\ifdim\newhdsize>\hdsize \hdsize=\newhdsize \fi%
\let\next\getsize \fi%
\next%
}%
\def\lookend{\afterassignment\sublookend\let\looknext= }%
\def\sublookend{\relax%
\ifx\looknext\cr %
\let\looknext\relax \else %
   \relax
   \ifx\looknext\end \global\endsizetrue \fi%
   \let\looknext=\lookend%
    \fi \looknext%
}%
%
%
\def\tablelet#1{%
   \tableLETtokens=\expandafter{\the\tableLETtokens #1}%
}%
\catcode`\@=12
%

%
\def\OFFSET{\hoffset=6.pt\voffset=40.pt}

\OFFSET
\def\PLANCK{\rrr\PLANCK{L.E. Ib\'a\~nez, \plb126 (1983) 196;
J.E. Bjorkman and D.R.T. Jones, \nup259 (1985) 533.}}

\def\ZNZM{\rrr\ZNZM{A. Font, L.E. Ib\'a\~nez and F. Quevedo,
\plt217 (1989) 272.}}

\def\DSW{\rrr\DSW{M. Dine, N. Seiberg and E. Witten, \nup289 (1987) 589;
W. Lerche, B.E.W. Nilsson and A.N. Schellekens, \nup289 (1987) 609.}}

\def\FLIPP{\rrr\FLIPP{I. Antoniadis, J. Ellis, J.S. Hagelin and
D.V. Nanopoulos, \plt205 (1988) 459, \plt213 (1988) 56.}}

\def\FLIPPT{\rrr\FLIPPT{I. Antoniadis, J. Ellis, R. Lacaze and
D.V. Nanopoulos, \plt268 (1991) 188; S. Kalara, J.L. Lopez and
D.V. Nanopoulos, \plt269 (1991) 84.}}

\def\IN{\rrr\IN{L.E. Ib\'a\~nez and H.P. Nilles, \plb169 (1986) 354.}}

\def\MAGN{\rrr\MAGN{S. Ferrara, N. Magnoli, T.R. Taylor
and G. Veneziano, \plt245 (1990) 409.}}

\def\LUE{\rrr\LUE{D. L\"ust, {\it ``Duality Invariant Effective String
Actions and Automorphic Functions for (2,2) String Compactifications''},
      talk presented at the Workshop
  on String Theory, Trieste 1991,
      preprint CERN-TH.6143/91;
    {\it ``Duality Invariant Effective String Actions and Minimal
      Superstring Unification''},
      talk presented at the EPS-Conference, Geneva 1991,
      preprint CERN-TH.6259/91.}}

\def\VALEN{\rrr\VALEN{L.E. Ib\'a\~nez, {\it ``Topics in String
  Unification''}, talk given at the Workshop on Electroweak Physics
    Beyond the Standard Model, Valencia 1991, preprint CERN-TH.6342/91.}}

\def\KAPLU{\rrr\KAPLU{V. Kaplunovsky, \nup307 (1988) 145.}}

\def\SUSYUN{\rrr\SUSYUN{
S. Dimopoulos, S. Raby and F. Wilczek, Phys. Rev.
D24 (1981) 1681;
                 L.E. Ib\'a\~nez and G.G. Ross, \plb105 (1981) 439;
S. Dimopoulos and H. Georgi, \nup193 (1981) 375;
                          M. Einhorn and D.R.T. Jones, \nup196 (1982)
                      475.}}

\def\AMALDI{\rrr\AMALDI{
G. Costa, J. Ellis, G.L. Fogli, D.V. Nanopolous and F. Zwirner, \nup297
   (1988) 244;
J. Ellis, S. Kelley and D.V. Nanopoulos, \plt249 (1990) 441;
\plb260 (1991) 131;
P. Langacker, {\it ``Precision tests of the standard model",}
Pennsylvania preprint UPR-0435T, (1990);
U. Amaldi, W. de Boer and H. F\"urstenau, \plt260 (1991) 447;
P. Langacker and M. Luo, Phys.Rev.{\bf D44} (1991) 817;
R. Roberts and G.G. Ross, preprint RAL-92-005 (1992).
}}

\def\IMNQ{\rrr\IMNQ{L.E. Ib\'a\~nez, H.P. Nilles and F. Quevedo,
\plt187 (1987) 25; L.E. Ib\'a\~nez, J. Mas, H.P. Nilles and
F. Quevedo, \nup301 (1988) 157.}}

\def\SCHELL{\rrr\SCHELL{A.N. Schellekens, \plt237 (1990) 363.}}

\def\GQW{\rrr\GQW{H. Georgi, H.R. Quinn and S. Weinberg, Phys. Rev.
Lett. ${\underline{33}}$ (1974) 451.}}

\def\GINS{\rrr\GINS{P. Ginsparg, \plt197 (1987) 139.}}

\def\DFKZ{\rrr\DFKZ{J.P. Derendinger, S. Ferrara, C. Kounnas and
F. Zwirner,{\it ``On loop corrections to string effective field theories:
         field-dependent gauge couplings and sigma-model anomalies'',}
        preprint CERN-TH.6004/91, LPTENS 91-4 (revised version) (1991).}}

\def\LOUIS{\rrr\LOUIS{J. Louis, {\it
         ``Non-harmonic gauge coupling constants in supersymmetry
         and superstring theory'',} preprint SLAC-PUB-5527 (1991);
         V. Kaplunovsky and J. Louis, as quoted in J. Louis,
         SLAC-PUB-5527 (1991).}}

\def\DIN{\rrr\DIN{J.P. Derendinger, L.E. Ib\'a\~nez and H.P Nilles,
        \nup267 (1986) 365.}}

\def\DHVW{\rrr\DHVW{L. Dixon, J. Harvey, C.~Vafa and E.~Witten,
         \nup261 (1985) 651;
        \nup274 (1986) 285.}}

\def\DKLB{\rrr\DKLB{L. Dixon, V. Kaplunovsky and J. Louis,
         \nup355 (1991) 649.}}

\def\DKLA{\rrr\DKLA{L. Dixon, V. Kaplunovsky and J. Louis, \nup329 (1990)
            27.}}

\def\ALOS{\rrr\ALOS{E. Alvarez and M.A.R. Osorio, \prv40 (1989) 1150.}}

\def\FILQ{\rrr\FILQ{A. Font, L.E. Ib\'a\~nez, D. L\"ust and F. Quevedo,
           \plt245 (1990) 401.}}

\def\CFILQ{\rrr\CFILQ{M. Cvetic, A. Font, L.E.
           Ib\'a\~nez, D. L\"ust and F. Quevedo, \nup361 (1991) 194.}}

\def\FILQ{\rrr\FILQ{A. Font, L.E. Ib\'a\~nez, D. L\"ust and F. Quevedo,
           \plt245 (1990) 401.}}

\def\DUAGAU{\rrr\DUAGAU{
           H.P. Nilles and M.
           Olechowski, \plt248 (1990) 268; P. Binetruy and M.K.
           Gaillard, \plt253 (1991) 119; J. Louis,
    {\it ``Status of Supersymmetry Breaking in String Theory''},
           SLAC-PUB-5645
           (1991);
            D. L\"ust and T.R. Taylor, \plt253 (1991) 335;
           B. Carlos, J. Casas and C. Mu\~noz, \plt263 (1991) 248;
           S. Kalara, J. Lopez and D. Nanopoulos,
       {\it ``Gauge and Matter Condensates in Realistic String
       Models''},
            preprint CTP-TAMU-69/91.}}

\def\CFILQ{\rrr\CFILQ{M. Cvetic, A. Font, L.E.
           Ib\'a\~nez, D. L\"ust and F. Quevedo, \nup361 (1991) 194.}}

\def\FIQ{\rrr\FIQ{A. Font, L.E. Ib\'a\~nez and F. Quevedo,
        \plt217 (1989) 272.}}

\def\FLST{\rrr\FLST{S. Ferrara,
         D. L\"ust, A. Shapere and S. Theisen, \plt225 (1989) 363.}}

\def\FLT{\rrr\FLT
{S. Ferrara, D. L\"ust and S. Theisen, \plt233 (1989) 147.}}

\def\IBLU{\rrr\IBLU{L.E. Ib\'a\~nez and D. L\"ust,
          \plt267 (1991) 51.}}

\def\GAUGINO{\rrr\GAUGINO{J.P. Derendinger, L.E. Ib\'a\~nez and H.P. Nilles,
            \plb155 (1985) 65;
         M. Dine, R. Rohm, N. Seiberg and E. Witten, \plb156 (1985) 55.}}

\def\GHMR{\rrr\GHMR{D.J. Gross, J.A. Harvey, E. Martinec and R. Rohm,
         \prl54 (1985) 502; \nup256 (1985) 253; \nup267 (1986) 75.}}

\def\ANT{\rrr\ANT{I. Antoniadis, K.S. Narain and T.R. Taylor,
        \plt267 (1991) 37.}}

\def\WITTEF{\rrr\WITTEF{E. Witten, \plb155 (1985) 151.}}

\def\SCHELL{\rrr\SCHELL{A.N. Schellekens, ``Superstring
construction'', North Holland, Amsterdam (1989).}}

\def\FLIP{\rrr\FLIP{I. Antoniadis, J. Ellis, J. Hagelin and
D.V. Nanopoulos, \plt231 (1989) 65 and references therein. For a
recent review see J. Lopez and D.V. Nanopoulos,
                  Texas A\& M preprint CTP-TAMU-76/91 (1991).}}

\def\OTH{\rrr\OTH{D. Bailin, A. Love and S. Thomas, \plb188 (1987)
193; \plt194 (1987) 385; B. Nilsson, P. Roberts and P. Salomonson,
\plt222 (1989) 35;
J.A. Casas, E.K. Katehou and C. Mu\~noz, \nup317 (1989) 171;
J.A. Casas and C. Mu\~noz, \plt209 (1988) 214, \plt212 (1988) 343
J.A. Casas, F. Gomez and C. Mu\~noz, \plt251 (1990) 99;
A. Chamseddine and J.P. Derendinger, \nup301 (1988) 381;
A. Chamseddine and M. Quiros, \plb212 (1988) 343, \nup316 (1989) 101;
T. Burwick, R. Kaiser and H. M\"uller, \nup355 (1991) 689;
Y. Katsuki, Y. Kawamura, T. Kobayashi, N. Ohtsubo,
Y. Ono and K. Tanioka, \nup341 (1990) 611.}}

\def\SCHLUS{\rrr\SCHLUS{For a review, see e.g. J. Schwarz, Caltech
preprint CALT-68-1740 (1991); D. L\"ust,  CERN preprint TH.6143/91.
}}

\def\ILR{\rrr\ILR{ L.E. Ib\'a\~nez, D. L\"ust and G.G. Ross,
\plt272 (1991) 251.}}

\def\ANTON{\rrr\ANTON{I. Antoniadis, J. Ellis, S. Kelley and
D.V. Nanopoulos, \plt271 (1991) 31.}}

\def\HIGH{\rrr\HIGH{D. Lewellen, \nup337 (1990) 61;
J.A. Schwartz, Phys.Rev. D42 (1990) 1777.}}

\def\HIGHK{\rrr\HIGHK{A. Font, L.E. Ib\'a\~nez and F. Quevedo,
\nup345 (1990) 389; J. Ellis, J. Lopez and D.V. Nanopoulos,
\plt245 (1990) 375.}}

\def\LLR{\rrr\LLR{C.H. Llewellyn-Smith, G.G. Ross and
J.F. Wheater, \nup177 (1981) 263.}}

\def\YO{\rrr\YO{ L.E. Ib\'a\~nez, {\it``Some topics
in the low energy physics from superstrings''} in proceedings of
the NATO workshop on ``Superfield Theories", Vancouver,
Canada. Plenum Press, New York (1987).}}

\def\SDUAL{\rrr\SDUAL{A. Font, L.E. Ib\'a\~nez, D. L\"ust
and F. Quevedo, \plb249 (1990) 35.}}

\def\MALLOR{\rrr\MALLOR{For a recent review see L.E. Ib\'a\~nez,
{\it ``Beyond the Standard Model (yet again)''}, CERN preprint
TH.5982/91, to appear in the Proceedings of the 1990 CERN
School of Physics, Mallorca (1990).}}

\def\FIQS{\rrr\FIQS{A. Font, L.E. Ib\'a\~nez, F. Quevedo and
A. Sierra, \nup337 (1990) 119.}}

\def\INO{\rrr\INO{K. Inoue et al., Prog.Theor.Phys. {\bf 68} (1982) 927;
L.E. Ib\'a\~nez, \nup218 (1983) 514;
L.E. Ib\'a\~nez and C. L\'opez, \plb126 (1983) 54; \nup233 (1984) 511;
L. Alvarez-Gaume, J. Polchinsky and M. Wise, \nup221
(1983) 495.}}

\def\DEG{\rrr\DEG{A. Font, L.E. Ib\'a\~nez, H.P. Nilles and
F. Quevedo, \nup307 (1988) 109.}}

\def\ZSIE{\rrr\ZSIE{A. Casas, A. de La Macorra,
M. Mondragon and C. Mu\~noz, \plt247 (1990) 50;
Y. Katsuki et al., \nup341 (1990) 611.}}

\def\CAN{\rrr\CAN{P. Candelas, X. de la Ossa, P. Green and
L. Parkes, \plt258 (1991) 118; \nup359 (1991) 21.}}

\def\HIGH{\rrr\HIGH{ A. Font, L.E. Ib\'a\~nez and F. Quevedo,
\nup345 (1990) 389.}}

\def\ELN{\rrr\ELN{J. Ellis, J.L. L\'opez and D.V. Nanopoulos,
\plt245 (1990) 375.}}

\def\KIM{\rrr\KIM{ L.E. Ib\'a\~nez, J.E. Kim, H.P. Nilles and
F. Quevedo, \plt191 (1987) 282; L.E. Ib\'a\~nez, J. Mas,
H.P. Nilles and F. Quevedo, \nup301 (1988) 157.}}

\def\CASMU{\rrr\CASMU{ J.A. Casas, E.K. Katehou and C. Mu\~noz,
\nup317 (1989) 171; J.A. Casas and C. Mu\~noz, \plt214 (1988) 63.}}

\def\FIQSA{\rrr\FIQSA{ A. Font, L.E. Ib\'a\~nez, F. Quevedo and
A. Sierra, \nup331 (1990)  421.}}

\def\CLASS{\rrr\CLASS{ J.A. Casas, M. Mondragon and C. Mu\~noz,
\plt230 (1989) 63.}}

\def\KIMB{\rrr\KIMB{ J. E. Kim, \plt207 (1988) 434; Seoul preprint
SNUHE 90/01 (1990).}}

\def\ILLT{\rrr\ILLT{ L.E. Ib\'a\~nez, W. Lerche, D. L\"ust and
S. Theisen, \nup352 (1991) 435.}}

\def\KATSUKI{\rrr\KATSUKI{Y. Katsuki, Y. Kawamura, T. Kobayashi, Y. Ono
and K. Tanioka, \plt218 (1989) 169.}}

\def\KAC{\rrr\KAC{ A. Font, L.E. Ib\`a\~nez and F. Quevedo,
\nup345 (1990) 389;
J. Ellis, J. L\'opez and D.V. Nanopoulos,  \plt245 (1990) 375.}}

\def\MODSPACE{\rrr\MODSPACE
        {S. Ferrara, C. Kounnas and M. Porrati, \plt181 (1986) 263;
         M. Cvetic, J. Louis and B. Ovrut, \plt206 (1988) 227.}}

\def\SCHELL{\rrr\SCHELL{For a review see: A.N. Schellekens, {\it
``Superstring Constructions''}, North Holland, Amsterdam (1989).}}

\def\CREMMER{\rrr\CREMMER{E. Cremmer, S. Ferrara, L. Girardello and
          A. Van Proeyen, \nup212 (1983) 413.}}

\def\DFKZ{\rrr\DFKZ{J.P. Derendinger, S. Ferrara, C. Kounnas and F. Zwirner,
         {\it ``On loop corrections to string effective field theories:
         field-dependent gauge couplings and sigma-model anomalies'',}
        preprint CERN-TH.6004/91, LPTENS 91-4 (revised version) (1991),
          \plt271 (1991) 307.}}

\def\LOUIS{\rrr\LOUIS{J. Louis, {\it
         ``Non-harmonic gauge coupling constants in supersymmetry
         and superstring theory'',} preprint SLAC-PUB-5527 (1991);
         V. Kaplunovsky and J. Louis, as quoted in J. Louis.}}

\def\KL{\rrr\KL{V. Kaplunovsky and J. Louis, paper to appear.}}

\def\DIXON{\rrr\DIXON{L. Dixon, talk given at the Berkeley conference,
            May 1991.}}

\def\FKLZ{\rrr\FKLZ{S. Ferrara, C. Kounnas, D. L\"ust and F. Zwirner,
           \nup365 (1991) 431.}}

\def\LAUER{\rrr\LAUER{
J. Lauer, J. Mas and H.P. Nilles, \plt226 (1989) 251;
            \nup351 (1991) 353;
     E.J. Chun, J. Mas, J. Lauer and H.-P. Nilles, \plt233 (1989) 141.}}

\def\MOD{\rrr\MOD
        {R.~Dijkgraaf, E.~Verlinde and H.~Verlinde, \cmp115 (1988) 649;
        {\it ``On moduli spaces of conformal field theories with $c\geq
        1$'',} preprint THU-87/30;
           A. Shapere and F. Wilczek, \nup320 (1989) 669.}}

\def\DUALALL{\rrr\DUALALL{
       V.P. Nair, A. Shapere, A. Strominger and F. Wilczek,
           \nup287 (1987) 402;   B. Sathiapalan, \prl58 (1987) 1597;
     J.J. Atick and E. Witten, \nup310 (1988) 291;
              R. Brandenberger and C. Vafa, \nup316 (1989) 391;
            A. Giveon, E. Rabinovici and G. Veneziano, \nup322 (1989)
            167;
            M. Dine, P. Huet and N. Seiberg, \nup322 (1989) 301;
            J. Molera and B. Ovrut, \prv40 (1989) 1146;
            M. Duff, \nup335 (1990) 610;
            E. Alvarez and M.A.R. Osorio, \prv40 (1989) 1150;
            W. Lerche, D. L\"ust and N.P. Warner, \plt231 (1989) 417;
                    M. Cvetic, J. Molera and B. Ovrut, \plt248 (1990)
          83.}}

\def\DUAL{\rrr\DUAL{K. Kikkawa and M. Yamasaki, \plb149 (1984) 357;
      N. Sakai and I. Senda, Progr. Theor. Phys. {\bf 75} (1986) 692.}}

\def\CAOV{\rrr\CAOV{G. Lopes Cardoso and B. Ovrut, \nup369 (1992) 351;
                                {\it ``Sigma Model Anomalies,
           Non-Harmonic Gauge and Gravitational
           Couplings and String Theory,''}
           preprint UPR-0481T (1991).
            }}

\def\KAPLUG{\rrr\KAPLUG{V. Kaplunosvky, unpublished notes.}}

\def\LOLU{\rrr\LOLU{J. Louis and D. L\"ust, work in progress.}}

\def\CHSW{\rrr\CHSW{P. Candelas, G. Horowitz, A. Strominger and
           E. Witten, \nup258 (1985) 46.}}

\def\FERRA{\rrr\FERRA{S. Ferrara, {\it ``Geometry and Quantum Symmetries
of Superstring Vacua,''} lecture given at the International School
of Subnuclear Physics, Erice 1991,
preprint CERN-TH.6213/91.}}

\def\GS{\rrr\GS{M.B. Green and J.H. Schwarz, \plb149 (1984) 117.}}

\def\SCHEWA{\rrr\SCHEWA{A.N. Schellekens and N.P. Warner, \nup287
(1987) 317.}}

\def\LM{\rrr\LM{D. L\"ust and C. Mu\~noz, {\it ``Duality-Invariant
Gaugino Condensation and One-loop Corrected K\"ahler Potentials
in String Theory''}, CERN-TH.6358/91 (1991).}}

\def\ANOMA{\rrr\ANOMA{G. Moore and P. Nelson,
\prl53 (1984) 1519;
L. Alvarez-Gaum\'e and P. Ginsparg, \nup262 (1985) 439;
J. Bagger, D. Nemeschansky and S. Yankielowicz, \nup262 (1985) 478;
A. Manohar, G. Moore and P. Nelson, \plb152 (1985) 68;
W. Buchm\"uller and W. Lerche, Ann. Phys. {\bf 175} (1987) 159.}}

\def\WITTEF{\rrr\WITTEF{E. Witten, \plb155 (1985) 151.}}

\def\boxit#1{\vbox{\hrule\hbox{\vrule\kern3pt\vbox{\kern4pt#1\kern4pt}
\kern3pt\vrule}\hrule}}
\def\lozenge{\boxit{\hbox to 1.1pt{%
             \vrule height 1pt width 0pt \hfill}}}


\def\pubnum{
\hbox{CERN-TH.6380}}

\def\pdate{January 1992}

\titlepage
\title
{Duality Anomaly Cancellation, Minimal String Unification and
the Effective Low-Energy Lagrangian of 4-D Strings}
\vskip-.8cm
\author{{\bf Luis E. Ib\'a\~nez}}
\andauthor{{\bf Dieter L\"ust}\foot{Heisenberg Fellow}}
\CERN
\vskip-.8 cm
\abstract
{We present a systematic study of the constraints coming from
target-space duality and the associated
duality anomaly cancellations on orbifold-like 4-D strings.
             A prominent role is played by the modular weights
of the massless fields. We present a general classification of
all possible modular weights of massless fields in Abelian
orbifolds. We show that the cancellation of modular anomalies
strongly constrains the massless fermion content of the theory,
in close analogy with the standard ABJ anomalies.
We emphasize the validity of this approach not only for
(2,2) orbifolds but for (0,2) models with and without
Wilson lines.
As an application
one can show that one cannot build a ${\bf Z}_3$ or ${\bf Z}_7$ orbifold
whose massless charged sector with respect to the (level one) gauge group
$SU(3)\times SU(2)
\times U(1)$ is that of the minimal supersymmetric standard model,
since any such model would necessarily have
duality anomalies.
A general study of those constraints for Abelian orbifolds is
presented. Duality anomalies are also related to the computation
of string threshold corrections to gauge coupling constants. We
present an analysis of the possible relevance of those threshold
corrections to the computation of $\sin^2\theta_W$ and $\alpha_3$
for all Abelian orbifolds. Some particular {\it minimal} scenarios,
namely those based on all  ${\bf Z}_N$ orbifolds except ${\bf Z}_6$
and ${\bf Z}_8'$,
are ruled
out on the basis of these constraints. Finally we discuss the
explicit dependence of the SUSY-breaking soft terms on the
modular weights of the physical fields. We find that those terms
are in general not universal. In some cases specific relationships
for gaugino and scalar masses are found.}

\endpage
%

\chap{ Introduction}

Four-dimensional superstrings \SCHELL\
constitute at the moment the best
candidates for unification of all interactions. In trying to
use these theories to describe the observed physics, it is of
fundamental importance to obtain the effective low-energy
field theory of each given 4-D string. Much progress along
these lines has been achieved in the last few years and there is
at present a good knowledge of the form of the effective
Lagrangian for the case of orbifold-like 4-D strings.

One of the general features found is that the kinetic term
of the matter fields is not canonical but presents
a $\sigma $-model structure, which depends on the moduli fields
of the orbifold variety. Furthermore, the effective Lagrangian is
in general invariant under certain duality symmetries acting on the
moduli space. In many cases, in particular orbifolds,
these symmetries always
contain the subgroup $SL(2,{\bf Z})^3$ \MOD,
i.e. three copies of the
modular group acting on the moduli of each of the three orbifold
complex planes. The required target space modular invariance provides
a connection between the effective Lagrangian and the theory of
modular functions \FLST.
The effective tree-level Lagrangian of orbifolds
has been found to be invariant under these symmetries and
the Yukawa couplings to transform as modular functions of definite
modular weight \refs{\LAUER,\FLT}.
The same invariance has been checked in
one-loop string computations \refs{\DKLB,\ANT}
and indeed it is expected on
general grounds that these modular symmetries will survive
non-perturbative effects and that they may be broken only
spontaneously, but not explicitly \FILQ. This is important because
it implies constraints on the possible form of non-perturbative
string effects, as remarked in refs.\refs{\FLST,\FILQ,\MAGN,\CFILQ}.
In the present paper we analyze several different contexts in which
the $SL(2,{\bf Z})^3$ symmetry of orbifold 4-D strings constrains
the low-energy effective field theory. The  crucial ingredient
in these constraints is the cancellation of ``duality anomalies''.

The structure of the paper is as follows. In section two, after
discussing the duality-invariant       structure of the
effective low-energy Lagrangian of 4-D strings, we concentrate
on the orbifold case. We classify all the possible modular
weights for matter fields in arbitrary Abelian ${\bf Z}_N$
\refs{\DHVW,\IMNQ}
and ${\bf Z}_N\times{\bf Z}_M$ \ZNZM\
orbifolds. We also show how the available ranges
for modular weights depend on the Kac--Moody level of the gauge algebras.

In section three we discuss the constraints on the massless particle
content of 4-D strings coming
from cancellation of duality anomalies
\multref\DFKZ{\LOUIS\CAOV}\IBLU.
Indeed, the $SL(2,{\bf Z})^3$ duality transformations induce
chiral $U(1)$ rotations on the massless fermions. The associated
current has to be anomaly-free in order to preserve the duality
symmetries. The anomaly may be cancelled in two possible ways:
either through a sort of Green--Schwarz mechanism \GS\ or through
contributions coming  from the superheavy string spectrum. In some
particular cases only the first mechanism is available  and
the particular universal (gauge group independent) structure of
the Green--Schwarz mechanism forces the massless fermion content
of the theory to be remarkably constrained.
This would have very limited applications if it were only true
for the few existing Abelian (2,2) orbifolds. In fact we emphasize
that it also applies to the millions of possible (0,2) Abelian
orbifolds one can construct with or without Wilson lines.
This we have explicitely checked in a large number
of such (0,2) models.
We present a general
study of such constraints in Abelian orbifolds. As an example
of the power of the duality anomaly cancellation conditions we
are able to show that one cannot possibly build a ${\bf Z}_3$
or ${\bf Z}_7$
4-D string whose massless charged sector with respect to
$SU(3)\times SU(2)\times U(1)$ is that of the minimal
supersymmetric standard model. We also discuss constraints
from the cancellation of mixed duality-gravitational anomalies.

The mixed duality-gauge anomaly is related through supersymmetry
to one-loop moduli-dependent corrections
\refs{\KAPLU,\DKLB}  to the gauge
coupling constants of the effective field theory.
In many models the one-loop threshold
effects, coming from massive string modes,
provide an additional moduli dependence
to the running of the
gauge coupling constants. In section four we study the possible
role of these threshold effects in rendering values for
$\sin^2\theta_W$ and $\alpha_3$ consistent with the experimental
results. Indeed, if we restrict ourselves to the minimal
particle content of the supersymmetric standard model, we find
that there is no possible (0,2) ${\bf Z}_N$ orbifold apart from
${\bf Z}_6$, ${\bf Z}_8'$ yielding
the correct values for those quantities unless, the gauge groups
are realized at higher Kac--Moody level. On the other hand,
agreement can be obtained
in ${\bf Z}_N\times {\bf Z}_M$ orbifolds
(even with lowest Kac--Moody levels).
This is a generalization of the analysis of ref.\ILR\  to include
the three orbifold complex planes and making use of the
general classification of modular weights presented in section two.
For completeness we also discuss the case in which one modifies
the minimal particle content of the standard model to include
extra massless multiplets.

In section five we discuss some constraints on SUSY-breaking
soft terms from duality-invariant actions. We do not in fact
rely on any specific assumption about how supersymmetry is
broken but simply assume that the auxiliary fields of the
marginal deformation fields (the dilaton $S$ and the three
complex untwisted moduli $T_i$) are non-vanishing and provide
the seed for SUSY-breaking. This is indeed what happens in all
supersymmetry-breaking scenarios discussed up to now.
If this is the case one can write down certain constraints
amongst the physical gaugino masses depending on the
modular weights of the massless fields. Similar though more
model-dependent expressions may be found for the soft
SUSY-breaking scalar masses. One important point      is the
lack of universality of these soft terms, unlike the usual
assumptions made in the standard minimal low-energy
supersymmetry models. Some final comments and conclusions are
left for the sixth section.

\chap{Duality-Invariant Effective Field Theory}
\sect{General supersymmetric 4-D string compactications}

Let us define the basic string degrees of freedom which appear in
the low-energy effective $N=1$ supergravity action \CREMMER\
of four-dimensional
strings
and which are needed
for our discussion.  In the following we will consider compactifications
of the ten-dimensional heterotic string \GHMR\
on a six-dimensional
(smooth) Calabi--Yau space \CHSW\ or on a six-dimensional orbifold.
(We will in general {\it not assume} that
we deal with (2,2) compactifications.)
First, there are moduli fields which describe the allowed deformations
of a given background. In the context
of the underlying conformal field theory the moduli are the coupling
constants related to the truly marginal operators which can be added
to the two-dimensional world-sheet action without destroying the
conformal invariance of the theory. These moduli parameters correspond
in the effective low-energy theory to chiral matter fields $M_i(x)$
(we will denote both the chiral superfields and their scalar components
by the same symbols) with
completely flat potential, i.e. with
undetermined vacuum expectation values ($vev's$)
(at least if one disregards possible non-perturbative (infinite genus)
string effects).
The moduli take their values in the so-called moduli space ${\cal M}$,
$M_i\in{\cal M}$, whose metric is determined by the two-point
function of the associated marginal  operators.
Locally, the moduli space
is a smooth K\"ahler
manifold. However it is very important to realize that there
may exist discrete reparametrizations of the form
$$M_i\rightarrow\tilde M_i(M_i)\eqn\repar
$$
which change the geometry of the internal space
but     leave the massive string spectrum and the
underlying conformal field theory invariant.
These are the famous target space duality transformations
\refs{\DUAL,\MOD,\DUALALL,\CAN}
which form
the discrete duality group $\Gamma$. The duality invariance is an
entirely stringy phenomenon and is not present in the field
theory compactifications. The most prominent example is the
$R\rightarrow 1/R$ duality symmetry \DUAL\ for closed string
compactification on a circle originating from the existence
of topologically
stable winding states. The target space duality invariance
implies that certain points in ${\cal M}$ have to be identified,
which means that the moduli space of the conformal field theory
is a K\"ahler orbifold and not a smooth K\"ahler manifold:
${\cal M}_{\rm CFT}={\cal M}_{\rm field~theory}/\Gamma$.

In addition to the moduli fields there are in general chiral matter fields
$A_\alpha(x)$ with non-flat potential such that $\langle A_\alpha
\rangle=0$.
Furthermore there are $N=1$ vector supermultiplets with
gauge bosons corresponding to a
gauge group $G=\prod G_a$. In general, both the matter fields
$A_m$ as well as the moduli fields $M_i$ are gauge non-singlets
and build representations $\sum \underline R_a^\alpha$,
$\sum \underline R_a^i$ respectively.
It follows that at least some part of the gauge group is generically
spontaneously broken by the moduli $vev's$. Only at special
points in the moduli space the unbroken part of the gauge group
gets enhanced. (Very often these points are given by orbifold
points of ${\cal M}$.)
Finally there is the gravitational
supermultiplet together with the chiral dilaton--axion
field commonly denoted by $S$. This field can be equivalently
described by a linear supermultiplet.

Let us now describe the string tree-level
low-energy action involving the moduli fields.
Since     all $S$-matrix elements computed in the underlying
conformal field theory are invariant under the discrete
target space duality transformations, it follows that also
the corresponding effective action of the moduli fields must be duality
invariant. (For recent summaries on these issues see
refs.\multref\LUE{\VALEN}\FERRA.)
The kinetic energy of the moduli
is given by a 4-dimensional supersymmetric non linear $\sigma$-model
with the moduli space ${\cal M}$ as target space and is therefore
determined by the K\"ahler potential
$K(M_i,\bar M_i)$ leading to the
K\"ahler metric $K_{ij}(M_i,\bar M_i)=\partial K/\partial M_i\partial
\bar M_j$: ${\cal L}_{\rm kin}=K_{ij}
\partial_\mu M_i\partial^\mu\bar M_j$.
Under the discrete target space duality transformations, being
just discrete symmetries of ${\cal M}$, the K\"ahler metric
transforms as
$$K_{ij}(M_i,\bar M_i)\rightarrow
f_{ik}(M_i)^{-1}\bar f_{jl}(\bar M_i)^{-1}
K_{kl}(M_i,\bar M_i)
\eqn\metrans
$$
where the holomorphic Jacobian is given as
$f_{ik}(M_i)={\partial\tilde M_i\over \partial M_k}$.
On the other hand, the K\"ahler potential is in general not
invariant under the discrete duality transformations. They
will act as (U(1)) K\"ahler transformations on $K$, i.e.
$$K(M_i,\bar M_i)\rightarrow K(M_i,\bar M_i)+g(M_i)+\bar g(\bar M_i)
,\eqn\kptrans
$$
where $g(M_i)$ is a holomorphic function of the moduli fields.

The kinetic energy of the matter fields is in general a moduli
dependent function.
Expanding around the classical vacuum $\langle A_\alpha\rangle=0$,
the K\"ahler potential for the matter fields in lowest order in
$A_\alpha$ looks like:
$$K^{\rm matter}=K_{\alpha\beta}^{\rm matter}
(M_i,\bar M_i)A_\alpha\bar A_\beta+\dots\eqn\mattkp
$$
The discrete reparametrizations \repar\ in general induce
a change of the matter K\"ahler potential
$K_{\alpha\beta}(M_i,\bar M_i)$ of the form
$$K_{\alpha\beta}
\rightarrow h_{\alpha\gamma}(M_i)^{-1}h_{\beta\delta}
(\bar M_i)^{-1}K_{\gamma\delta}.
\eqn\mattkptr
$$
It follows that also the matter
fields possess a non-trivial transformation behaviour under $\Gamma$
transformations in order to obtain  duality-invariant
kinetic energy terms for the matter fields:
$$A_\alpha\rightarrow h_{\alpha\beta}(M_i)A_\beta.\eqn\matrans
$$

Some of the moduli are related to
the geometrical and topological data
of the six-dimensional complex space.
Specifically, these moduli fields split into two classes: first
there are the moduli $T_i$ ($i=1,\dots ,h_{(1,1)}$),
which are associated with the deformations of the
K\"ahler class. Second, we deal with deformations of the
complex structure of the compact space parametrized by moduli
$U_m$, ($m=1,\dots ,h_{(2,1)}$). ($h_{(1,1)}$ and $h_{(2,1)}$ are
the non-trivial Hodge numbers of the compact space.)
The moduli space factorizes into two distinct subspaces:
${\cal M}={\cal M}_T\otimes {\cal M}_U$. Thus the moduli
K\"ahler potential splits into the
sum $K=K(T_i,\bar T_i)+K(U_m,\bar U_m)$.
It follows that
the change of the K\"ahler potential under duality transformations
is given by two different holomorphic functions $g(T_i)$ and
$g(U_m)$.
There is however an important difference between the discrete
reparametrizations of the $T$ and $U$ fields. Specifically, the
discrete transformations acting on the complex structure moduli $U_i$
do not change at all the geometry of the underlying compact space;
these invariances are of a field-theoretical nature. On the other hand,
the moduli fields $T_i$ are related with the size of the internal
compact space. The existence of possible discrete reparametrization
invariances in the $T$--field sector is an entirely stringy phenomenon
and can be viewed as the generalization of the $R\rightarrow 1/R$
duality symmetry. Thus the stringy duality group is given by $\Gamma_T$
and ${\cal M}_{\rm CFT}={\cal M}_{\rm field~theory}/\Gamma_T$.
Considering only the overall modulus $T$ whose real part
determines the overall size of the compact space, the K\"ahler
potential for $T$ becomes in the large radius limit
\WITTEF
$$K=-3\log(T+\bar T).\eqn\overallt
$$
However,
for small values of the $T$--fields, formula \overallt\
gets in general corrected by world-sheet instanton effects. These
corrections determine the explicit form of the duality group $\Gamma_T$.

Considering the particular case of (2,2) compactifications,
the matter fields are in a one-to-one correspondence with the  moduli.
Specifically, there exist $h_{(1,1)}$ matter fields
$A_i$
in the ${\underline{27}}$ representation and $h_{(2,1)}$ matter fields
$A_m$ in the ${\underline{\bar{27}}}$ representation of $E_6$.
(Note that some of these ``matter'' fields may actually have a completely
flat potential such that they are also (Wilson) line moduli
whose non-zero $vev's$ break the gauge group $E_6$ and destroy the (2,2)
superconformal symmetry, see the discussion in section 3.6.)
For the compactifications on a smooth Calabi--Yau space,
the K\"ahler metric of these matter fields is related to
the moduli metric in a special way \DKLA:
$$\eqalign{&
K_{ij}^{\rm matter}=e^{-(K(T_i)-K(U_m))/3}K_{ij}(T_i,\bar T_i)
,\cr &
K_{mn}^{\rm matter}=e^{-(K(U_m)-K(T_i))/3}K_{mn}(U_m,\bar U_m).\cr}
\eqn\special
$$
It follows that the matter metrics transform under target space duality
transformations with a combined reparametrization and K\"ahler
transformation, i.e.
$$\Gamma_T:\quad K_{ij}^{\rm matter}\rightarrow |e^{-g(T_i)/3}|^2
f_{ik}(T_i)^{-1}\bar f_{jl}(\bar T_i)^{-1}K_{kl}^{\rm matter}\eqn\comb
$$
and likewise for $K_{mn}^{\rm matter}$. The transformation eq.\comb\
implies that the matter fields transform like
$$\Gamma_T:\quad A_i\rightarrow e^{g(T_i)/3}
f_{ij}(T_i)A_j.\eqn\comba
$$
Also note that in the large radius limit both types of
matter metrics in eq.\special\ display
a particularly simple, universal
dependence on the overall $T$--field:
$$K^{\rm matter}\sim{1\over T+\bar T}.\eqn\matovera
$$

\sect{(0,2) Abelian orbifolds}

In the following we will discuss symmetric
${\bf Z}_M$ and ${\bf Z}_M\times {\bf Z}_N$ orbifolds. This
discussion will be in a certain sense also more general
than the above Calabi--Yau considerations since, it
will not be restricted to the case of (2,2) orbifolds. All
our formulas will be valid also for (0,2) models with non-standard
gauge embeddings and/or with the presence of Wilson lines. These
compactifications include some examples that are of phenomenological
interest since
the gauge group can be different from $E_6\times E_8$. In fact,
there exist models with standard model gauge group $G=SU(3)\times
SU(2)\times U(1)$ and three generations plus additional vector-like
matter fields \multref\KIM{\FIQSA}\CASMU.
Every orbifold of this type has three complex planes,
and each orbifold
twist $\vec\theta=(\theta_1,\theta_2,\theta_3)$ acts either
simultaneously on two or all three planes.
In the following we will consider the dependence of the effective action
on the untwisted moduli fields. The twisted moduli fields,
whose non-trivial $vev's$ blow up the orbifold singularities leading
to a smooth Calabi--Yau space, will be treated as matter fields.
Furthermore, the twisted moduli are generically only present for the
unique (2,2) embedding of each orbifold.
The untwisted moduli split into $T$ and $U$ fields.
All orbifolds have $h_{(1,1)}=3,5$ or 9 respectively,
where the three generic
$T_i$--fields ($i=1,2,3$)
describe the size of the three complex planes. The possible
additional fields correspond to the relative shape of the three planes.
On the other hand, the number of
$U$-fields, with $h_{(2,1)}=0,1$ or 3,
depends on whether the orbifold twists are compatible with the
freedom of varying the complex structure of the three subtori.
Specifically, there are six different
cases of untwisted orbifold moduli spaces \MODSPACE:
$$(1)\qquad h_{(1,1)}=3,
\,\,\, h_{(2,1)}\!=\!0,1,3:\,\,\,
{\cal M} \! = \! \Biggl\lbrack{SU(1,1)
\over U(1)}\Biggr\rbrack^3_T \!\! \otimes
\Biggl\lbrack{SU(1,1)
\over U(1)}\Biggr\rbrack^{h_{(2,1)}}_U.
\eqn\casea
$$
$$\eqalign{(2)\qquad h_{(1,1)}=5,
\,\,\, &h_{(2,1)}\!=\!0,1:\cr
&{\cal M}=\Biggl\lbrack
{SU(1,1)\over U(1)}\otimes
{SU(2,2)\over SU(2)\times SU(2)\times U(1)}
\Biggr\rbrack_T\otimes\Biggl\lbrack{SU(1,1)\over U(1)}
\Biggr\rbrack^{h_{(2,1)}}_U.\cr}
\eqn\caseb
$$
$$(3) \qquad h_{(1,1)}=9,\,\,\, h_{(2,1)}=0:\,\,\,
{\cal M}
=\Biggl\lbrack{SU(3,3)
\over SU(3)\times SU(3)\times U(1)}\Biggr\rbrack_T .\eqn\casec
$$
The corresponding K\"ahler potentials are given by
$$
(1)\qquad
h_{(1,1)}=3,\,\,\, h_{(2,1)}\!=\!0,1,3:\,\,\,
K=-\sum_{i=1}^3\log(T_i+\bar T_i)-\sum_{m=1}^{h_{(2,1)}}
\log(U_m+\bar
U_m).
\eqn\kcasea
$$
$$\eqalign{
(2)\qquad h_{(1,1)}=5,\,\,\, &h_{(2,2)}=0,1:\cr
&K=-\log(T_1+\bar T_1)-\log\det(T_{ij}+\bar T_{ij})
-\sum_{m=1}^{h_{(2,1)}}\log(U_m+\bar U_m). \cr}
\eqn\kcaseb
$$
$$
(3) \qquad h_{(1,1)}=9,\,\,\, h_{(2,1)}=0:\,\,\,
K=-\log\det(T_{ij}+\bar T_{ij}).\eqn\kcasec
$$
(In order to keep contact with the previous notation we define
$T_i=T_{ii}$ ($i=2,3$ for case (2) and $i=1,2,3$ for case (3).)
For the overall modulus field,
which is obtained by identifying the three generic $T$--fields,
i.e. $T=T_1=T_2=T_3$, and setting $T_{ij}=0$ ($i\neq j$),
the above formulae exactly reproduce $K=-3\log(T+\bar T)$
(cf. eq.\overallt).
The correspondence with the Abelian orbifolds is the following
$$\eqalign{(1)\qquad h_{(1,1)}=3,\,\,\, &h_{(2,1)}=3:\,\,\,
{\bf Z}_2\times {\bf Z}_2
\cr &h_{(2,1)}=1:\,\,\, {\bf Z}_6,{\bf Z}_8,{\bf Z}_4\times {\bf Z}_2,
{\bf Z}_6\times {\bf Z}_2,{\bf Z}_{12}\cr
& h_{(2,1)}=0: \,\,\, {\bf Z}_7,{\bf Z}_8',
{\bf Z}_{12}',{\bf Z}_6'\times {\bf Z}_2,{\bf Z}_3\times {\bf Z}_3,
\cr &\qquad\qquad {\bf Z}_6
\times {\bf Z}_3,{\bf Z}_4\times {\bf Z}_4,{\bf Z}_6\times {\bf Z}_6\cr
(2)\qquad h_{(1,1)}=5,\,\,\, &h_{(2,1)}=1:\,\,\,  {\bf Z}_4\cr
\,\,\,&h_{(2,1)}=0:\,\,\,  {\bf Z}_6'\cr
(3)\qquad h_{(1,1)}=9,\,\,\, &h_{(2,1)}=0:\,\,\, {\bf Z}_3\cr}\eqn\orbi
$$
The corresponding \ZNZM\
twist vectors $\vec\theta$, which define each of the
above ${\bf Z}_N$ or ${\bf Z}_N\times{\bf Z}_M$ orbifold, are
shown in the the second column of table 1.
Note that
a necessary (but not sufficient) condition
for the presence of
the complex structure modulus $U_m$ is that the
corresponding $m^{\rm th}$ complex plane is left unrotated
by at least one of the orbifold twists. An analogous statement
is also true for the $T$--fields:
for case (2) ((3)), the two (three) complex planes which are
associated with the four (nine) moduli fields $T_{ij}$
are rotated by all of the orbifold twists.
Notice that the above classification of orbifold moduli also
applies to the ${\bf Z}_N\times {\bf Z}_M$ orbifolds in the presence
of discrete torsion described in ref.\ZNZM.

The target space duality groups corresponding to the above six different
cases have the following form (up to possible permutation symmetries):
$$
(1)\qquad h_{(1,1)}=3,\,\,\, h_{(2,1)}=0,1,3:\,\,\,
\Gamma=\lbrack SL(2,{\bf Z})\rbrack^3_T\times
\lbrack SL(2,{\bf Z})\rbrack^{h_{(2,1)}}_U
.\eqn\gcasea
$$
$$
(2)\qquad h_{(1,1)}=5,\,\,\, h_{(2,1)}=0,1:\,\,\,
\Gamma=\lbrack SU(2,2,{\bf Z})\times SL(2,{\bf Z})\rbrack_T
\times \lbrack SL(2,{\bf Z})\rbrack^{h_{(2,1)}}_U
.\eqn\gcaseb
$$
$$
(3)\qquad h_{(1,1)}=9,\,\,\, h_{(2,1)}=0:\,\,\,
\Gamma=SU(3,3,{\bf Z}).\eqn\gcasec
$$
Considering only the three generic `diagonal' fields $T_i$ with
K\"ahler potential as shown in eq.\kcasea,
the
target space duality group (up to permutations)
is simply given by the product of three
modular groups, $\Gamma_T=\lbrack SL(2,{\bf Z})\rbrack^3$,
acting on the three moduli as
$$T_i\rightarrow{a_iT_i-ib_i\over ic_iT_i+d_i},\eqn\mod
$$
with $a_i,b_i,c_i,d_i\in{\bf Z}$ and $a_id_i-b_ic_i=1$.

The K\"ahler potential of the matter fields $A_\alpha$ (which include,
as already mentioned,
also possibly existing twisted moduli) has a particularly
simple dependence on the three generic moduli $T_i$\foot{We omit to
discuss the dependence on the off-diagonal moduli since these
fields will not enter the threshold contributions discussed in the next
sections. Alternatively, we can regard the fields $T_{ij}$ ($i\neq j$)
also as matter fields since the K\"ahler potentials
\kcaseb\ and \kcasec\ expanded to lowest order
in these fields have the form of eq.(2.24).}
and on the three model-dependent fields $U_m$:
$$K^{\rm matter}=\delta_{\alpha\beta}
\prod_{i=1}^3(T_i+\bar T_i)^{n^i_\alpha}
\prod_{m=1}^{h_{(2,1)}}(U_m+\bar U_m)^{l^m_\alpha}A_\alpha\bar A_\beta\eqn
\mattkpor
$$
Thus each matter field $A_\alpha$ is characterized by $3+h_{(2,1)}$
rational numbers which we will collect for convenience into two
vectors $\vec n_\alpha=(n_\alpha^1,n_\alpha^2,n_\alpha^3)$ and
$\vec l_\alpha=l_\alpha^m$ ($m=1,\dots ,h_{(2,1)}$).
Invariance of the matter kinetic energies under the target space
modular group $\Gamma=\lbrack SL(2,{\bf Z})\rbrack^3_T\times\lbrack
SL(2,{\bf Z})\rbrack^{h_{(2,1)}}_U$ requires the following
transformation behaviour of the matter fields (up to constant matrices):
$$A_\alpha\rightarrow A_\alpha \prod_{i=1}^3(ic_iT_i+d_i)^{n_\alpha^i}
\prod_{m=1}^{h_{(2,1)}}(ic_mU_m+d_m)^{l_\alpha^m}.
\eqn\matorbitr
$$
Therefore, in the following, we will call the numbers
$n_\alpha^i$ and $l_\alpha^m$ the {\it modular weights} of
the matter fields.
Comparing with the formulae
for smooth (2,2) Calabi--Yau
spaces we recognize that the orbifold K\"ahler potential eq.\mattkpor\
is in general different from eqs.\special,\matovera\ using the
moduli K\"ahler potential eq.\kcasea.
However for the special case $\vec n_\alpha=-\vec e_i$ ($\vec e_i$ are
the three
unit vectors) the orbifold matter kinetic energy reproduces
the large radius limit of Calabi--Yau spaces eq.\matovera\
upon identification of the three $T$--fields, i.e. $T=T_1=T_2=T_3$.
As we will discuss now, this situation is exactly true for the untwisted
matter fields.

Let us now discuss what the modular weight vectors $\vec n_\alpha$
and $\vec l_\alpha$ are in the case of Abelian orbifolds. One can
immediately derive those from eq.\mattkpor \ and the knowledge
of the corresponding metrics of the matter fields (the pure gauge
and gravitational fields    are modular invariant). The metric of
untwisted and twisted matter fields were computed (to first
order in the matter fields) in ref.\DKLA. For
the untwisted matter fields associated to the $j^{\rm th}$ complex plane,
$A^{\rm untw}=A_j$, one finds
$$n_j^i=-\delta_j^i, \qquad l_j^i=-\delta_j^i.\eqn\untweigh
$$
States in the twisted sectors are created by the twist fields
$\sigma_{\vec\theta}(\bar z,z)$ which are associated with
an order $N$ twist vector $\vec\theta=(\theta^1,\theta^2,\theta^3)$
($0\leq\theta^i<1$, $\sum_{i=1}^3\theta^i=1$) acting
non-trivially on two or all three complex planes.
The knowledge of the metric of these fields gives the
result for the modular weights
$$
\eqalign{&n_{\alpha }^i\ =\ -(1\ -\ \theta ^i),\qquad
l_{\alpha }^i\ =\ -(1\ -\ \theta ^i),\qquad \theta ^i\not= 0 \cr
&n_{\alpha }^i\ =\ l_{\alpha }^i\ =\ 0,\qquad \theta ^i\ =\ 0.\cr}
                \eqn\tmodwei
$$
In the twisted sector of the theory, there are also massless
states, which correspond to oscillators acting on the twisted vacuum.
This is the case of the twisted moduli in (2,2) orbifolds, but there are
also plenty of other twisted oscillator states both in (2,2) and (0,2)
models without that geometric interpretation.
The associated vertices contain excited twisted fields
$\tau _{\theta _i} ({\bar z},z)$  obtained upon operator
product expansions of ground state twist operators with
$$
\eqalign{&\partial _z X^i\ =\ \sum _{m\in {\bf Z}}\ \alpha _{m+\theta _i}
^i\ z^{-m-1-\theta _i}       \cr
&\partial _z {\overline X}^i\ =\ \sum _{m\in {\bf Z}}\
{\tilde {\alpha }}_{m-\theta _i}^i\ z^{-m-1+\theta _i}. \cr }
\eqn\dxdx
$$
A generic twisted oscillator state will then look like
$$\prod _i^3\prod _j^3\prod _{m_i}  \prod _{n_j}
(\alpha _{m_i+\theta _i}^i)^{p_m^i}\ ({\tilde {\alpha }}^j
_{n_j-\theta _j})
^{q_n^j}\ |\sigma >\bigotimes |{\vec P}+{\vec {\delta }}>
\eqn \oscil
$$
where $\vec {\delta }$ is the shift vector which describes the
embedding of the twist into the $E_8\times E_8$ lattice.
In ref.\DKLA\ the metric for a particular type of oscillator
states (the twisted moduli associated to the reparation of the
orbifold singularities in (2,2) models) was computed.
Examining those results and using symmetry arguments, one sees
that                an oscillator ${\alpha }^i$ changes the
modular weight in the $i^{\rm th}$
direction of the twisted ground state by $-1$.
(An explicit check of this is given in ref.\KL.)
This is the case irrespective of the order  $m$   of the oscillator.
Due to the commutation relationships $ [ {\tilde {\alpha }}_{m+\theta _i}
^i,\alpha _{n-\theta _j}^j] \ =\ \delta _{ij}\delta _{m,-n}(m+\theta _i)$
one concludes that the ${\tilde {\alpha }}$ oscillators contribute
to the modular weight with the opposite sign.
The modular weights of oscillator states thus are
$$
  \eqalign{&n_\alpha^i=-(1-\theta^i +p^i-q^i    ),\qquad
l_\alpha^i=-(1-\theta^i +q^i-p^i),\qquad \theta^i\neq 0\cr
&n_\alpha^i=l_\alpha^i=0,\qquad\theta^i=0\cr}
\eqn\tweight
$$
where $p^i=\sum _{m} p_m^i, q^i=\sum _{n} q_n^i$ count the number
of oscillators of each chirality.

Sometimes one is specially interested in the dependence on the
overall modulus $T=T_1=T_2=T_3$. The overall
modular weights of matter fields
with respect to this diagonal $SL(2,{\bf Z})$ modular group take
particularly simple forms. It is just given by $n=\sum_{i=1}^3n^i$
and from the previous equations one obtains
$$  \eqalign{
&n\ =\ -1  \qquad {\rm (untwisted)}  \cr
&n\ =\ -2\ -\ p\ +\ q  \qquad {\rm (twisted\ states\ with\ all\ planes\
rotated)}\cr
&n\ =\ -1\ -\ p\ +\ q\qquad {\rm
(twisted\ states \ with \ only \ two \ planes \ rotated )}\cr }
\eqn \unmod
$$
where $p=\sum _{i} p^i, q=\sum _{i} q^i$.

An important question is what the possible modular weights
of the massless particles are. From the above discussion,
it is clear that
this depends on the number of oscillators present in our state.
Important constraints on the maximal possible oscillator number
can be obtained from the mass formula for the (left-moving) twisted
states:
$${M_L^2\over 8}=N_{\rm osc}+h_{KM}+E_0-1.\eqn\massf
$$
Here $N_{\rm osc}$ is the fractional oscillator number and
$E_0$ is the ground state energy of the twist field $\sigma_{\vec\theta}$
given as
$$E_0=\sum_{i=1}^3{1\over 2}|v^i|(1-|v^i|),\eqn\grouen
$$
where $\vec v=v_i$ is a slightly redefined $N^{\rm th }$ order
twist vector with
$0\leq |v^i|<1$, $\sum_{i=1}^3v^i=0$.
Finally $h_{KM}$ is the contribution
to the conformal dimension of the matter
fields from the left-moving (gauge) $E_8\times E_8$ part.
In the case of level one non-Abelian gauge groups one
can    represent  this sector  in terms of a shifted lattice
and one just has
            $$h_{KM}=({\vec P}\ +\ {\vec\delta })^2/2.
\eqn \hkm
$$
More generally one can construct Abelian orbifolds with higher
Kac--Moody level algebras \HIGH. In the general case
               one can write down a general expression for $h_{KM}$
in terms of the Casimirs and levels of the relevant gauge
Kac--Moody algebras involved. For a
primary field in the representation $\underline R$ of the
gauge group $G$ associated to a Kac--Moody algebra of level $k$,
one has
 $h_{KM}={C(\underline R)\over C(G)+ k}$,
where $C(\underline R)$ ($C(G)$) is the quadratic Casimir
of the $\underline R$ (adjoint) representation of $G$;
$C(\underline R)$ is related to
$T(\underline R)$ by $C(\underline R){\rm dim}(\underline R)
=T(\underline R){\rm dim}G$ ($T(\underline R)$ is the index of the
representation $\underline R$).
In general there will be a gauge group with several group factors $G_a$
(as in the phenomenologically relevant case
$SU(3)\times SU(2)\times U(1)$) and each massless particle
transforming like $({\underline R}_1,{\underline R}_2,....)$
will have the conformal dimension
$$
h_{KM}\ =\ \sum_a {{C({\underline R}_a)}\over {C(G_a) +
k_a}}.
\eqn \hkm
$$

Notice that the contribution of the KM sector to the
conformal dimension decreases as the level increases.
On the contrary it increases as the dimension of the representation
increases, which causes the absence of massless
particles with large dimensionality in string models \refs{\HIGH,\ELN}.
In phenomenological applications, the conformal dimension of
the standard model particles will be relevant.
For convenience of the reader we show in table 2   the separate
contribution of each group to the standard model particles
$KM$ conformal dimension.
We also show the total contribution to $h_{KM}$ for two choices
of $KM$ levels consistent with GUT-like gauge coupling boundary
conditions at the string scale. They correspond to level
one and level two non-Abelian gauge factors. Notice that in the
presence of extra gauge interactions (such as e.g. extra $U(1)$'s)
there would be additional contributions to $h_{KM}$ and hence
the results in the table give a {\it lower bound} to the $KM$
conformal dimension of each standard model particle.

For a massless state with $M^2=0$ it now follows that the maximum
oscillator number is constrained by
$$N_{\rm osc}\ \leq (1\ -\ E_0\ -h_{KM}).  \eqn \maxosc
$$

Since for a given orbifold the value of $E_0$ is fixed, it is clear that
the smaller $h_{KM}$ is, the bigger $N_{\rm osc}$ will be and hence
the range of allowed modular weights. In fact, in the KM level one
case, the smallest
possible contribution to $h_{KM}$ is obtained in orbifolds
for ${\vec P}=0$ in which case $h_{KM}=({\vec {\delta }})^2/2$.
In any given Abelian orbifold one can check that the shortest
(world-sheet) modular invariant shift possible has length
$({\vec {\delta }})^2=({\vec v})^2$. Thus irrespective of the
gauge group one finds
$$
N_{\rm osc}\ \leq \ (1\ -\ {1\over 2}\sum _i\ |v_i|).
\eqn \bound
$$
In a twisted sector of order $N$ the largest possible (negative)
value for the modular weight will be obtained when there are
$p$ oscillators with the lowest possible $1/N$ fractional
modding. This number $p$ of oscillators will thus be bounded
by
$$
p\ \leq \ N\ (1\ -\ {1\over 2}\sum _i \ |v_i|). \eqn \ppp
$$
{}From this equation one can obtain the maximum number $p_{\rm max}$ of
oscillators $\alpha ^i$ present in any possible twisted sector of
Abelian orbifolds. This is shown in the second column of
table 3.
The maximum possible number $q_{\rm max}$ of ${\tilde {\alpha }}^i$
oscillators is more constrained. Such oscillators only appear
in complex planes $i$ in which $\theta _i \geq 1/2$. Only in this case
their contribution to the mass formula
may allow for massless particles. The third column of table 3 shows
$q_{\rm max}$ for the different twisted sectors. For the ${\bf Z}_3$
orbifold sectors no such type of oscillators are present in the massless
sector. For the rest of the twists with three rotated planes,
one ${\tilde {\alpha }}$ oscillator  may be present along one particular
       complex plane in each case. Each of these oscillators
contribute to the mass formula $1-\theta_1$ which is always
bigger than $1/N$. This is why only one        such oscillator
might be present for such type of twists. For twisted
sectors with one unrotated plane (last four cases in the table),
${\tilde {\alpha }}$ oscillators
are generically present in one of the rotated planes.
Of course the above arguments only guarantee that the relevant
state can be massless, but this does not mean that it is present
in the massless spectrum since it has also to survive the
generalized GSO projection of the orbifold symmetries.

{}From the second and third columns of table 3           and
eq.\tweight \ one can obtain the allowed range for the modular
weights in each of the three complex dimensions. In the case
of the overall modulus $T$, using eq.\unmod , it is easy to find
the allowed values of the modular weights. This is shown in
the fourth column of table 3. The extreme values for $n$ assume the
minimum value for $h_{KM}=\sum _i|v_i|^2/2$ corresponding to
level one non-Abelian factors. Notice that each particular
${\bf Z}_N$ or ${\bf Z}_N\times {\bf Z}_M$
orbifold is constructed by joining
in a (world-sheet) modular invariant way all the twists
generated by the vectors in the first column of table 3.
We thus observe that the modular weights with respect to the
diagonal modulus vary in an absolute range $-9\leq n\leq 4$
for Abelian ${\bf Z}_N$ and ${\bf Z}_N\times{\bf Z}_M$ orbifolds.

If the value of the Kac--Moody conformal dimension of the
relevant massless field is larger than the minimal value
discussed above, the range of possible modular weights
is substantially decreased. Consider in particular the
phenomenologically interesting case of possible massless
states describing the standard model fields. For the
standard choice of the $KM$ levels, $3/5k_1=k_2=k_3=1$,
we know we have $h_{KM}=3/5$ for $Q,U,E$-type states
and $h_{KM}=2/5$ for $L,D,H,\bar H$-type states (see table 2).
Plugging those values in eq.\maxosc , one obtains the much
more restricted values for the diagonal modulus modular weights
shown in columns 5 and 6 of table 3.
We thus see that in Abelian orbifolds
the range of overall modular weights is given by
$-3\leq n_{Q,U,E}\leq 0$ and $-5\leq n_{L,D,H,\bar H}\leq 1$
(for $3/5k_1=k_2=k_3=1$). All these constraints on the
modular weights will play an important role in the
discussions in the following sections.

\chap{Cancellation of Target Space Modular Anomalies}
\sect{The $\sigma$-model and duality anomalies in general 4-D strings}

In this section we will discuss some loop effects in four-dimensional
string theories. As is well known, gauge and gravitational anomalies
destroy the quantum consistency of ``ordinary'' field theories.
The requirement of the absence of these anomalies puts severe
constraints on the form of the massless fermionic spectrum of the
theory. Considering four-dimensional string theories, gauge and
gravitational anomalies must also be absent, leading to
the same constraints on the massless particle spectrum as compared to
field theory. They come as an automatic consequence of world-sheet
modular invariance \SCHEWA.
An important exception of this statement is given
by an anomalous $U(1)$ gauge group. In fact, the anomalous
triangle diagram with external $U(1)$ gauge bosons can arise
in string theory. The quantum consistency of the theory is still
preserved since the $U(1)$ anomaly can be cancelled by the so-called
Green--Schwarz mechanism \GS. This mechanism leads to a mixing \DSW\
between the $U(1)$ gauge boson and the axion (dilaton) field, such
that the anomalous variation of the field theory triangle diagram
is cancelled by a proper change of the dilaton field under $U(1)$
transformations.

Apart from these gauge anomalies, supersymmetric non-linear
$\sigma$-models exhibit another type of anomalies, namely the so-called
$\sigma$-model anomalies \ANOMA. These involve, similar to the local
gauge anomalies, triangle diagrams where some of
the external fields are given by composite $\sigma$-model
connections, which do not correspond to
dynamical, propagating degrees of freedom.
In the string case, the continuous
$\sigma$-model symmetries are generically broken by explicit couplings
among the massless fields, e.g. by the tree-level Yukawa couplings
or by one-loop effects. This explicit breakdown originates from the
influence of the heavy string spectrum, which is not invariant under
continuous changes of the moduli.
Therefore possible
$\sigma$-model anomalies would not
constrain the field theory spectrum further.
However, in string theory, the massive spectrum, i.e. momentum together
with `winding' states, is invariant under the discrete duality
transformations. In fact,
the $\sigma$-model anomalies
due to the massless fermions
also induce a non-invariance of the corresponding one-loop effective
action under target space duality transformations.
These ``stringy'' duality anomalies must be cancelled
since one knows that the duality symmetries are preserved in
any order of string perturbation theory.
This is in sharp contrast with normal field theory. As we
will discuss in the following, for a broad class of string
compactifications the requirement of the
absence of the target space duality anomalies provides new
constraints on the massless string spectrum which were not
discussed before. In particular, we will show that for some
orbifold compactifications the
minimal supersymmetric standard model spectrum would lead to
target space duality anomalies and has therefore to be discarded.

Specifically, consider the supersymmetric non-linear $\sigma$-model
of the moduli $M_i$ with target space ${\cal M}$ coupled to  gauge and
matter fields as described in the last section.
At the one-loop level one encounters
two types of triangle diagrams with two gauge bosons of the gauge
group $G=\prod G_a$\foot{Analogously, there is also a mixed
gravitational, $\sigma$-model anomaly with two  gravitons
and one composite K\"ahler/curvature connection as external legs
(see section 3.5).}
and several moduli fields as external legs and massless gauginos and
charged (fermionic) matter
fields circulating inside the loop: first the coupling of the moduli
to the charged fields
contains a part
described by a composite K\"ahler connection, proportional to $K(M_i,
\bar M_i)$, which couples to gauginos as well as to chiral matter
fields $A_\alpha^{R_a}$. (We assume here that the moduli are gauge
singlets.)
It
reflects the (tree-level) invariance of the theory under K\"ahler
transformations. Second, there is a coupling between the moduli
and the $A_\alpha^{R_a}$'s
by the composite
curvature (holonomy) connection. It
originates from the non-canonical
kinetic energy $K_{\alpha\beta}^{\rm matter}$, eq.\mattkp,
of the matter fields $A_\alpha^{R_a}$.
These two anomalous contributions
lead, together with the tree-level part which is given by the
dilaton/axion field $S$,
to the following (non-local)
one-loop effective supersymmetric Lagrangian
\refs{\DFKZ,\LOUIS,\CAOV}:
$$\eqalign{
{\cal L}_{\rm nl}&=
\sum_a\int d^2\theta{1\over 4}W^a
W^a\Biggl\lbrace S
-{1\over 16\pi^2}
{1\over 16}\lozenge^{-1}\bar{\cal D}\bar{\cal D}{\cal D\cal D}
\biggl(\bigl\lbrack C(G_a)\cr&
-\sum_{\underline R_a}
T(\underline R_a)\bigr\rbrack K(M_i,\bar M_i)+
2\sum_{\underline R_a}T(\underline R_a)
\log\det K_{\alpha\beta}^{\rm matter}(M_i,\bar M_i)\biggr)\Biggr\rbrace
+{\rm h.c.}\cr}
\eqn\nl
$$
Here $W^a$ is the Yang--Mills superfield.
Writing this expression in components it leads
to a non-local contribution to the $CP$ odd term $F_{\mu\nu}
\tilde F_{\mu\nu}$ and
to a local contribution
to the gauge coupling constant.
Now, it is easy to recognize that ${\cal L}_{\rm nl}$, eq.\nl,
is not invariant under K\"ahler transformation and reparametrizations
which act non-trivially on
$K(M_i,\bar M_i)$, $K_{\alpha\beta}^{\rm matter}$
respectively. These non-invariances correspond to the two types of
$\sigma$-model anomalies, namely to the K\"ahler and curvature (holonomy)
anomalies. It follows that ${\cal L}_{\rm nl}$ is
not invariant under the discrete target space duality transformations
\repar\ since these act exactly as combined K\"ahler transformations
and reparametrizations. Using eqs.\kptrans\ and \mattkptr,
the change of ${\cal L}_{\rm nl}$ under duality transformations
is given by  the local expression
$$\eqalign{
\delta(
{\cal L}_{\rm nl})&=-{1\over 16\pi^2}\sum_a\int d^2\theta{1\over 4}W^a
W^a
\biggl(\bigl\lbrack C(G_a)-\sum_{\underline R_a}
T(\underline R_a)\bigr\rbrack g(M_i)\cr&+
2\sum_{\underline R_a}T(\underline R_a)
\log\det h_{\alpha\beta}(M_i)^{-1}\biggr)+{\rm h.c.}\cr}
\eqn\gaugevar
$$
It follows that the duality anomalies
must be cancelled by adding
new terms to the effective action.
Specifically, there are two ways of cancelling these anomalies.
In the first one \DFKZ,\CAOV\ the $S$ field may transform
non-trivially under duality transformations and cancels in this
way some part or all of the duality non-invariance of eq.\nl.
This non-trivial transformation behaviour
of the $S$--field is completely analogous to the
Green--Schwarz mechanism for the case of an anomalous $U(1)$ gauge group
and leads to a mixing between the moduli and the $S$--field in the
$S$--field K\"ahler potential. Note that the Green--Schwarz
mechanism does not only cancel the anomalies associated to the
discrete duality transformations, but in fact all continuous
$\sigma$-model anomalies.

Second, the duality anomaly can be cancelled
by adding to eq.\nl\ a term,
which is related to the gauge group
dependent threshold contribution
due to the massive string states.
The threshold contributions are given in
terms of automorphic functions of the target space duality group,
which have the required transformation behaviour only under the
discrete duality transformations but not under all continuous
reparametrizations and K\"ahler transformations. The explicit
breaking of the continuous $\sigma$-model symmetries
by the one-loop threshold effects follows from the fact that
the heavy mass spectrum is not invariant under continuous
reparametrizations of the moduli, but only under the discrete
duality transformations.

It is clear that the part of the $\sigma$-model anomaly which is
removed by the Green--Schwarz  mechanism is universal, i.e.
gauge group independent. (Otherwise one would need for each
gauge group factor an extra $S$--field which is not present in string
theory.)  Thus for cases where there are no moduli dependent
threshold contributions from the massive states
the $\sigma$-model anomaly coefficients have to coincide for each
gauge group factor $G_a$. As we will discuss,
this particularly interesting constraint arises
for many orbifold compactifications where the moduli
dependent threshold contributions are absent because of an enlarged
$N=4$ supersymmetry in the massive spectrum.

Before we come to the orbifolds, let us briefly evaluate
the one-loop effective action eq.\nl\ for the case of
(2,2) Calabi--Yau compactifications with gauge group $G=E_8\times E_6$
and with $h_{(1,1)}$ matter fields in the ${\underline{27}}$
representation of $E_6$ and with $h_{(2,1)}$ matter fields  in the
${\underline{\bar{27}}}$ representation.
Using the special relation between
the K\"ahler metrics of the moduli and matter fields as given in
eq.\special\ one derives \refs{\DIXON,\FKLZ}
$$\eqalign{
{\cal L}_{\rm nl}&=
\sum_a\int d^2\theta{1\over 4}W^a
W^a\Biggl\lbrace S
-{1\over 16\pi^2}
{1\over 16}\lozenge^{-1}\bar{\cal D}\bar{\cal D}{\cal D\cal D}
\biggl(\bigl\lbrack C(G_a)-\cr &
T(\underline R)({5\over 3}h_{(1,1)}
+{1\over 3}h_{(2,1)})\bigr\rbrack K(T_i,\bar T_i)
\cr &+\bigl\lbrack C(G_a)-
T(\underline R)
({5\over 3}h_{(2,1)}+{1\over 3}h_{(1,1)})\bigr\rbrack K(U_m,\bar U_m)
\cr &+2T(\underline R)\bigl\lbrack
\log\det K_{ij}(T_i,\bar T_i)+\log\det K_{mn}(U_m,\bar U_m)\bigr\rbrack
\biggr)\Biggr\rbrace +{\rm h.c.} ,\cr}
\eqn\gauges
$$
with $a=E_6,E_8$ and $C(E_8)=30$, $C(E_6)=12$, $T({\underline{27}})
=T({\underline{\bar{27}}})=3$.
Looking for example at the K\"ahler class moduli $T_i$
one observes that one needs in principle
two types of automorphic functions:
the first one provides a duality
invariant completion of $K(T_i,\bar T_i)$, where the second one
is needed to cancel the duality anomaly
coming from $\log\det K_{ij}(T_i,\bar T_i)$.
These two types of automorphic functions can be, at least formally,
constructed for all (2,2) Calabi--Yau compactifications \FKLZ.

It is interesting to compute eq.\gauges\ for the large radius limit
of Calabi--Yau compactifications as a function of the overall
$T$--field. With eq.\overallt\ one obtains
$$
{\cal L}_{\rm nl}=
\sum_a\int d^2\theta{1\over 4}W^a
W^a\biggl\lbrace S
-{1\over 16\pi^2}
{1\over 16}\lozenge^{-1}\bar{\cal D}\bar{\cal D}{\cal D\cal D}
b_a\log(T+\bar T)\biggr\rbrace +{\rm h.c.}
\eqn\gaugesla
$$
We see that in the large radius limit the $\sigma$-model anomaly
is simply determined by the $N=1$ $\beta$-function coefficient
$b_a=-3C(G_a)+(h_{(1,1)}+h_{(2,1)})T(\underline R)$.

\sect{Target space modular anomalies in Abelian orbifolds}

Let us now return to the
(2,2) and (0,2) orbifold compactifications.
We will discuss the $\sigma$-model anomalies
as a function of the diagonal moduli $T_i$ ($=1,2,3$)
and $U_m$ ($m=1,\dots ,h_{(2,1)}$).
Recall that the K\"ahler metrics of the matter fields, eq.\mattkpor,
are determined in terms of the
modular weights $n_\alpha^i$ and $l_\alpha^m$.
The moduli K\"ahler potential is given in eq.\kcasea.
Then the tree-level Yang--Mills action, together with
one-loop anomalous  contribution of the massless gauginos and matter
fermions, takes the form
$$          \eqalign{
{\cal L}_{\rm nl}=  &
\sum_a\int d^2\theta{1\over 4}W^a
W^a\biggl\lbrace S
-{1\over 16\pi^2}
{1\over 16}\lozenge^{-1}\bar{\cal D}\bar{\cal D}{\cal D\cal D}
\cr &\biggl\lbrack\sum_{i=1}^3{b'}_a^{i}\log(T_i+\bar T_i)
+\sum_{m=1}^{h_{(2,1)}}{b'}_a^{m}
\log(U_m+\bar U_m)\biggr\rbrack\biggr\rbrace+{\rm h.c.}\cr}\eqn\gaugeor
$$
The anomaly coefficients ${b'}_a^{i}$ and ${b'}_a^{m}$ look like
\LOUIS,\DFKZ
$$\eqalign{{b'}_a^i&=-C(G_a)+\sum_{\underline R_a}
T(\underline R_a)(1+2n_{\underline R_a}^i),\cr
{b'}_a^m&=-C(G_a)+\sum_{\underline R_a}T(\underline R_a)
(1+2l_{\underline R_a}^m).\cr}\eqn\bprimes
$$
Considering only the dependence on the overall modulus $T=T_1=T_2=T_3$
the Yang--Mills Lagrangian becomes
$$
{\cal L}_{\rm nl}=
\sum_a\int d^2\theta{1\over 4}W^a
W^a\biggl\lbrace S
-{1\over 16\pi^2}
{1\over 16}\lozenge^{-1}\bar{\cal D}\bar{\cal D}{\cal D\cal D}
b_a'\log(T+\bar T)\biggr\rbrace +{\rm h.c.} \eqn\gaugeoro
$$
with
$$\eqalign{b_a'&=-3C(G_a)+\sum_{\underline R_a}T(\underline R_a)
(3+2n_{\underline R_a})\cr &=b_a
+2\sum_{\underline R_a}T(\underline R_a)(1+n_{\underline R_a}),
\cr}\eqn\overabprim
$$
where $n$ is
the overall modular weight: $n
=\sum_{i=1}^3n^i$.
We recognize that the gauginos and matter fields with
$n_{\underline R_a}=-1$ contribute to the anomaly coefficient $b_a'$
in the same way as to the $N=1$ $\beta$-function coefficient $b_a$.
Since all untwisted matter fields have $n=-1$, the whole
untwisted sector of orbifolds behaves like the large radius limit
of Calabi--Yau compactifications (cf. eq.\gaugesla).
The deviation from this behaviour is caused by the twisted orbifold
fields, whereas world-sheet instantons modify the large radius
limit for the case of Calabi--Yau compactifications.

As already mentioned, the target space modular anomaly, i.e. the non-invariance
of the gauge coupling constant eq.\gaugeor\ under
$T_i\rightarrow{a_iT_i-ib_i\over ic_iT_i+d_i}$, can be cancelled in two
different ways. First, the universal, gauge group independent part
of the anomaly is removed by the Green--Schwarz mechanism, which
induces the following modular transformation behaviour
of the dilaton field \DFKZ:
$$S\rightarrow
S-{k_a\over 8\pi^2}\sum_{i=1}^3\delta^i_{GS}\log(ic_iT_i+d_i).
\eqn\diltrans
$$
For modular transformations on the $U_m$ fields, an analogous
transformation holds. Here the $\delta^i_{GS}$ is
the gauge group
independent ``Green--Schwarz''
coefficient which describes the one-loop mixing between
the $S$--field and the modulus $T_i$.

Second, the target space modular anomaly can be
alternatively cancelled by a local contribution to ${\cal L}_{\rm nl}$,
which is related to the one-loop threshold contributions to the
gauge coupling constants. This topic will be discussed in section 4.
Let us here consider the
interesting case  provided by those orbifold compactifications
for which a modulus $T_i$  does not appear in the threshold
corrections.
This exactly happens \DKLB\ if $all$ orbifold twists $\vec\theta$,
which define a particular ${\bf Z}_M$ or ${\bf Z}_M
\times{\bf Z}_N$ orbifold, act non-trivially on the corresponding
$i^{\rm th}$ complex plane of the underlying six-torus, i.e.
$\theta_i\neq 0$ for all $\vec\theta$. Then the contribution from the
momentum and winding states with $T_i$  dependent masses
to the threshold corrections is absent, since this part of the massive
spectrum is organized into $N=4$ supermultiplets. Furthermore,
the massive spectrum in the twisted sectors does not depend on the
untwisted moduli.
Thus in this case
the target space modular anomaly associated to $T_i$
has to be completely cancelled by the Green--Schwarz mechanism. This
implies that $\delta _{GS}^i={b'}_a^i/k_a$ for all gauge group factors
$G_a$. Since
$\delta_{GS}^i$ is gauge group independent, it immediately follows
that the anomaly coefficients \bprimes\ must take the same values for
all group factors of the gauge group $G=\prod_aG_a$:
$${ {{b'}_a^i}\over {k_a}} ={{{b'}_b^i}\over {k_b}}= {{{b'}_c^i}\over
{k_c}} =...   \eqn\bequal $$
This requirement of the complete cancellation of the $\sigma$-model
anomalies by the Green--Schwarz mechanism will provide very strong conditions
on the possible spectrum of gauge non-singlet states, specifically on
the allowed representations $\underline R_a$ in connection with the
possible modular weights $n^i_{\underline R_a}$ as we will discuss
now in the following. In the next section we will return to the
threshold effects and their implications on string unification scenarios.

Let us first give a list \refs{\ZNZM,\DKLB}
of those orbifold compactifications
for which there is complete $\sigma$-model anomaly cancellation without
additional threshold effects (see also the twist vectors $\vec\theta$
in the second column of table 1).
First for the ${\bf Z}_3$ and ${\bf Z}_7$ orbifolds,
all three complex planes are rotated by all orbifold twists.
Therefore there is complete modular anomaly cancellation with respect to
$T_1$, $T_2$ and $T_3$, and the
${b'}_a^i$ coefficients have to
coincide for all three planes $i=1,2,3$.
The ${\bf Z}_4$, ${\bf Z}_6'$,
${\bf Z}_8$, ${\bf Z}_8'$, ${\bf Z}_{12}$
and ${\bf Z}_{12}'$ orbifolds have two completely rotated planes,
such that the ${b'}_a^i$ coefficients have to
coincide for $i=2,3$. Finally, for the ${\bf Z}_6$ orbifold, the moduli
$T_1$, $T_2$ and $U_1$ appear in the threshold corrections, and
only the modular anomaly associated to $T_3$ is completely cancelled by
the Green--Schwarz mechanism. On the other hand, for all ${\bf Z}_M\times
{\bf Z}_N$ orbifolds each complex plane is left unrotated by at least
one orbifold twist \ZNZM. Therefore one does not find the complete
modular anomaly cancellation for any of these compactifications.
It is interesting to note that if there exists a complex structure
modulus $U_m$ ($h_{(2,1)}\neq 0$), the threshold corrections always depend
on this modulus. Therefore the anomaly coefficients ${b'}_a^m$ are
never constrained
by the requirement of complete modular anomaly cancellation.
The reverse situation   is true for possible complex planes with
off-diagonal fields $T_{ij}$. These fields never appear in the threshold
corrections, since the corresponding complex planes are always rotated by
all orbifold twists.

It is instructive to check how
the modular anomaly cancellation conditions
for completely rotated  planes are satisfied for already
known orbifold constructions. This check involves the determination
of the modular weights of the charged fields
as described in section 2. For example it is easy to show that
for the symmetric
(2,2) ${\bf Z}_3$ orbifold with standard embedding of the twist
into $E_8\times E_8$, the anomaly coefficients agree for all gauge
group factors: ${b'}_{E_8}^i={b'}_{E_6}^i={b'}_{SU(3)}^i=-30$,
$i=1,2,3$. The modular anomalies with respect to the gauge groups
$E_8$ and $E_6$ of
additional (2,2) orbifolds were considered in ref.\DFKZ.
In the appendix A we also discuss the case of the (2,2) ${\bf Z}_7$
orbifold in some detail.

It is important to stress that the modular anomaly cancellation
has to work not only for the few (2,2) cases but also for the
numerous (0,2) constructions with non-standard embeddings and/or
additional background fields such as discrete Wilson lines.
In general, these orbifolds yield gauge groups different from
$E_6\times E_8$. As a non-trivial (0,2) example we explicitly
discuss in the appendix B
a ${\bf Z}_4$ orbifold with gauge group $G=E_6\times SU(2)_1\times
U(1)\times
E_7\times SU(2)_2$ \KATSUKI. As required from our general arguments
on modular anomaly cancellation, we find for the anomaly coefficients
of the second and third plane:
${b'}_{E_6}^i={b'}_{SU(2)_1}^i={b'}_{E_7}^i={b'}_{SU(2)_2}^i=-6$,
$i=2,3$.
We also checked for (0,2) ${\bf Z}_3$ orbifolds with discrete
Wilson lines that the anomaly coefficients of all gauge group factors coincide
for each of the three complex planes.

Instead of looking at already explicitly constructed orbifold
compactifications we now want to demonstrate that the requirement of
having no target space modular anomalies leads to very strong, and
so far undiscussed, constraints on the possible spectrum of massless
fields for orbifolds with completely rotated planes.
Specifically, we want to investigate the important question
whether an orbifold with the standard model gauge group
$G=SU(3)\times SU(2)\times U(1)_Y$, plus exactly the massless particle
content of the minimal supersymmetric standard model, can be
free of target space modular anomalies and can thus describe a consistent
compactification scheme.
The massless spectrum of the hypothetical minimal orbifold
is characterized by the modular weights $n_\alpha^i$ of the standard
model chiral fields where $\alpha=Q_g,U_g,D_g,L_g,E_g,H,\bar H$
($g=1,2,3$) in an
obvious notation. Then the anomaly coefficients of
the minimal model take the following form:
$$\eqalign{{b'}_3^i&=3+\sum_{g=1}^3(2n_{Q_g}^i+n_{U_g}^i+n_{D_g}
^i),\cr
{b'}_2^i&=5+\sum_{g=1}^3(3n_{Q_g}^i+n_{L_g}^i)+n_{H}^i+n_{\bar H
}^i,\cr
{b'}_1^i&=11+\sum_{g=1}^3({1\over 3}n_{Q_g}^i
+{8\over 3}n_{U_g}^i+{2\over 3}n_{D_g}^i+n_{L_g}^i+2
n_{E_g}^i)+n_H^i+n_{\bar H}^i.\cr}\eqn\bprstand
$$
The condition of vanishing target space modular anomalies
of the minimal orbifold compactifications now reads
${b'}_3^i/k_3={b'}_2^i/k_2={b'}_1^i/k_1$, where $i$ corresponds to
those complex planes that are rotated by all orbifold twists.
For the case of $U(1)_Y$, our normalization
of the $U(1)$ charges used in ${b'}_1^i$ is chosen in such a way that
$k_1=5/3$ corresponds to the standard grand unification value
of the Kac--Moody level. The above anomaly cancellation condition
turns into two independent equations, which the modular weights of the
standard model fields have to satisfy. For our purposes, it will
be convenient to define the following two linear combinations
which have to vanish:
$${A'}^{i}={k_2\over k_1}{b'}_1^i-{b'}_2^i=0,\qquad
{B'}^{i}={b'}_1^i+{b'}_2^i-{k_1+k_2\over k_3}{b'}_3^i=0.\eqn\abpr
$$
Whether these two equations have any solutions crucially depends
on the distribution of the allowed modular weights of the standard
model fields, which were classified in section 2.
Of course, similar constraints may be obtained for other extended
gauge groups and particle contents.

\sect{The case of general (0,2) ${\bf Z}_3$ orbifolds.}

The strongest constraints arise for the ${\bf Z}_3$
orbifolds where eq.\abpr\ must be satisfied  with
respect to all three complex planes, and where the choice of
possible values for the modular weights is very limited.
Indeed there exist \refs{\KIM,\CASMU,\FIQSA} several ${\bf Z}_3$
examples with standard model gauge group and three chiral families,
but all of these possess in addition some extra (non-chiral) matter
fields. Thus, can there exist  anomaly free ${\bf Z}_3$ orbifolds
with just the minimal particle content?
In the previous section we gave an extensive description of how
one determines the allowed modular weights for massless particles.
Let us give the result for the standard model particles for the
${\bf Z}_3$ orbifold case.
The first possibility for allowed modular weights is given by
$\vec n_\alpha=(-1,0,0),(0,-1,0),(0,0,-1)$. These three choices
correspond to untwisted chiral fields. The twist vector $\vec\theta$
of the ${\bf Z}_3$ orbifold has the form $\vec\theta=(1/3,1/3,1/3)$
($\vec v=(1/3,1/3,-2/3)$). The conformal dimensions as a function
of the Kac--Moody levels are given in table 2 for the standard model
particles. Then, using eq.\maxosc, we obtain that for
$k=3/5k_1=k_2=k_3=1$ there are no oscillators allowed for the
standard model particles. For $k=2$ the fields $Q_g$ and $U_g$
still must not have any oscillator, whereas the other fields
may possess one twisted oscillator. Finally, for $k=3$ all
fields except $Q_g$ may possess one oscillator, and for $k>3$
we obtain $p_{\rm max}=1$ for all standard model particles.

Let us first investigate the most
common case of level one Kac--Moody algebras,
$k_1=5/3$, $k_2=k_3=1$.
Then, apart from the ``untwisted'' modular weights, the only
additional ``twisted'' possibility for the standard model fields is
to have $\vec n_\alpha=-(2/3,2/3,2/3)$. (The corresponding
overall modular weight $n=-2$ is displayed  in table 3.)
Thus, in total, the range of allowed modular weights consists
of four possibilities. With the help of a computer program one can
now check whether the two
equations \abpr, together with eq.\bprstand, have any simultaneous
solutions
using the four different allowed values for the
modular weight vectors of each of the standard model
particles. In doing this we find that there are no solutions at all.
It is interesting to note that also the single equation
${b'}_2^i-{b'}_3^i=0$ cannot be satisfied using the allowed
modular weights. In this way the uncertainty of
the $U(1)$ normalization factor $k_1$ is eliminated.
Thus we have obtained the striking result that for the ${\bf Z}_3$
orbifold with level one gauge groups the target space modular
anomalies cannot be cancelled by the standard model particles.
In this way we have ruled out the minimal level one
${\bf Z}_3$ compactifications by general consistency arguments.
The requirement of target space anomaly freedom forces us to
introduce additional fields.

At higher levels the range of allowed modular weights is broader.
Specifically, if there can be one twisted oscillator one obtains the
following additional allowed modular weight vectors:
$\vec n_\alpha=-(5/3,2/3,2/3),-(2/3,5/3,2/3),-(2/3,2/3,5/3)$.
In this case we find that for $k\ \geq 2$ the anomaly cancellation
condition
eq.\abpr\ can be satisfied in various ways. Therefore we cannot prove
minimal ${\bf Z}_3$ models with higher level Kac--Moody algebras
to be inconsistent. Orbifold models with gauge groups realized
at a higher Kac--Moody level were constructed in ref.\HIGH\ (an example
is briefly discussed in section 3.6). It is however
         complicated
to explicitly construct models with higher level standard model
gauge group. In addition we will show in the next
section on threshold corrections, that the minimal ${\bf Z}_3$
compactifications cannot meet the phenomenological requirement
of proper coupling constant unification, and are therefore ruled out
on physical grounds.

Let us investigate more what the spectrum of additional massless states
has to look like in order to satisfy the anomaly matching conditions.
General classes of (0,2) ${\bf Z}_3$ orbifold models with gauge group
$SU(3)\times SU(2)\times U(1)^n\times G_{\rm hidden}$
and three quark lepton generations plus extra vector-like matter
were constructed
in refs.\refs{\KIM,\FIQSA}. This was done by adding two quantized
Wilson lines on the four different basic embeddings of the
${\bf Z}_3$ orbifold.
For example, in ref.\KIM\ a three generation model with gauge group
$G=SU(3)\times SU(2)\times U(1)^8\times SO(10)_{\rm hidden}$ was
constructed. It is straightforward to show that the conditions
of modular anomaly cancellation are satisfied in this model.
Specifically, the untwisted states with non-trivial $SU(3)\times SU(2)
\times SO(10)$ quantum numbers are given by
$3\lbrack (\underline 3,\underline 2,\underline 1)
+({\underline{\bar 3}},\underline 1,\underline 1)
+(\underline 1,\underline 2,\underline 1)
+(\underline 1,\underline 1,{\underline{16}})\rbrack$, whereas the
twisted states are of the form $3\lbrack 5({\underline{\bar 3}},
\underline 1,\underline 1)
+4(\underline 3,\underline 1,\underline 1)
+12(\underline 1,\underline 2,\underline 1)
\rbrack$. Then one obtains ${b'}_{SU(3)}={b'}_{SU(2)}={b'}_{SO(10)}=-18$.

More generally, it was shown that one can construct, just within
the ${\bf Z}_3$ orbifold, of order 50000 such three generation models
but only nine different classes are inequivalent \CLASS. Furthermore,
these nine classes divide into two types, according to the
lepton and quark content of the untwisted sector. Five of the
classes have in that sector  three copies of particles
with quantum numbers
$$
(\underline 3,\underline 2)\ +\ 2(\underline 1,\underline 1) \eqn\uno
$$
whereas the other four models have three copies of fields
transforming like
$$
(\underline 3,\underline 2)\ +\ ({\underline{\bar 3}},\underline 1)\
+\ (\underline 1,\underline 2)         \eqn\dos
$$
under $SU(3)\times SU(2)$. Models belonging two these two  classes
always had some specific properties. In particular, irrespective
of the structure of the model and of which  other gauge factors are
present, the number of $SU(2)$ doublets exceeds by 12 the number of
$SU(3)$ triplets. This meant in practice the impossibility of
getting the minimal Higgs content since in the minimal model
the number of $SU(2)$ doublets exceeds the number of $SU(3)$
triplets only in the number of Higgs pairs.

Duality anomaly cancellation arguments allow us to understand
the origin of these peculiar properties of these three generation
models. Consider modular invariance with respect to the overall
modulus $T=T_1=T_2=T_3$. In the type of models of
refs.\refs{\KIM,\CASMU,\FIQSA}, the
twisted oscillator states are always $SU(3)\times SU(2)$ singlets
and hence the possible modular weights of quarks, lepton-doublets
and Higgses are $n=-1$ (untwisted) and $n=-2$ (twisted).
Cancellation of modular anomalies
requires
$$
{b'}_2\ -\ {b'}_3\ =\ 9\ -\ 6\ -\ {1\over 2}\ (N_3^U\ -\ N_2^U)\ +
\ {1\over 2}\ (N_3^T\ -\ N_2^T)\ =\ 0,
\eqn \pec
$$
where $N_3^T$ and $N_3^U$ are the numbers of triplets in the
twisted and untwisted sectors respectively (and equivalently for
the doublets). In the models with untwisted content as in eq.\uno\
one has $N_3^U=6$ and $N_2^U=9$ whereas in the second type
one has $N_3^U=9$ and $N_2^U=12$. Thus from eq.\pec\
one gets in both cases
$$
N_2^T\ -\ N_3^T\ =\  9 ,
$$
and there are altogether $9+3=12$ more doublets than triplets in
all models. This gives an understanding of this pattern previously
found in
a model-by-model basis. Similar arguments should be
applicable to other ${\bf Z}_3$ three-generation models obtained
through   the addition of three Wilson lines instead of two
(for examples see ref.\KIMB). The determination of the complete
massless spectrum of these models is extremely painful and
using duality anomaly cancellation arguments one can obtain
immediately information about the twisted spectrum by simply
knowing the untwisted sector (which is trivial to obtain)  and
using equations analogous to eq.\pec .

\sect{${\bf Z}_7$ and other orbifolds}

Now let us discuss the ${\bf Z}_7$ compactifications \refs{\IMNQ,\ZSIE}
which provide the
second class of orbifolds with three completely rotated planes.
Again we will find that the target space modular anomalies cannot
be cancelled, assuming a minimal particle content together
with the standard model gauge group $SU(3)\times SU(2)\times U(1)_Y$.
The distribution of allowed modular weight vectors for the standard model
fields is as follows. For untwisted fields there are again the three
possibilities $\vec n_\alpha=(-1,0,0)+{\rm perm}$. The ${\bf Z}_7$
twist vector is given as $\vec\theta=(1/7,2/7,4/7)$. Then, for states
without oscillators, one obtains in the twisted sectors three
different  allowed modular weight vectors: $\vec n_\alpha=-(6/7,5/7,3/7)
+{\rm perm}$.
For simplicity, we restrict the discussion from now on
to the level one
case $3/5k_1=k_2=k_3=1$.
This implies (see table 3) that the fields $Q_g,U_g,E_g$ are not
allowed to have any twisted oscillator contribution. Thus, the
range of allowed modular weights consists     of the six above
mentioned choices. On the other hand, looking at
the first
twisted sector, the fields $D_g,L_g,H,\bar H$ may maximally have
two oscillators associated to the first complex plane or one
oscillator in the direction of the second plane. For the remaining
twisted sectors an analogous  statement is true. Thus we
obtain nine additional allowed modular weight vectors for these type
of fields: $\vec n_\alpha=-(13/7,5/7,3/7),-(20/7,5/7,3/7),
-(6/7,12/7,3/7)+{\rm perm}$.
With these 15 allowed choices for $D_g,L_g,H,\bar H$ and the
6 possibilities for $Q_g,U_g,E_g$ eq.\abpr\ has not a single
solution. This again rules out any minimal ${\bf Z}_7$ compactification
at the lowest Kac--Moody level. However, at higher level we expect
to find a variety of consistent solutions.

Concerning the other (level one) cases,
we have summarized our results about the
cancellation of the modular anomalies in the third column of table 1.
Unfortunately, the ${\bf Z}_3$ and ${\bf Z}_7$ orbifolds
provide the only two classes of models where
the minimal standard model spectrum can be ruled out on the
basis of modular anomaly cancellation requirements alone. As already
discussed, all other models have at least one complex plane which
is left unrotated by one of the orbifolds twists. Then the
anomaly cancellation condition eq.\abpr\ cannot be applied with
respect to this particular complex plane. The consequence of the
corresponding
loss of one, two,
or even all three equations is that for the
remaining conditions one always finds modular weights
of the standard model particles which are allowed by the
particular model under investigation. Thus, in order to obtain further
constraints on these orbifolds, additional ``phenomenological''
boundary conditions are needed. This subject
will be discussed in  section 4.

\sect{The mixed gravitational anomalies}

The above discussion  was mainly concerned with the presence
of mixed K\"ahler (and curvature) gauge anomalies. There will
also be in general mixed triangular anomalies involving two
gravitons instead of two gauge bosons. The computation of those
anomalies is exactly parallel to the gauge anomalies, the only
difference being that now all massless fermions (including the
gravitino and fermionic partners of the dilaton and moduli)
will contribute. The analogous anomaly coefficient ${b'}_{\rm grav.}$
can be computed to be
$$
{b'}_{\rm grav}^i\ =\ 21 \ +\ 1\ +\ \delta _M^i \ -\ {\rm dim}\ G\
+\ \sum _\alpha\ (1\ +\ 2n_\alpha^i),
\eqn \gravi
$$
where the sum runs over all the massless charged chiral fields.
(The mixed gravitational anomalies
were also considered by \refs{\KAPLUG,\CAOV,\LOLU}. However in ref.\CAOV\
the contribution of the gravitino is apparently missing.)
The contribution of the gravitino to the anomaly is given by 21,
the dilatino contributes 1 (since it has no
modular weight) and $\delta _M^i$ represents the contribution of
modulinos to the anomaly, which is model-dependent; ${\rm dim}\ G$ is the
dimension of the complete gauge group. Notice that for a completely
rotated orbifold plane one has
$$
{b'}_{\rm grav}^i\ =\ 24\ {{{b'}_{\rm gauge}^i}\over {k}}
\eqn \bgrav
$$
for any gauge group
(as an example we check eq.\bgrav\ for the (2,2) ${\bf Z}_7$
orbifold in the appendix A and for a (0,2) ${\bf Z}_4$ example in
appendix B).
This fact leads to further constraints
on the massless fermionic spectrum, although the computation
of ${b'}_{\rm grav}$ requires a complete knowledge of the massless
spectrum including all singlets. In the case of the overall modulus
$T$ the anomaly coefficient is
$$
{b'}_{\rm grav}\ =\ 66\ +\ \delta _M\ -\ 3({\rm dim}\ G)\ +
\ \sum _\alpha\ (3\ +\ 2n_\alpha) .
\eqn \bgravv
$$
It is perhaps interesting to remark that, since in four dimensions
there are no pure gravitational anomalies, these mixed $\sigma$ model
gravitational anomalies are the only ones that are sensitive to
singlet particles, they are our only          control on the
possible existence of a ``hidden sector'' in the theory.
Consider for example the ${\bf Z}_3$ and ${\bf Z}_7$ orbifolds.
One can get an idea of the size of the ``hidden sector'' of the theory
by using   eqs.\bgrav,\bgravv\
which yield
$$\eqalign{
3({\rm dim}\ G_{\rm hidden})-\sum _{\alpha={\rm hidden}}(3+2n_\alpha)\ &=\ \cr
-24b_{\rm gauge} '+66+\delta _M
&-3({\rm dim}\ G_{\rm visible})+\sum _{\alpha={\rm visible}}(3+2n_\alpha)
\cr }\eqn \hidd   $$
where $b_{\rm gauge}'$ corresponds to any of the gauge groups.
One can also make use of the gravitational anomaly cancellation in
order to check in an easy manner if the computed massless
spectrum of a given (0,2) ${\bf Z}_3$ or ${\bf Z}_7$ orbifold
is correct. Consider for example the class of three-generation
${\bf Z}_3$ orbifold models with gauge group
$SU(3)\times SU(2)\times U(1)^n\times G_{\rm hidden}$ discussed
above \refs{\KIM,\CASMU,\FIQSA}.
The computation of the massless sector can be extremely
cumbersome (especially in the case with three Wilson lines)
and it is very easy to miss some singlet states.
In these models one has (one can check $\delta _M=-9$ for ${\bf Z}_3$)
$$
\sum _TN_T+3\sum _{\rm osc}N_{\rm osc}\ =\
-24b_{\rm gauge}'+57-3{\rm dim}\ G+\sum _UN_U
\eqn\check
$$
where $U,T,$osc denote untwisted, twisted and twisted oscillator states.
The quantities in the right-hand side are very easy to compute
and eq.\check\ gives us a handle on the singlets of the model.
Using this equation, we have checked that the spectrum of some
models in the literature is incomplete. Thus, for example, one
can immediately check that the massless spectrum of the
three-generation $SU(3)\times SU(2)\times U(1)^{13}$ model
of ref.\KIMB\  is missing 75 states in the twisted sector.

\sect{The effect of symmetry-breaking flat directions on the
modular anomalies}

All the above discussion applies directly to the case of (0,2)
orbifold 4-D strings with gauge group of ${\rm rank} \geq 16$, i.e.
standard Abelian orbifold models. However, there are string
solutions with smaller rank which are continuously connected
to the above models by giving non-vanishing $vev$'s to some
marginal deformations. In the effective field theory, these marginal
deformations correspond to fields $C$ with completely flat potential
such that $<C>$ is an arbitrary parameter of the theory.
An example of that are the Calabi--Yau  limits of the unique (up to
discrete torsion)    (2,2) version of each orbifold, which are
obtained by giving $vev$'s to the twisted moduli of each model.
However it is important to remark that for a given orbifold there
is a unique (2,2) version but $many$ $thousands$ of consistent
(0,2) embeddings for which the ``repaired'' Calabi--Yau limit
does not exist. Still, these models, as remarked in ref.\DEG\ do
also possess other marginal deformations not necessarily
associated with the geometry of the orbifold. When giving $vev$'s to
these marginal deformations one generically lowers the rank of the
gauge group and some chiral fields $A$, $B$
get massive through renormalizable
Yukawa couplings corresponding to a superpotential of the form
$W\sim CAB$.
This possibility is in fact often used in
model-building in order to reduce the number of massless fields,
which tends to be quite large in generic 4-D strings.

An obvious question is whether the anomaly constraints
described above apply also to these ``deformed'' lower-rank
models. Of course, modular anomalies must also cancel for
these models; the question is  whether one can impose constraints,
e.g. as in eqs.\bequal ,\bgrav\ to the massless sector.
The answer to that question
depends on the type of flat direction considered. If the flat direction
involves $vev$'s only for untwisted fields in completely
rotated planes, the answer is immediately
yes. These types of flat directions
were  discussed  in refs.\DEG\  and \ILLT . A simple example is
provided by the standard (2,2) ${\bf Z}_3$ orbifold. The gauge
group is $E_6\times SU(3)\times E_8$ and there are three
copies of $({\underline{27}},\underline 3)$
in the untwisted sector and 27 copies of
$({\underline{27}},\underline 1)$
in the twisted sector. One can easily check \refs{\DEG,\ILLT}
that there is a flat
direction involving non-vanishing $vev$'s for three scalars inside
one of the
$({\underline{27}},\underline 3)$'s
which breaks $E_6\times SU(3)\rightarrow
SO(8)\times U(1)^2$. At the same time, some untwisted fields
acquire a mass through trilinear Yukawas, and one is left with
untwisted and twisted matter fields with $SO(8)$ quantum numbers:
$$
\eqalign{
U\ \ \ :\ \ &3\  (\underline 8_v\ +\ \underline 8_s\ +\
\underline 8_c\ +\ \underline 1)    \cr
T\ \ \ :\ \ &27\ (\underline 8_v\ +\
\underline 8_s\ +\ \underline 8_c\ +\ \underline 1\ +\
\underline 1\ +\underline 1).\cr }
\eqn\uflat
$$
Consider now the modular anomalies with respect to the overall
modulus $T$. We know that for the ${\bf Z}_3$ orbifold case
one has $b_{E_6}'=b_{E_8}'=b_{SU(3)}'= -90$. For the ``deformed''
model obtained after the flat direction is taken one finds
$$
b_{SO(8)}'\ =\ -3\ C_{SO(8)}\ +\ 3(1+1+1)\ -\ 27(1+1+1)\ =\ -90 ,
\eqn\sooch
$$
i.e., the constraints of eqs.\bequal\ and \bgrav\ still hold.
This is in fact not surprising, since the massive spectrum
in untwisted sectors
corresponding to completely rotated complex planes have
an  $N=4$ structure. When the marginal deformation appears,
the fields which get a mass form complete $N=4$ multiplets,
which give vanishing contribution to $b'$ such that the value of $b'$
is not changed by $<C>\neq 0$.
Thus for these types of flat directions the constraints in
eqs.\bequal\ and \bgrav\ still hold.

Sometimes, taking one of these flat directions gives rise to a
new orbifold model with gauge groups realized
at a higher Kac--Moody level \HIGH.
In these cases the $k$ factors in eq.\bequal\
are important. A ${\bf Z}_3$
example of this type was constructed in ref.\HIGH. This is a model
with gauge group $SU(3)^3\times SO(14)\times U(1)^3$ and untwisted and
twisted matter fields transforming like
$$
\eqalign{
U\ &:\ 3\lbrack (\underline 3,\underline 3,\underline 3)
+(\underline 1,\underline 1,\underline 1)\
+\ (\underline 1,{\underline {14}})
+(\underline 1,{\underline{64}})\rbrack  \cr
T\ &:\ 27\lbrack ({\underline{\bar 3}},\underline 1,\underline 1)
+(\underline 1,{\underline{\bar 3}},\underline 1)
+(\underline 1,\underline 1,{\underline{\bar 3}})\rbrack .\cr }
\eqn \alto
$$
One can easily check that for the overall modulus one has
$b'_{SU(3)}=b'_{SO(14)}=-9$. One can also check that
there is an untwisted
flat
direction in which, giving vacuum expectation values to some
appropriate components of the
$(\underline 3,\underline 3,\underline 3)$ fields, one arrives
at a model with gauge group $SU(3)\times SO(14)\times U(1)^3$
in which the $SU(3)$ algebra is realized at level $k=3$ whereas
$SO(14)$ is still at level $k=1$.
The new model particle content has
untwisted and twisted matter fields transforming like
$$
\eqalign{
U\ &:\ 3\lbrack ({\underline{10}},\underline 1)
+(\underline 1,\underline 1)\ +\ (\underline 1,{\underline{14}})
+(\underline 1,{\underline{64}})\rbrack  \cr
T\ &:\ 81 ({\underline{\bar 3}},\underline 1) \cr }
\eqn \bajo
$$
where the $10$-plet corresponds to a three-index completely
symmetric $SU(3)$ representation. The reader can easily check
that
$$
b'_{SU(3)}\ =\ k_3\ b'_{SO(14)}\ =\ 3\ b'_{SO(14)}\ =\ -27,
\eqn \bal
$$
verifying eq.\bequal . Thus along this type of flat directions,
it is the ratio $b'/k$ which remains invariant due to the $N=4$
supersymmetry of the massive fields
and not $b'$
itself.

For other types of flat directions, namely for untwisted fields
that do not belong to a completely rotated plane and for
twisted fields, the fields getting a mass do not form in general
complete $N=4$ multiplets and do contribute to the $b'$s.
Therefore the anomaly cancellation conditions \bequal\ and \bgrav\
will no longer hold if one computes the anomaly coefficients
by taking into account only the contribution of the remaining massless
fields. (Of course, including also the fields which get their mass
by $<C>\neq 0$, the anomaly conditions are always satisfied.)
However, in this case
one can show \refs{\IBLU,\LM}
that the contribution of the mass field $C$ to
the one-loop threshold corrections exactly cancels the target space
anomalies.

\chap{Gauge Coupling Unification and Threshold Effects}

In the last section we have obtained very strong constraints on
the massless spectrum of those two classes of orbifold compactifications,
for which all three complex planes are rotated simultaneously by all
orbifold twists. However, this situation is certainly not the
generic case since most orbifolds have at least one complex
plane on which a particular twist is not acting. On the other hand,
for the orbifolds with unrotated planes very useful restrictions
on the form of the massless spectrum, in particular on the possible
modular weights, can be obtained from the requirement of
the proper unification of the running gauge coupling constants.
Again we will show that large classes of orbifolds are ruled out
by this ``phenomenological'' requirement when one assumes
the particle content of the minimal supersymmetric standard model.
In ref.\ILR\ we already discussed the gauge coupling running in
minimal orbifold compactifications. The analysis presented in this
section should be viewed as the continuation and elaboration
of our earlier work in ref.\ILR.
We will comment on the specific relation between the present analysis
and ref.\ILR\ later in the text.
Threshold corrections and gauge coupling unification
within the flipped $SU(5)$ string model \FLIPP\ were considered
in ref.\FLIPPT.

For orbifolds with unrotated complex planes,
the threshold contributions to the gauge coupling constant due to
the massive momentum and winding states  are of the following
form \DKLB:
$$\eqalign{
\Delta_a(T_i,U_m)&
=-{1\over 16\pi^2}\sum_{i=1}^3({b'}_a^i-k_a\delta_{GS}
^i)
\log |\eta(T_i)|^4\cr &
-{1\over 16\pi^2}\sum_{m=1}^{h_{(2,1)}}
({b'}_a^m-k_a\delta_{GS}^m)
\log |\eta(U_m)|^4,\cr}
\eqn\thresor
$$
where $\eta(T)=e^{-\pi T/12}\prod_{n=1}^\infty
(1-e^{2\pi nT})$ is the
Dedekind function.
Using the well-known modular transformation properties of the Dedekind
function, it is easy to check that the threshold corrections,
together with the non-trivial transformation property of the
$S$--field, exactly cancel the anomalies under the discrete
target space modular transformations.

In the following we will neglegt the $U_m$ dependence of the threshold
corrections.
This limitation will not affect our main conclusions.
In particular, one can check that in orbifolds with two completely
rotated planes the contribution from
the standard model fields to
the anomaly coefficients of the $T_i$ and $U_m$
moduli agree. Thus the possible contribution of the $U$-fields can be
absorbed by a simple redefinition of the $T$--fields.
Using eq.\thresor\ for the moduli-dependent threshold corrections,
the one-loop  running gauge coupling constants take the following form
(up to a small field independent contribution):
$${1\over g_a^2(\mu)}={k_a\over g^2_{\rm string}}+{b_a\over
16\pi^2}\log{M_{\rm string}^2\over\mu^2}-{1\over 16\pi^2}
\sum_{i=1}^3({b'}_a^i-k_a\delta_{GS}^i)\log\lbrack(T_i+\bar T_i)
|\eta(T_i)|^4\rbrack.\eqn\running
$$
Here $M_{\rm string}$ is the typical string scale, which is
of the order of the Planck mass. Its precise value, using the
${\overline{MS}}$ scheme,
is determined by
the universal string coupling constant $g_{\rm string}$ as \KAPLU\
$M_{\rm string}=0.7\times g_{\rm string}\times 10^{18}{\rm GeV}$.
The unification mass scale $M_X$ where two gauge
group coupling constants become equal, i.e. ${1\over k_ag_a^2(M_X)}=
{1\over k_bg_b^2(M_X)}$, becomes using eq.\running :
$${M_X\over M_{\rm string}}=\prod_{i=1}^3
\lbrack
(T_i+\bar T_i)|\eta(T_i)|^4\rbrack^{{b_b^{i}{'}k_a-{b'}_a^ik_b
\over 2(b_ak_b-b_bk_a)}}
.\eqn\unific
$$
Note that since we are interested only in the difference of two
gauge couplings, the universal Green--Schwarz term is irrelevant for $M_X$.
It is clear that the unification scale $M_X$ does not depend on
those moduli $T_i$ which correspond to a completely rotated complex
plane, since in this case one has ${b'}_a^ik_b=b_b^{i}{'}k_a$
(cf. eq.\bequal). Thus for the ${\bf Z}_3$ and ${\bf Z}_7$ orbifolds
it immediately follows that $M_X=M_{\rm string}$. For the remaining
classes of models with at least one unrotated plane one generically
obtains either $M_X<M_{\rm string}$ or $M_X>M_{\rm string}$. Which of
the two possibilities is realized depends on the modular weights
of the charged fields and also on the values of the $T_i$
fields. In our previous paper \ILR\ we  considered the
threshold dependence on the overall modulus $T=T_1=T_2=T_2$. Identifying
the three moduli in this way, one concludes that $M_X>M_{\rm string}$
if only charged states with overall modular weights $n=-1$ (e.g.
as in the ${\bf Z}_2\times {\bf Z}_2$ orbifold)
contribute to
$b'$. On the other hand, leaving the three fields $T_i$ as
independent parameters, $M_X<M_{\rm string}$ is in principle
possible even in this case if, for example, $\vec n_\alpha=(-1,0,0)$ and
$T_1>>T_2,T_3$, such that the dominant contribution to the
threshold effects comes from $T_1$.

As a useful illustration of the fact that both $M_X>M_{\rm string}$ and
$M_X<M_{\rm string}$ can actually occur in known
orbifold compactifications,
we discuss in the appendix B
a (0,2) ${\bf Z}_4$ orbifold with gauge group
$G=E_6\times SU(2)_1\times U(1)\times E_7\times SU(2)_2$. As discussed
in the previous section, this orbifold requires complete modular
anomaly cancellation with respect to $T_2$ and $T_3$.
Thus the threshold corrections entirely depend only on $T_1$.
Specifically one finds the following unification scenario:
The unification of $E_6$ with $SU(2)_2$ occurs at $M_X<M_{\rm string}$.
(The precise value of $M_X$ depends on the value of $T_1$.)
Second, $M_X=M_{\rm string}$ for the unification of $E_6$ with
$SU(2)_1$ and of $E_7$ with $SU(2)_2$. Finally, $E_6$ and $E_7$
as well as $SU(2)_1$ and $E_7$ unify at $M_X>M_{\rm string}$,
whereas the coupling constants of $SU(2)_1$ and $SU(2)_2$
do not meet at all ($b_{SU(2)_1}=b_{SU(2)_2}$).

Now we want to investigate the question whether the ${\bf Z}_N$ or
${\bf Z}_N\times {\bf Z}_M$ orbifold compactifications can
lead to the correct unification of the three coupling constants
of the standard model gauge group $SU(3)\times SU(2)\times U(1)_Y$,
taking into account the threshold correction of the massive string
excitations. Our analysis will be based on the experimentally
measured values of the strong coupling constant and the weak
mixing angle: $\alpha_3^{\rm exp}=0.115\pm 0.007$,
$\sin^2\theta_W^{\rm exp}=0.233\pm 0.0008$.
Considering the
effect of the spectrum of the minimal
supersymmetric standard model on the one-loop renormalization
group equations \SUSYUN\ (without any threshold corrections)
with a SUSY threshold close
to the weak scale, one finds that the quoted results for $\alpha_3$
and $\sin^2\theta_W$ are in very good agreement with a unification
mass $M_X/M_{\rm string}\simeq 1/50$ \AMALDI . In ref.\ILR\ we showed,
looking only at the overall modulus $T$, that
this substantial discrepancy of scales can be made consistent by
the inclusion of the string threshold correction, provided
the modular weights of the standard model particles satisfy certain,
very constrained conditions. Now we will check in which classes of
orbifolds the required conditions on the modular weights can be
met taking into account also the constraints from the complete
target space modular cancellation associated to completely
rotated complex planes. We will also comment on the viability
of some possible extensions of the minimal particle content.

Making use of eq.\unific\
one gets for the value of the electroweak angle $\theta_W$
and for the strong coupling constant $\alpha_3$,
after some standard algebra:
$$\eqalign{
{\sin^2\theta _W}(\mu) & =
{{k_2}\over {k_1+k_2}}\cr & - {{k_1}\over {k_1+k_2}}
{{\alpha_e(\mu)}\over {4\pi}}\biggl(  A \log({{M_{\rm string}^2}
\over {\mu ^2}}) +\sum_{i=1}^3  {A'}^i
\log\lbrack(T_i+\bar T_i)|\eta (T_i)|^4\rbrack\biggr),\cr
{1\over {\alpha _3(\mu )}}& = {{k_3}\over {(k_1+k_2)}}
\biggl({1\over {\alpha_e(\mu)}} \cr &-{1\over {4\pi }} B
\log({{M_{\rm string}^2}\over {\mu ^2}})+
{{1}\over {4\pi }}\sum_{i=1}^3{B'}^i
\log\lbrack(T_i+\bar T_i)|\eta (T_i)|^4\rbrack\biggr),\cr}  \eqn \alfs
$$
where $A
={{k_2}\over {k_1}}b_1-b_2$ and
$B= b_1+b_2-{{(k_1+k_2)}\over {k_3}}b_3$.
${A'}^i$ and ${B'}^i$ have the same expressions after replacing
$b_a\rightarrow {b'}_a^i$ (see eq.\abpr).

Let us first compute $\alpha_3^0$ and $\sin^2\theta_W^0$, which we
define to be the values of the strong coupling constant and the weak
mixing angle, respectively,
without any string threshold corrections,
i.e. ${A'}^i,{B'}^i=0$. From now on we restrict the discussion
to the Kac--Moody level one case $k_3=k_2=3/5k_1=1$. For the spectrum
of the minimal supersymmetric standard model one has $A=28/5$ and
$B=20$. Then the large value for the string unification mass
leads to $\alpha_3^0=0.20$ and $\sin^2\theta_W^0=0.218$
\refs{\ANTON,\ILR},
in gross
conflict with the experimental data. It is useful to consider also
the effect of including extra, light (with mass at the weak scale)
chiral fields in non-minimal compactifications. Specifically,
we compute $\alpha_3^0$ and $\sin^2\theta_W^0$, including various
combinations of extra $SU(3)$ octets $O$ and $SU(2)$ triplets $T$, being
singlets under the other two gauge groups, plus possible extra
non-exotic standard model particles. The results of these
computations are displayed in table 4. None of the choices
is in agreement with the experimentally required values for
$\alpha_3$ and $\sin^2\theta_W$. (Various other
extensions of the minimal unification scenario can be tried out,
but we found
that none of them meets the experimental requirements.) For later use,
it is convenient to introduce the following parameter $\gamma^0$,
which indicates the deviation of the `bare' string predictions
from the experimentally correct values:
$$ \gamma^0 = {5\over 3}{\alpha _e}\left( {(1/ \alpha_3^0)
-(1/\alpha_3^{\rm exp})\over \sin^2\theta_W^0-
\sin^2\theta_W^{\rm exp}}\right).\eqn \bbet
$$
The relevant values for $\gamma^0$ are again shown in table 4.
(There, $\gamma^0\sim 0$ means that $\alpha_3^0$ is in good agreement
with experiment, whereas $\sin^2\theta_W^0$ is several
standard deviations away.)

Having convinced ourselves that the minimal string unification as well
as various extensions with light particles fail to reproduce the
experimentally known parameters without or with only tiny
string threshold corrections, we can think of two kinds of alternatives
(of course, the following two  possibilities can also be realized
simultaneously): First, some of the
additional fields are not light but sit at some
intermediate mass scale with or without intermediate unification
of the standard model gauge groups.
In fact it was shown in ref.\ANTON\ that this is a viable
approach for case j of our table 4. Secondly, insisting on
having only light fields, the effects of string threshold corrections
accommodate for the correct values of $\alpha_3$ and $\sin^2\theta_W$.
This generically requires the compactification scale with be
relatively large compared to the string scale \ILR. In addition,
the parameters ${A'}^i$ and ${B'}^i$, and thus the modular weights
of the light fields, are strongly constrained.

Let us discuss the second approach of sizable threshold corrections.
As discussed, the ${\bf Z}_3$ and ${\bf Z}_7$ orbifolds do not
lead to threshold corrections at all.
Therefore, considering a particular
non-minimal ${\bf Z}_3$ or ${\bf Z}_7$ orbifold with extra particles,
it will not lead to consistent unification of the coupling
constants unless $\alpha_3^0$ and $\sin^2\theta_W^0$ agree
with the experimental values. Thus, all cases displayed in table 4
cannot be used for these two types of orbifolds, unless one
introduces an intermediate scale.

For the remaining cases we assume that only one
modulus contributes to eqs.\alfs. This situation
is realized for the following three cases:

\noindent (i) If the orbifold has two completely rotated planes.
Then the entire moduli dependence of the threshold correction is given
by $T_1$.

\noindent (ii) If one assumes that $T_1>>T_2,T_3$ considering
an anisotropic, `squeezed' orbifold.

\noindent (iii) If one considers an isotropic orbifold
identifying the three moduli, i.e.
$T=T_1=T_2=T_3$.
(For orbifolds with two rotated planes cases (i) and (iii) are completely
equivalent since $T_2$ and $T_3$ do not contribute to eqs.\alfs.)
This possibility was
discussed       in ref.\ILR. In order to realize the minimal string
unification scenario, we found a unique solution for the overall modular
weights of the standard model particles, assuming that the range
of the
overall modular weights is between $-3$ and $-1$,
$-3\leq n_\alpha\leq -1$,
and also assuming generation universality:
$$\eqalign{
&n_Q=n_D=-1,\ n_U=-2,\ n_L=n_E=-3,\cr
&n_H+n_{\bar H}=-5,-4.\cr}\eqn\overallsol
$$
In addition, the overall compactification scale has to be
relatively large: ${\rm Re}T\sim 16$.
Repeating this analyis by taking into account the allowed values
of the overall modular weights of the standard model fields as
shown in table 3, and abandoning the assumption of generation
universality, we obtain that minimal unification can be realized
with the restriction to one overall modulus $T$ for the following
cases: ${\bf Z}_8'$, ${\bf Z}_{12}$, ${\bf Z}_{12}'$, ${\bf Z}_2\times
{\bf Z}_6$, ${\bf Z}_6\times {\bf Z}_6$. This result is
summarized in the $4^{\rm th}$ column of table 1.
For the other cases
minimal unification is either impossible or one has to consider
a squeezed orbifold.

In addition one has also to fullfill the requirements from the
modular anomaly cancellation associated to the completely rotated planes.
Specifically, for all orbifolds with two
completely rotated planes,
we have three sets of equations, which the modular
weights of the standard model particles have to satisfy. Namely,
there are first the conditions for complete modular anomaly cancellation
with respect to the second and third complex plane, i.e.
$$A^{2,3}{'}=B^{2,3}{'}=0.\eqn\abtpr
$$
Eq.\alfs\ is then a function of the
single modulus $T_1$.
Since $\log\lbrack(T_1+
\bar T_1)|\eta(T_1)|^4\rbrack<0$ for all possible values of $T_1$,
the correct value for $\sin^2\theta_W$ is obtained provided the following
condition is met:
${A'}^1<0$ (${A'}^1>0$) if
$\sin^2\theta_W^0<\sin^2\theta_W^{\rm exp}$
($\sin^2\theta_W^0>\sin^2\theta_W^{\rm exp}$).
Analogously,
for $\alpha_3^0>\alpha_3^{\rm exp}$
($\alpha_3^0<\alpha_3^{\rm exp}$) one requires ${B'}^1<0$ (${B'}^1
>0$).
Additional information about the allowed values of ${A'}^1$ and
${B'}^1$ can be obtained
if one eliminates the explicit $T_1$ dependence
by combining the two equations in \alfs. In this way one finds
that the ratio $\gamma={B'}^1/{A'}^1$ has to coincide with
the `experimental' parameter $\gamma^0$, which is defined in eq.\bbet.
For the
minimal string unification scenario we then obtain the following
third condition on the modular weights of the standard model particles:
$${A'}^1<0,\qquad {B'}^1<0,\qquad 2.7\leq \gamma
={{{B'}^1}\over {{A'}^1}} \leq3.7\eqn\unifcon
$$
All three coupling constants meet (at the one-loop level) at $M_X=
O(10^{16}{\rm GeV})$ if $\gamma=B/A=25/7$.

Let us check these three conditions for all orbifolds with two
completely rotated planes. Our results are summarized in the
$5^{\rm th}$ column of table 1. For example,
the ${\bf Z}_4$ orbifold compactifications allow for the following
modular weight vectors of the standard model particles (for Kac--Moody
level one): The untwisted choices are $\vec n_\alpha=(-1,0,0),(0,-1,0),
(0,0,-1)$. In the first twisted sector with $\vec\theta_1=(1/2,1/4,1/4)$
the modular weights for states without any oscillator
take the form $\vec n_\alpha=-(1/2,3/4,3/4)$. For the particles
$D_g,L_g,H,\bar H$ one can allow for a single twisted oscillator
either in the second or third plane. The corresponding modular
weights are $\vec n_\alpha=-(1/2,7/4,3/4),-(1/2,3/4,7/4)$.
Finally, in the second twisted sector with $\vec\theta_2=(0,1/2,1/2)$
all standard model particles must not have any oscillator
that one obtains in this sector $\vec n_\alpha=-(0,1/2,1/2)$.
Thus, in total there are five different choices for the modular
weights of the nine fields
$Q_g,U_g,E_g$ and seven possibilities for the eight fields
$D_g,L_g,H,\bar H$.
With the help of a computer program it is immediately checked
that eqs.\abtpr\ and \unifcon\ do not possess a single solution
using this range of modular weights. Thus all ${\bf Z}_4$
orbifold compactifications (with Kac--Moody level one)
do not pass the constraints
of a minimal string unification scenario.

The distribution of the allowed modular weights of the standard
model particles for the remaining cases with two completely
rotated planes, ${\bf Z}_6',{\bf Z}_8,{\bf Z}_8',
{\bf Z}_{12},{\bf Z}_{12}'$, is easily determined. Interestingly enough,
none of these compactifications, except ${\bf Z}_8'$,
is consistent with
the minimal string unification scenario.
On the other hand, for ${\bf Z}_8'$
the allowed values of the modular weights
for the standard model fields provide numerous solutions to eqs.\abtpr,
\unifcon. Thus minimal unification is in principle possible
for an isotropic, but relatively large ${\bf Z}_8'$ orbifold.
A specific solution of eqs.\abtpr,\unifcon\ has for example the following
form:
$\vec n_{Q_{1,2,3}}=(0,-1,0)$, $\vec n_{D_{1,2,3}}=(-1,0,0)$,
$\vec n_{U_1}=(0,-1/2,-1/2)$,
$\vec n_{U_{2,3}}=(-3/4,-15/8,-3/8)$,
$\vec n_{L_1}=(-14/8,-3/8,-7/8)$,
$\vec n_{L_{2,3}}=(-14/8,-7/8,-3/8)$,
$\vec n_{E_1}=(-1,0,0)$,
$\vec n_{E_{2,3}}=(-3/4,-15/8,-3/8)$,
$\vec n_H=(-1/2,-3/4,-3/4)$,
$\vec n_{\bar H}=(-14/8,-3/8,-7/8)$.
This choice requires that ${\rm Re}T_1\sim 24$.

Now turn to the ${\bf Z}_6$ orbifold with only one rotated plane
(the third). As discussed before, minimal unification is not
possible for the isotropic case with $T_1=T_2$. Thus we demand
$T_1>>T_2$, and the modular weights $n_\alpha^2$
are no longer constrained, since the threshold corrections
depend only very little on $T_2$. (This fact comes from
the $T$-dependence of the Dedekind function.) Thus there
are only two sets of equations to be satisfied. In turn,
minimal string unification is now possible. In fact there
are numerous solutions of eq.\unifcon\ together with the
anomaly requirement
$A^{3}{'}=B^{3}{'}=0$. Just for illustration, one randomly chosen
solution with generation universality
looks like $\vec n_{Q_{1,2,3}}=\vec n_{D_{1,2,3}}=(0,-1,0)$,
$\vec n_{U_{1,2,3}}=(-2/3,0,-1/3)=\vec n_{E_{1,2,3}}=(-2/3,0,-1/3)$,
$\vec n_{L_{1,2,3}}=(-5/3,0,-1/3)$,
$\vec n_H=\vec n_{\bar H}=(0,0,-1)$.
In addition to satisfying the constraints on the modular weights,
minimal string unification also requires a moderately large value
for ${\rm Re}T_1$. For example, the above solution
corresponds to ${\rm Re}T_1\sim 10$.
Let us remark   at this point that in all the relevant examples
we checked that the value of ${\rm Re}T_1$
was not too big and dominated the tree-level coupling constant.

Finally, for all ${\bf Z}_N\times {\bf Z}_M$ orbifolds,
the constraints from modular anomaly cancellation are not anymore
present, and
minimal string unification is possible making again the
squeezing assumption $T_1>>T_2,T_3$. (See the last column in table 1.)
For example, a possible
set of modular weights that satisfies eq.\unifcon\ has the following
form, considering the
${\bf Z}_2\times {\bf Z}_2$ orbifold:
$\vec n_{Q_{1,2,3}}=(-1/2,-1/2,0)$, $\vec n_{U_{1,2,3}}
=\vec n_{L_{1,2,3}}
=\vec n_{E_{1,2,3}}=\vec n_H=(-1,0,0)$,
$\vec n_{D_{1,2,3}}=\vec n_{\bar H}=
(0,-1,0)$.
In addition, the
${\bf Z}_2\times {\bf Z}_6$ and
${\bf Z}_6\times {\bf Z}_6$ orbifolds also allow for minimal
string unification with $T_1=T_2=T_3$ (see the $5^{\rm th}$
column of table 1).

The conclusion of the above analysis is that ${\bf Z}_N$ orbifolds
cannot accommodate a model in which the only charged fields
with respect to $SU(3)\times SU(2)\times U(1)$ are those of the
minimal supersymmetric standard model except
${\bf Z}_6$ and ${\bf Z}_8'$.
A ${\bf Z}_N\times {\bf Z}_M$
orbifolds can also yield adequate coupling unification
in general. Thus we are able to rule out the existence and
possible experimental consistency of large classes of
4-D strings based on (0,2) orbifolds.

The above specific analysis was done assuming the low
energy particle content of the minimal supersymmetric
standard model, but it may obviously be straightforwardly
applied to any extended model. In particular, one can repeat
the analysis e.g. for any of the extended models of table 4
in which the massless spectrum is enlarged by including
some extra chiral multiplets \refs{\PLANCK,\ANTON}.
In order for any of those
models to yield consistent unified couplings, one must
have e.g. ${A^1}'/{B^1}'\sim \gamma ^0$ and adequate
cancellation of modular anomalies in the rotated planes
(if there are any). The signs of the $A'$ and $B'$ also have
to be appropriate. We refrain from presenting a complete
analysis of all the models in table 4. It is however worth
remarking that the conditions for adequate coupling unification
through threshold effects are in some cases more easy to fulfill
than in others. For example, it is easier to make
the extended models b and j in
table 4 agree with experiment
than the minimal model itself. The extended model b
contains extra fields $O=(\underline 8,\underline 1,0)$ and
$T=(\underline 1,\underline 3,0)$ as well as
an extra pair of multiplets transforming like right-handed
electrons. The massless fields $O$ and $T$ appear in possible
grand unified models in which the (e.g. $SU(5)$) symmetry
is broken by a  Higgs field in the adjoint \PLANCK. This requires an
original string model with higher Kac--Moody level. Thus,
using for simplicity only the overall modulus $T$, one finds
adequate values for $\sin^2\theta _W$ and $\alpha _3$, using  only
the most common overall modular weights $-1,-2$. One possible solution is
found for ${\rm Re}T=7.0$ and generation-independent modular weights
$$
\eqalign{
n_Q\ &=\ n_L\ =\ n_E\ =\ n_{\bar H}\ =\ -1,     \cr
n_U\ &=\ n_D\ =\ n_H\ =\ -2    \cr }
\eqn \otee
$$
with the modular weights of the extra particles $(O+T+E+{\bar E})$
equal to $-1$.  The same happens with model j, which was considered in
ref.\ANTON. This includes extra fields transforming like conjugate
pairs $Q+{\bar Q}+D+{\bar D}$, which may be available at Kac--Moody
level one.  For example, one can find good agreement with experiment
for ${\rm Re}T=9.0$ and flavor-independent modular weights
$$
\eqalign{
n_U\ &=\ n_D\ =\ n_E\ =\ -1,\cr
n_Q\ &=\ n_L\ =\ n_H\ =\ n_{\bar H}\ =\ -2 \cr }
\eqn \qqdd
$$
with the modular weights of the extra $(Q+{\bar Q}+D+{\bar D})$ fields
equal to $-1$.  Many other solutions giving up flavor-independence
and considering the three complex moduli exist for both models.
The important point is that in either case, one does not need to
have massless fields with modular weight $\leq -3$ to obtain
good agreement. Although we have not done a complete
scan for all ${\bf Z}_N$ orbifolds, this seems to imply that,
unlike what happens with the minimal model, one can have
various
${\bf Z}_N$    models with correct $\sin^2\theta _W$ and $\alpha _3$
results at the price of adding some particular extra massless
fields. In addition, isotropic ${\bf Z}_N\times {\bf Z}_M$ orbifolds,
described by an overall modulus $T$, should also be  allowed.
This is a different approach to that in \ANTON\  where
two extra mass scales are introduced in order to get good
agreement for the model j. In the present approach no extra
mass scales are needed and the string threshold effects
render the results consistent with experiment.

\chap{ Constraints on SUSY-breaking soft terms from duality
invariant actions.}

We will now turn to a different type of application of the target-space
duality symmetries in 4-D orbifold strings.
It is well known that in low-energy supergravity theories \MALLOR ,
if supersymmetry breaking occurs in a $hidden$ $sector$ of
the theory, soft-susy breaking terms are generated in the
non-singlet $observable$ $sector$. Since the effective low
energy Lagrangian from strings is an $N=1$ supergravity
theory, the same type of soft susy-breaking terms will be
generated in this case. The possible soft terms will
include i) gaugino Majorana masses $M_a$,
ii) soft scalar mass terms of the type $m_{\alpha }^2\phi _{\alpha }
\phi ^{*}_{\alpha }$ for each of the scalars in the theory,
and iii) soft scalar couplings proportional to each of the
superpotential terms.
In principle there are as many different soft terms as independent
particles and/or couplings     present. Simplifying  assumptions
reduce the number of independent soft terms by imposing some
hypothetical symmetry which might be present at the GUT/Planck
scale. The presence of these soft terms reflects itself into
the SUSY-particle mass spectrum at low energies. If the idea of
low-energy supersymmetry is correct, the latter should be
amenable to experimental tests in future accelerators. It would
thus be very important  to find constraints on the pattern of
SUSY-breaking soft terms in effective low-energy
Lagrangians from strings.

This might look hopeless, since we know that in string theory
supersymmetry breaking necessarily has to be a non-perturbative
phenomenon and we are far from a non-perturbative
understanding of string theory. This view is, however,
unnecessarily pessimistic. One does not need            to understand
all the details of a symmetry-breaking process in order
to obtain important physical information. The most obvious
example of this is the standard model itself: we do not know
how the symmetry breaking process $SU(2)_L\times U(1)_Y\rightarrow
U(1)_{em}$ takes place yet; that does not stop us from  computing
all the extremely successful predictions of the standard model.
Simply assuming that there is a certain doublet scalar field
(fundamental or composite), which gets a $vev$ $v\not= 0$, is
enough to obtain most of the relevant information in the
standard model, we do not need to know how or why it happens so.
In the same way one might hope to get some physical
information about the susy-breaking soft terms without
knowing all the non-perturbative dynamics from which they originate.
In particular, as we now describe,
one can get some interesting information
about those terms in effective low-energy theories from
orbifold-like 4-dimensional strings.

Most supersymmetry-breaking scenarios discussed up to now
\refs{\GAUGINO,\DIN,\FILQ,\MAGN,\DUAGAU,\CFILQ,\LM} in
the context of strings assume that the ``seed'' for SUSY-breaking
is provided by the auxiliary fields of the dilaton $S$ and the
moduli $T_i$. On the other hand, as we have discussed in the
previous sections, the matter field kinetic terms depend on the
moduli $T_i$ in a way parametrized by the modular weights.
As we will now discuss, this fact makes      the soft SUSY-breaking
terms depend in a known way on the modular weights and allow
us to draw some constraints on them.

\sect{Gaugino masses}

Let us consider a $(0,2)$ type of 4-D string based on an
Abelian ${\bf Z}_N$ or ${\bf Z}_N\times {\bf Z}_M$ as the ones discussed in
prev
sections. We want to address now the  question of the soft
gaugino masses $M_a$         in this type of theories.
The value of those masses is given in a general $N=1$ supergravity
Lagrangian by \CREMMER
$$
M_a\ =\ \sum _{\alpha}\ f_a^{\alpha }(\phi_\alpha)
\ K_{\alpha\bar\beta}^{-1}
\ G_{\bar\beta} . \eqn \mmm  $$
$f_a^\alpha$ is the derivative of the gauge kinetic function $f_a$
with respect to the        field $\phi _\alpha$,
$K_{\alpha\beta}^{-1}$
is the inverse K\"ahler metric and $G_\alpha=\partial G/\partial
\phi_\alpha$
is the auxiliary field
of $\phi _\alpha$, where $G(\phi_\alpha,\bar\phi_\alpha)=
K(\phi_\alpha,\bar\phi_\alpha)+\log |W(\phi_\alpha)|^2$ is the
$N=1$ K\"ahler function.
The sum runs over all massless chiral fields and
$a$ labels each gauge group factor.
In principle the $f$-function
could also depend on the charged matter fields but the presence
of those fields in $f$ is severely constrained by gauge
invariance. Thus we will discuss only the $S$--field and gauge singlet
moduli dependence of the gauge kinetic function.
Furthermore,       as discussed in the previous
section, one expects for orbifolds
the $T_i$-dependent piece to be the
leading one-loop contribution to $f$ due to the exponential
large-$T$ behaviour   of the $\eta $ functions.
Thus the gauge kinetic function has the general form
$$f_a(S,M_i)=k_aS+{1\over 16\pi^2}\Delta(M_i).\eqn\gaugekina
$$
Here, $\Delta(M_i)$ is the one-loop threshold contribution
from the massive string excitations given in terms of
automorphic functions of the corresponding duality  group $\Gamma$.
The moduli dependence of the holomorphic gauge kinetic function
was already noted in \IN\ in the context of truncating the
ten-dimensional effective Lagrangian of the heterotic string.
Then the gaugino masses take the form
$$M_a=k_a
(K_{S\bar S}^{-1}G_{\bar S}+K_{S\bar M_i}^{-1}G_{\bar M_i})
+{1\over 16\pi^2}{\partial\Delta_a(M_i)\over\partial M_i}
(K_{M_i\bar S}^{-1}G_{\bar S}+K_{M_i\bar M_j}^{-1}G_{\bar M_j}).\eqn
\gaugmassa
$$
Here we have allowed for a mixing between the $S$--fields and the
moduli in kinetic energy which is expected to occur beyond the
tree-level \refs{\IN,\DFKZ}. It is important to realize that the above
expression for the gaugino masses is not target space duality-invariant
since ${\partial\Delta_a(M_i)\over\partial M_i}$
transforms inhomogeneously under duality
transformations. Duality invariance is restored when one also includes
the moduli dependent one-loop contribution of the massless fields
to the effective Yang-Mills Lagrangian.
This is
just a SUSY counterpart of the non-holomorphic contribution
to the gauge kinetic terms discussed in section 3.
Specifically, the non-local Lagrangian eq.\nl\
together with the threshold
corrections leads to the following expression for the
gaugino masses valid in general
compactification schemes:
$$\eqalign{M_a&=k_a
(K_{S\bar S}^{-1}G_{\bar S}+K_{S\bar M_i}^{-1}G_{\bar M_i})\cr&
+{1\over 16\pi^2}\biggl({\partial\Delta_a(M_i)\over\partial M_i}-
2\bigl\lbrack C(G_a)
-\sum_{\underline R_a}
T(\underline R_a)\bigr\rbrack
{\partial K(M_i,\bar M_i)\over\partial M_i}\cr&+
4\sum_{\underline R_a}T(\underline R_a)
{\partial\log\det K_{\alpha\beta}^{\rm matter}(M_i,\bar M_i)\over
\partial M_i}
\biggr)
(K_{M_i\bar S}^{-1}G_{\bar S}+K_{M_i\bar M_j}^{-1}G_{\bar M_j}).\cr}\eqn
\gaugmassa
$$

Now let us compute above expressions for orbifold compactifications.
The holomorphic gauge kinetic function is
determined by the threshold corrections eq.\thresor\ and
has the form \FILQ,\DKLB ,\DFKZ\ (by the same reasons as in the
previous chapter we neglect the $U_m$ dependence of the threshold
corrections in the following):
$$f_a(S,T_i)=k_aS-{1\over 16\pi^2}\sum_{i=1}^3({b'}_a^i-k_a\delta_{GS}^i)
\log \eta(T_i)^4.
\eqn\gaugekin
$$
Then one finds for the gaugino masses
$$
\eqalign{M_a &= f_a^{S}
( K_{S\bar S}^{-1}G_{\bar S}+
K_{S\bar T_i}^{-1}G_{\bar T_i})
+f_a^{T_i}(
K_{T_i\bar S}^{-1}G_{\bar S}+K_{T_i\bar T_j}^{-1}G_{\bar T_j})
 =  \cr
&= k_a
( K_{S\bar S}^{-1}G_{\bar S}+
K_{S\bar T_i}^{-1}G_{\bar T_i})+
{1\over {16\pi ^3}}
({b'}_a^i-k_a\delta_{GS}^i)G_2(T_i)(
K_{T_i\bar S}^{-1}G_{\bar S}+K_{T_i\bar T_j}^{-1}G_{\bar T_j}),\cr}
\eqn \mma
$$
where $G_2(T)=-4\pi{\partial\eta(T)\over\partial T}/\eta(T)$
is the holomorphic
Eisenstein function. Again we notice that the above
expression is $not$ modular invariant since the function
$G_2(T)$ transforms in an ``anomalous way" under modular
transformations:
$G_2(T)\rightarrow (icT+d)^2G_2(T)\ -\ 2\pi ic(icT+d)$.
The gaugino mass is made modular invariant by the
further contribution from the $massless$ fields of the theory,
which is proportional to $2{b'}_i^a/(T_i+\bar T_i)$.
This induces a replacement of ${b'}_a^iG_2(T_i)$ in eq.\mma\
by the modified weight two (non-holomorphic) Eisenstein
function ${b'}_a^i{\hat G}_2(T_i)$ which transforms as
${\hat G}_2\rightarrow {\hat G}_2(icT_i+d)^2$
and admits an expansion
$$
{\hat G}_2(T_i,\bar T_i )\ =\ G_2(T_i)\ -\ {{2\pi }\over
{T_i+\bar T_i}}\ \simeq\ {{\pi ^2}\over 3}\
-\ {{2\pi}\over {T_i+\bar T_i
 }} \ -\ 8\pi ^2 e^{-2\pi T_i}\ -\ ....
\eqn \gga
$$

We do not know the physical $vev$'s of the relevant fields
$S,T_i$, nor the auxiliary fields which break supersymmetry
$G_{\bar S}, G_{\bar T_i}$ nor the higher-loop
expressions for the inverse K\"ahler metrics. However our ignorance
can be parametrized
in terms of some unknown parameters and eq.\mma\ can be rewritten:
$$
M_a\ =\ M^0(S,T_i)\ k_a\ +\ \sum _{i=1}^3\
{b'}_a^i\ {M'}^i(S,T_i).
\eqn \mmb
$$
Here $M^0$ is a gauge group independent parameter and all the
group dependence is included in the second term.
If we have $m$ independent gauge groups $a=1,2..,m$, one can
eliminate the universal piece by taking linear combinations
$$
\sum_{a =1}^m\ V_a\ {{M_a}\over {k_a}}
\ ;\qquad \sum _a^m\ V_a\ =\ 0 ,
\eqn \aaa $$
where the $V_a$ are appropriately chosen constants. If the number
of gauge groups is bigger than the number of free parameters, one
will get constraints amongst the $m$ physical gaugino masses.

All gaugino masses will be equal
(modulo the corresponding Kac--Moody level)
for the ${\bf Z}_3$ and ${\bf Z}_7$ orbifolds with three completely
rotated complex planes.
Also, if a large gauge group unifies the $m$ different
gauge groups, all the ${b'}_a^i$s will be equal and again
we will have universal boundary conditions for gaugino masses.
This is the case, for instance, of an $SU(5)$ symmetry unifying the
three interactions of the standard model and it is the usual
assumption that is taken in the minimal supersymmetric
standard model, i.e. universal gaugino masses.

In a   more general case,  there will be four  unknown
parameters $M^0,{M'}^1,{M'}^2,{M'}^3$, which will generically lead
to non-universal gaugino masses for any non-unified string
model. In certain situations one expects a reduction to two in
the number of free parameters. Indeed, if any of
the following three possibilities  is realized, only two
parameters, $M_0$ and $M'$, will be relevant (see also the discussion
about these three cases concerning the unification of the coupling
constants in section 4):

\noindent (i) If the particular orbifold considered has just one plane
that is left (for some particular twist) unrotated.
If the relevant plane is the $i^{\rm th}$ one, only ${M'}^i$
will be non-vanishing.

\noindent (ii) If the supersymmetry-breaking dynamics are such that
the modulus of a particular complex plane plays a leading role
(i.e. $|G_{T_1}|\gg |G_{T_2}|,|G_{T_3}|$), then only one of the
${M'}^i$s  will be relevant. This normally corresponds to $T_1>>T_2,T_3$.

\noindent (iii)
If we concentrate on the diagonal modulus one has, $T=T_1=T_2=T_3$
and hence one expects ${M'}^1={M'}^2={M'}^3\equiv M'$. This is
equivalent assuming an approximate symmetry
among the three complex planes leading to $G_{T_1}=G_{T_2}=
G_{T_3}$  for the auxiliary fields. In fact this is an
assumption that is often made in the supersymmetry-breaking
scenarios discussed up to now.

In these three cases, the number of free parameters will be
two, $M_0$ and $M'$, and hence we will get constraints on the
soft gaugino masses as long as the number of group factors is
larger than two. Physically the most interesting non-trivial
case is  that of three gauge group factors, which includes
the standard model and some of its more interesting extensions:
$SU(4)\times SU(2)_L\times SU(2)_R$, $SU(4)\times SU(2)_L\times
U(1)_R$, and $SU(3)_c\times SU(3)_L\times SU(3)_R$. Let us
first discuss  the case of the standard model group. One can
take two different linear combinations of gaugino masses
as in eq.\aaa  , for example, one with
$         {\vec  V  }_A=(k_1,k_2,-k_1-k_2)$ and the other with
${\vec V}_B=(k_2,-k_2,0)$, where $k_a$ are the Kac--Moody levels
of the standard model group (the reasons for this choice will
become  clear
below).                  Let us call these linear combinations
$M_B$ and $M_A$ respectively. Taking the ratio of the two, one finds:
$$
{\gamma }_M\ \equiv \
{{M_B}\over {M_A}}\ =\ {{M_1\ +\ M_2\ -{{(k_1+k_2)}\over {k_3}}
M_3}\over {{{k_2}\over  {k_1}} \ M_1\ -\ M_2}}\ =\
{{\sum _{i=1}^3\ {B'}^i\ {M'}^i}\over {\sum _{i=1}^3\ {A'}^i\
{M'}^i}} \ ,
\eqn \mam   $$
where ${A'}^i$ and ${B'}^i$ are the linear combinations of $b'$s
discussed in section 3 and the sum runs over the three orbifold
complex planes.         This is why we took the particular linear
combinations $M_A$ and $M_B$, since in this way we can connect
the gaugino masses to the threshold corrections discussed in
section 4. One can also rewrite the above equation as
$$
M_1\ (1-\gamma _M{{k_2}\over {k_1}})\ +\ M_2\ (1+\gamma _M)
\ -\ {{k_1+k_2}\over {k_3}}\ M_3\ =\ 0 .
\eqn \msem
$$

Consider first cases (i) and (iii) discussed
above, in which only one $M'$ variable is relevant; the explicit
$T_i$ dependence of $\gamma _M$ drops and one can write
$$
\gamma _M\ =\ {{\sum _{i=1}^3\ {B'}^i}\over {\sum _{i=1}^3 {A'}^i}}
\ =\ {{B'}\over {A'}}\ =\ \gamma
\eqn \gmmm
$$
where $\gamma $ was defined in section 3. Thus in that class of
models one has $\gamma _M=\gamma$ and one can obtain specific
constraints amongst gaugino masses if one has a knowledge of
the $b'$-coefficients, i.e. if one knows the massless spectrum and the
modular weights of the massless particles.

There are several cases in which $\gamma _M$ can be
easily computed without a detailed knowledge of each string model:

1) There are large classes of models in which one has
${b'}_a=b_a$. This is the case of
${\bf Z}_2\times {\bf Z}_2$ symmetric $(0,2)$ orbifolds, which can
also be constructed (at their multicritical $T_i$ values)
in terms of the free-fermion method. This will also be the case
of models with all the physically relevant particles either in
the untwisted sector or in twisted sectors with one
unrotated plane and no oscillator contribution.
In all these cases the overall modular weights
of the particles is $-1$ and hence ${b'}_a= b_a$.
In these cases one has
$$
\gamma _M\ =\ \gamma \ =\ {{B'}\over {A'}}\ =\ {{B}\over {A}}\ ,
\eqn \zdzd
$$
and one can compute  $\gamma _M$ if one just knows the massless
$SU(3)\times SU(2) \times U(1)$ spectrum of the theory.
For example, we discussed in the previous chapter how
one can enlarge the particle content of the minimal
SUSY-model by adding extra multiplets in order to get the
correct measured values for $\sin^2\theta _W$ and $\alpha _3$.
Since the $\beta$-functions for all these possibilities are known,
one can compute for each of them $\gamma _M=\gamma =B/A$.
The values of $B/A$       for some extended models of this type
may be found in the last column of table 4.
They               vary in a range $3\sim 6$.
For example, if one
considers a level one model with $3/5k_1=k_2=k_3=1$, the
last possibility (model j)  \ANTON ) with $\gamma _M=5$ would give rise
to the constraint among gaugino masses
$$
2 M_1\ -\ 6 M_2\ +\ {8\over 3} M_3\ =\ 0 \ .
\eqn \ell
$$
Notice that the relevant spectrum in computing the $b$'s is the
one present {\it just} {\it below} the Planck scale. Whether some
of the massless
particles at that level get an intermediate mass or not does not
modify the $T$-dependence of the $f$-function and hence does not
affect the string boundary conditions.

2) In other classes of models one can get information on $\gamma $
(and hence on $\gamma _M$) by imposing that the string threshold
corrections adequately explain the discrepancy between the
naive string result for $\sin^2\theta _W$ and $\alpha _3$ and
the experimental results \ILR . This was discussed in detail in the
previous section. We saw that, if we restrict ourselves to
one modulus, we need to have
$2.7\leq \gamma \leq 3.7$ for the minimal string unification scenario.
Furthermore, one has $\gamma =25/7$
if one wants the three couplings to join at a scale
$\simeq 10^{16}$ GeV  \ILR . Thus in this case one would get
a constraint
$$
M_1\ -\ 4\ M_2\ +\ {4\over 3}\ M_3\ =\ 0 \ .
\eqn \ilr
$$

Let us now discuss
the case (ii) in which the number of relevant
free parameters reduces to two.
Then, instead of eq.\gmmm, one has $\gamma_M={{B'}^1\over {A'}^1}=
\gamma$.
If in this case one also insists in explaining the needed
corrections to $\sin^2\theta _W$ and $\alpha _3$ purely in
terms of string threshold corrections one will again have
$2.7\leq \gamma =B_1/A_1 \leq 3.7 $, and $\gamma _M$ will also
be restricted by those limits.

The advantage of the situations discussed above is that we do not need
to know details about the specific string model nor the
supersymmetry-breaking mechanism  to get interesting
physical constraints. Notice that there is a possible limiting
situation in which $M^0   =0$.
In this case we would      be left with a unique
free parameter $M'$ and one has

$$
{{M_1}\over {b_1'}}\ =\ {{M_2}\over {b_2'}}\ =\ {{M_3}\over {b_3'}} .
\eqn \msi
$$
This situation is not in general expected. For example, in ref. \CFILQ\
it was found that for a typical modular invariant non-perturbative
potential the minima must have $G_S\not= 0$          for  the
potential to be stable.

All the above discussion can be straightforwardly generalized to
the other extended gauge models with three gauge factors.
Consider one such extended model with three gauge groups
labelled by an index $a =1,2,3$ in such a way that the three
$U(1)\times SU(2)\times SU(3)$ generators
($\alpha =1,2,3$) can be written as
$$
I_{\alpha}\ =\ \sum _{a =1}^3\ c_{\alpha  a }\ T_{a } \ .
\eqn \gen
$$
The relationship between the Kac--Moody levels is given by
$k_{\alpha }\ =\ \sum _{a } \ c_{\alpha  a }^2\ k_{a }$.
Define two linear combinations of gaugino masses, $M_A$ and $M_B$,
by taking
$$
\sum _{\alpha =1}^3\ \sum _{a =1}^3\ a_{\alpha }\ c_{\alpha ,a }^2
\ M_{a },
\eqn \com
$$
where $         {\vec V}_A=(k_1,k_2,-k_1-k_2)$ for $M_A$ and
$         {\vec V}_B=(k_2,-k_2,0)$ for $M_B$. Then one finds
$$
\gamma _M\  =\ {{M_1\ +\ M_2\ -\ {{(k_1+k_2)}\over {k_3}}\ M_3}
\over {{{k_1}\over {k_2}}\ M_1\ -\ M_2}}\ =
\ {{B_{\rm ext.}'}\over {A_{\rm ext.}'}}\ \equiv \ \gamma _{\rm ext.} \ ,
\eqn \ext  $$
where one defines
$$
{A'}_{\rm ext}\
=\   \sum _{\alpha =1}^3\ {V_A^\alpha\over k_\alpha}
\ \sum _{a }\ c_{\alpha a }^2\ {b'}_{a },\ \qquad
{B'}_{\rm ext}\
=\  \sum _{\alpha =1}^3\ {V_B^\alpha\over k_\alpha}
\  \sum _{a }\ c_{\alpha a }^2\ {b'}_{a} \ .
\eqn \aab
$$
We see in the above formulae that the value of $\gamma _M$ is determined
by the value of the ${b}'_a$-coefficients
of the extended theory which is
valid {\it at the string scale}. One can then use eq.\msem \  and the
relationship between the Kac--Moody levels of the standard model
and those of the extended theory to derive the relevant
constraint. The result is independent of the scale in which the
extended symmetry is broken down to the standard model.
For example, in the case of $SU(4)\times SU(2)_L\times SU(2)_R$
one gets ${A'}_{\rm ext}={b'}_{R}+\ {b'}_{L}-(k_R+k_L)/k_4 \ {b'}_{4}$
and ${B'}_{\rm ext}= (k_R+2/3k_4)/k_L\ {b'}_L\ -\ {b'}_R-2/3\ {b'}_4$.
These are
also the coefficients that are relevant for the threshold
corrections to the computations of $\alpha _3$ and
$\sin^2\theta _W$ for this particular extended gauge model.

Let us finally comment that all the above gaugino mass relationships
apply at the string scale. In order to make contact with the
low-energy physical quantities one has to take into account two
facts. First, one has to redefine the Planck mass gaugino fields in order
to reabsorve the non-minimal kinetic terms
(proportional to $f_a(M_{string})$). Second, one has to run down
the gaugino masses according to the renormalization group equations.
                 In fact it is well known that gaugino masses
are renormalized  \INO\
                 in exactly the same way as the corresponding
coupling constant and the ratio $M_a(\mu )/\alpha _a(\mu )$
is renormalization group invariant. The net effect of these two
points is the replacement  $M_a(M_{\rm string})
\rightarrow M_a(M_W)
    /\alpha _a(M_W       )$ in eq.\msem . In this way the corresponding
low-energy constraint is obtained.

\sect{ Scalar masses }

It is also useful to display the explicit dependence of the
soft SUSY-breaking scalar masses on the modular weights.
The scalar potential in the effective low-energy supergravity
action has the form \CREMMER
$$ V =  e^G \biggl\lbrace
\sum_{\alpha,\beta}         G_{\phi_\alpha}
G_{\bar\phi_\beta}  K_{\phi_\alpha\bar\phi_\beta}^{-1}
        - 3
  \biggr\rbrace ,\eqn\scalpotmatta
$$
where $K$ is the total K\"ahler potential and sum runs over
all charged massless chiral fields $\phi_\alpha$
(we neglect here D-terms which
are irrelevant in the present analysis).
                        One may describe the relevant non-perturbative
effects that break    supersymmetry through the introduction of a
non-vanishing (non-perturbative) superpotential for the moduli
fields of the string, $W_0(S,T_i)$ \refs{\GAUGINO,\FILQ}.
This seems reasonable since
supersymmetry breaking needs to be a field-theoretical phenomenon
occurring well below the string scale if it is going to be of
any use in solving the hierarchy problem. One usually assumes  also
       that supersymmetry breaking takes place in a ``hidden
sector'' of the theory which communicates with the ``observable
world'' only gravitationally. It is this ``hidden sector''
which will generate the non-perturbative superpotential $W_0$
and generate supersymmetry-breaking. This is precisely the
situation in the gaugino condensation mechanism abundantly
discussed in the literature
\refs{\GAUGINO,\DIN,\FILQ,\MAGN,\DUAGAU,\CFILQ,\LM}.
We will however simply assume that
supersymmetry breaking occurs somehow giving rise to non-vanishing
vacuum expectation values for the auxiliary fields of the
marginal-operator fields $S,T_i$.

Let us consider the scalar potential for orbifold compactifications.
One can expand
the scalar potential $V$ around the classical
values \CFILQ\  of the matter fields
$A_\alpha=0$:
$$
V\ =\ V_0(S,\bar S,T_i,\bar T_i)\ +\ \sum_\alpha{V_1}_\alpha
(S,\bar S,T_i,\bar T_i)
\ A_\alpha\bar A_\alpha \ +....
\eqn \vvv
$$
Writing $K\ =-\log(S+{\bar S})+K_0(         T_i,\bar T_i)\
+\ K^{\rm matter}(T_i,\bar T_i)A_\alpha\bar A_\alpha$,
see eq.\mattkpor, one finds
 $$\eqalign{  V_0= {e^{K_0}\over(S+\bar S)}
 &\Biggl\lbrace  (S+\bar S)^2\vert D_S W_0\vert ^2 +
               \sum _{i=1}^3   K_{0T_i\bar T_i}^{-1}\vert D_{T_i} W_0
 \vert^2                   -3\vert
                    W_0\vert^2  \Biggr\rbrace ,\cr
 {V_1}_\alpha= {e^{K_0}K^{\rm matter}\over (S+\bar S)}
 &\Biggl\lbrace
 (S+\bar S)^2\vert D_SW_0   \vert^2 +
\sum _{i=1}^3 K^{-1}_{0T_i\bar T_i}  \vert D_{T_i}
 W_0\vert^2                  \times \cr
 & \biggl\lbrack
  1- {K^{\rm matter}_{T_i\bar T_i}\over K_{0T_i\bar T_i}
 K^{\rm matter}}+
  {K^{\rm matter}_{T_i}
 K^{\rm matter}_{\bar T_i} \over (K^{\rm matter})^2
 K_{0T_i\bar T_i} }\biggr\rbrack
 -2\vert W_0      \vert^2 \Biggr\rbrace  .\cr} \eqn\scalpotmattb
 $$
With the explicit form of $K^{\rm matter}$ in terms of
the modular weights $n_\alpha^i$
one can easily check that the quantity in
brackets in ${V_1}_\alpha$ is equal to $1+n_\alpha^i$.
Rescaling the matter
fields in order to get properly normalized particles, one finally
gets a general expression for the soft scalar masses of the form
$$
m_\alpha^2\ =\ m_0^2(S,T_i,\bar T_i)\ + \ \sum _{i=1}^3\ n_\alpha^i\
{m'}^2(S,T_i,\bar T_i) .
\eqn \semesc
$$
The first term in the right-hand side is universal (i.e. does not
depend on the particular matter field $A_\alpha$ considered) and
one can easily check that
$$
m_0^2\ =\ m_{3/2}^2\ +\ V_0 ,
\eqn \cosm
$$
where $m_{3/2}^2=\exp K_0|W_0|^2$ is the gravitino mass squared and
$V_0$ is essentially the cosmological constant.
The second term in eq.\semesc\ depends on the modular weight $n_\alpha^i$
of the field $A_\alpha$
along each of the three complex planes and is in general not
universal. In eq.\semesc\ ${m'}^2$ is positive-definite, whereas
$m_0^2$ is not, since $V_0$ may be negative. Indeed, for simple
examples of $W_0$, the absolute minimum of the scalar potential
leads to $V_0<0$ \refs{\FILQ,\CFILQ}.
One can, on the other hand, find examples \CFILQ\
in which $V_0=0$ and hence $m_0^2=m_{3/2}^2$.
The mass$^2$ of matter fields may in general be negative (and hence
the potential unstable) unless certain conditions are fullfilled
(eq. 4.11 in ref.\CFILQ). We assume here that those conditions are
fullfilled or, more generally, that the minimum of the
potential is stable.
Equation \semesc\ shows that, contrary to the standard assumptions in
the minimal low-energy supersymmetric models \MALLOR  , the soft scalar
masses are $not$ $universal$ but depend on the modular
weights of the particles.
In particular, fields with higher (negative) modular weight
have smaller soft masses than those with smaller weight.
Since generically the modular weights of each individual physical
field are unknown, it is not easy to obtain definite constraints
on scalar masses in a model-independent way. If one considers
the simplified case of a single modulus, e.g.
the diagonal modulus $T=T_1=T_2=T_3$, one
can, as we did with the gaugino case, form ratios of scalar
masses combinations in which the explicit dependence on the
unknown $m_0^2,{m'}^2$ parameters drops. Thus, e.g. consider
the three $u$-like squarks ${\tilde u},{\tilde c},{\tilde t}$.
In the above scheme one would get
$$
{{m_{\tilde u}^2\ -\ m_{\tilde c}^2}\over
{m_{\tilde u}^2\ -\ m_{\tilde t}^2 }}
\ =\ {{n_u\ -\ n_c}\over {n_u\ -\ n_t }} ,
\eqn \fcnc
$$
where the $n$'s are the overall modular weights of each particle.
We know                                   from the
absence of flavor-changing neutral currents in kaon decays
that $m_{\tilde u}^2$ and $m_{\tilde c}^2$  must be almost degenerate
for the SUSY-GIM mechanism to work. This would suggest the
construction of string models in which both fields have similar
modular weights (or else SUSY-breaking mechanisms in which
${m'}^2\ll m_0^2$).
Notice that (unlike the gaugino case) in the case of scalar masses
the above non-universality effects may be present even if there is
a unification symmetry like $SU(5)$.

In  the absence of a knowledge of the modular weights of the
physical quark and lepton fields, one can try to form
combinations of scalar masses  which could somehow be related
to the particular combinations of modular weights included
in the quantities $A'$ and $B'$ discussed in previous sections.
Consider as an example the minimal particle content of the
SUSY standard model and take the following two
combinations of soft scalar masses
$$
\eqalign{
\sum _{g={\rm gen.}       }   (2{m_Q^g}^2-{m_U^g}^2-{m_E^g}^2)  =
&({5\over 8}\delta A+{1\over 8}\delta B)\ {m'}^2(S,T)  \cr
\sum _{g={\rm gen.}       }   (3{m_Q^g}^2-2{m_U^g}^2+{m_L^g}^2-
& {m_D^g}^2-{m_E^g}^2)  =\cr  &({5\over 4}\delta A-{1\over 4}\delta B-2
-(n_H+n_{\bar      H}))\ {m'}^2(S,T) \cr }
\eqn \esc
$$
where $\delta A\equiv A-A'$, $\delta B\equiv B-B'$.
The sum runs over the three quark-lepton generations.
Taking the ratio of the two equations one cancels the dependence on
$m'$ and gets a sum-rule in terms of $\delta A$,$\delta B$ and
the sum $n_H+n_{\bar H}$. If we have some information about those
quantities (e.g. from imposing that the string threshold
corrections yield the correct result for $\sin^2\theta _W$ and
$\alpha _3$) one can in principle check the consistency
of the model by checking the corresponding sum rule.

As in the gaugino case, all the above relationships apply at the
string scale. From that scale down to low energies $O(M_W)$
the masses evolve according to the renormalization group
equations \INO.
           If the particles have the same $SU(3)\times SU(2)
\times U(1)$ quantum numbers equations like \fcnc\ directly apply
at low energies (up to Yukawa couplings).

\chap{Final comments and conclusions}

In this paper we have analysed several implications of
the duality invariance of 4-D orbifold strings. One of the merits of
our argumentation is that it allows us to discard large classes
of models without needing to work in a (hopeless)
model-by-model basis.
An important point  is the necessary
                       cancellation of target-space one-loop
modular anomalies. We have shown how, in analogy with the ordinary
ABJ anomalies, such a cancellation strongly constrains the
massless fermion content of the theory. This allows us to prove
how certain massless fermion contents are incompatible with
cancellation of duality anomalies for certain orbifolds. This is the
case of generic (0,2) ${\bf Z}_3$ and ${\bf Z}_7$ orbifolds for
which duality anomaly cancellation forbids the existence of
models whose massless charged sector with respect to
$SU(3)\times SU(2)\times U(1)$ is that of the minimal
supersymmetric standard model. Such constraints do also exist
for other gauge groups and particle contents.

For other ${\bf Z}_N$ orbifolds the target-space modular anomaly
cancellation conditions are less restrictive. If those conditions are
combined with the additional requirement of getting adequate
gauge coupling unification results for $\sin^2\theta _W$ and
$\alpha _3$, the above minimal string scenario is ruled out
for all ${\bf Z}_N$ orbifolds except ${\bf Z}_6$ and ${\bf Z}_8'$.
A consistent
unification scenario with the massless content of the
minimal supersymmetric standard model is
also possible for ${\bf Z}_N\times {\bf Z}_M$.
Alternatively, one can modify the massless particle
content of the minimal supersymmetric standard model and
add additional vector-like fields; in this case the constraints
often become weaker, as discussed at the end of section 4.
Further alternatives include  the construction of higher
Kac--Moody level models or intermediate scale symmetry breaking
through flat (or quasi-flat) directions.

Whichever (if any) of     these possibilities is actually realized,
we think it fair to say that the simplest possibility of
direct string unification of the minimal supersymmetric standard model
does not appear to be an immediate possibility in
4-D (0,2) orbifold strings. Extra massless particles and/or
intermediate scales and/or distorted orbifold shape and/or higher
Kac--Moody levels seem to be required.

Another topic discussed in the paper is the soft SUSY-breaking
terms present in the low-energy effective Lagrangian
once supersymmetry is broken. Assuming that the auxiliary
fields of the dilaton $S$ and moduli $T_i$ are the seed of
supersymmetry breaking
one can obtain the explicit dependence
of the soft gaugino and scalar masses on the modular weights
of the massless chiral fields. This explicitly shows that in
generic string models the soft scalar masses are $not$ universal
but depend on the modular weight of the particle. Similarly,
one sees that (in non-unified models) the soft gaugino masses
are gauge group dependent. The departure from universality
of gaugino masses may be related in specific models to
the gauge coupling constant threshold effects. In some
cases specific mass relationships are found.

Most of the above topics were discussed in the context
of 4-D (0,2) orbifolds. Some general features (see beginning of
sections 2 and 3) are expected to hold in other classes
of 4-D strings. It is known that non-orbifold 4-D strings
do also have infinite discrete duality symmetries physically
rather similar (but mathematically different) to the
modular group of toroidal compactifications \CAN . It is difficult to say
which of the properties present in the toroidal compactifications
will still be present in more general (0,2) non-flat
compactifications. It is conceivable         that at least
some of the latter models may also have ``universal'' moduli
(analogous to the ``completely rotated'' moduli of orbifolds)
whose generalized duality anomalies may only be cancelled by
a Green--Schwarz mechanism. The cancellation of duality anomalies
with respect to such type of moduli would give rise to
constraints similar to those in eqs.\bequal\ and \bgrav .
One also expects the existence of other moduli explicitly
contributing to gauge threshold effects if there are string
massive modes whose mass depends on such moduli. The
relevant threshold effects will involve generalized
automorphic forms \FKLZ\ (appropriate generalizations of the Dedekind
function to non-toroidal strings). If this was the case most
of our discussion could be generalized to these compactifications.
Relationships amongst gaugino masses similar to the ones
discussed in section 5 would also be expected.

\bigskip
\bigskip
We acknowledge usefull discussions with A. Casas, W. Lerche, J. Louis,
C. Mu\~noz and F. Quevedo.

\vfill
\endpage

\appendix{A}{A ${\bf Z}_7$ example}

As a first example we consider now the computation of the
$\sigma $-model anomalies in the (2,2) ${\bf Z}_7$ orbifold. We
take this example because it explicitly shows the generic
appearance of multiple oscillator states (and hence of
high negative modular weights) in Abelian orbifolds. The
(2,2) ${\bf Z}_7$ orbifold \IMNQ ,\ZSIE\ has as gauge group
$E_6\times U(1)^2\times E_8$, and the massless charged
chiral fields are the following. The fields in the
untwisted sector transform as $3({\underline{27}}\ +\ \underline 1)$.
There are three
independent twisted sectors of order $\theta $, $\theta ^2$ and
$\theta ^4$, where $\theta $ denotes the order seven elementary twist
(the twisted sectors $\theta ^6$, $\theta ^5$ and $\theta ^3$
contain the corresponding antiparticles). Each one of these
three sectors contains seven copies of ${\underline{27}}$'s
(corresponding to the
seven fixed points of the orbifold). In addition there are
twisted oscillator states which are singlets under $E_6\times E_8$.
For example, in the $\theta $ sector one finds states with the
following oscillator number $N$ and creation operators
$\alpha ^i$, ${\tilde \alpha }^i$ :
$$
\eqalign{
N\ =\ 1/7 & \rightarrow \alpha ^1 \cr
N\ =\ 2/7 & \rightarrow \alpha ^2\ ,\ (\alpha ^1 )^2  \cr
N\ =\ 4/7 & \rightarrow { \alpha }^3 \ ,\ (\alpha ^2)^2\ ,\ ({
\tilde\alpha }^3)(\alpha ^1)\ ,\ (\alpha ^2)(\alpha ^1)^2\ ,\
(\alpha ^1)^4 \cr }
\eqn \oscisiete
$$
where the superindex of the oscillators denotes the relevant
complex plane in which the oscillator is acting. Similar oscillators
exist in the other $\theta ^2$ and $\theta ^4$ sectors with the
oscillator superindices permuted. With the above information, one
can easily check that, since all the three planes in the
${\bf Z}_7$ orbifold are completely rotated, one has
$$
{b'}_{E_6}^i\ =\ {b'}_{E_8}^i\ =\ {1\over 24 }\ {b'}_{\rm grav}^i.
\eqn\bsiete
$$
Let us check this result for the overall modulus and leave the
plane-by-plane case as an exercise for the reader. For the
$E_8$ group the result is, of course, ${b'}_{E_8}=-3\times 30=-90$.
For $E_6$ we only need to know that the
three untwisted ${\underline{27}}$'s
have modular weight $-1$ and that the twisted ones have modular
weights $-2$. Thus one again gets
${b'}_{E_6}=-3\times 12+3\times 3-21\times 3=-90$. The same can be
checked for the $U(1)^2$. The case of the gravitational anomalies
is trickier. In this case all singlet chiral fields contribute.
In particular, the structure of modular weights for the above oscillators
plays an important role. The field with oscillator number
$N=1/7$ has modular weight $-3$, the two fields with $N=2/7$
have modular weights $-3$,$-4$ and the five states with
$N=4/7$ have modular weights $-3,-4,-2,-5,-6$ respectively.
On the other hand the dimension of the gauge group is
${\rm dim}\ G=328$ and the contribution of the ``modulinos'' to the
gravitational anomaly is $\delta _M=-3$, since there are only
three untwisted moduli. Then the complete anomaly is with eq.\bgravv :
$$\eqalign{
{b'}_{\rm grav}\ &=\  63\    +\ 3\ -\ 3\  -\ 984 \cr
&+\ 3\times (27+1)\ -\ 3\times 7\times 27 \cr
&-\ 3\times 7\times (3+3+5+3+5+1+7+9) \                          \cr }
$$
where the second line gives the untwisted and twisted matter
contributions and the third line that of oscillator states.  The
reader can check that the total anomaly yields $-2160=-24\times 90$,
as it should. One may get the wrong impression that multiple
oscillator massless states appear only for gauge singlet particles.
This is in fact true for (2,2) orbifolds (other than ${\bf Z}_3,{\bf Z}_4
,{\bf Z}_6'$) but it certainly is not true for generic (0,2) models.
An easy example of this may be obtained by examining a (0,2) ${\bf Z}_7$
orbifold studied in ref.\DEG  . It is obtained by acting on one of
the $E_8$'s with the shift $\vec\delta={1\over 7}(1,1,2,-2,-2,0,0,0)$
instead of the (2,2) shift. The gauge group of the model is
$SU(7)\times SU(2)\times U(1)\times E_8$ and, for example, one
can check that in the $\theta ^2$ twisted sector there are
7-plets of $SU(7)$ with double oscillators and overall
modular weight $-4$.
We have also checked that in this (0,2) ${\bf Z}_7$ orbifold
one has $b'_7=b'_2=b'_1=b'_8=b'_{\rm grav}/24=-90$ and hence the
mixed duality anomalies are all cancelled by a Green--Schwarz
mechanism.

\appendix{B}{A ${\bf Z}_4$ example}

As a second example for the computation of the anomaly
coefficients and also for the determination of the unification masses
we present a (0,2) ${\bf Z}_4$ orbifold, which was constructed
in ref.\KATSUKI\ (model No.2 in table 2 of this article).
The ${\bf Z}_4$ twist vector  acting on the internal six-torus
is given by $\vec\theta=(1/2,1/4,1/4)$. Furthermore the model
is defined by the non-standard embedding of the twist into
the gauge group $E_8\times E_8'$. Specifically, the embedding is characterized
by the following shift vector $\vec\delta=(1/2,1/4,1/4,0^5;1/2,1/2,0^6)$.
This shift breaks the group $E_8\times E_8'$ to the gauge group
$G=\lbrack E_6\times SU(2)_1\times U(1)\rbrack\times\lbrack E_7\times
SU(2)_2\rbrack '$.

Let us now list the transformation properties and the corresponding
modular weight vectors of the matter fields (we will omit to list the
$U(1)$ quantum numbers).
In the untwisted sector
there are the following representations associated to the
first complex plane:
$$\phi_{U_1}=({\underline{27}},\underline 1;\underline 1,\underline 1)
+({\underline{\bar{27}}},\underline 1;\underline 1,\underline 1)
+(\underline 1,\underline 1;{\underline{56}},\underline 2).\eqn\repua
$$
The corresponding modular weight vector looks like
$$\vec n_{U_1}=(-1,0,0).\eqn\wua
$$
Secondly, there are the following untwisted fields associated to
the second and third complex plane:
$$\phi_{U_{2,3}}
=({\underline{\bar{27}}},\underline 2;\underline 1,\underline 1)
+(\underline 1,\underline 2;\underline 1,\underline 1).\eqn\repub
$$
Here the corresponding modular weights are
$$\vec n_{U_2}=(0,-1,0),\qquad \vec n_{U_3}=(0,0,-1).\eqn\wub
$$
In the first twisted sector with twist vector $\vec\theta_1=(1/2,1/4,
1/4)$, there are the following states with no twisted
oscillators:
$$\phi_{T_1}
=16(\underline 1,\underline 2;\underline 1,\underline 2).\eqn\repta
$$
The corresponding modular weight vectors are given as
$$\vec n_{T_1}=(-{1\over 2},-{3\over 4},-{3\over 4}).\eqn\wta
$$
Furthermore, we find in this sector the states with
a positive chirality oscillator with respect to either the second
or the third complex plane ($p_{2,3}=1$):
$$\phi_{T_1}=
16(\underline 1,\underline 1;\underline 1,\underline 2)^{p_2=1}+
16(\underline 1,\underline 1;\underline 1,\underline 2)^{p_3=1}.
\eqn\reptb
$$
Then the corresponding modular weight vectors have the form
$$\vec n_{T_1}^{p_2=1}=(-{1\over 2},-{7\over 4},-{3\over 4}),\qquad
\vec n_{T_1}^{p_3=1}=(-{1\over 2},-{3\over 4},-{7\over 4}).\eqn\wtb
$$
In the second twisted sector with $\vec\theta_2=(0,1/2,1/2)$ there
are the following states with no oscillators:
$$\phi_{T_2}=10({\underline{27}},\underline 1;\underline 1,\underline 1)
+6({\underline{\bar{27}}},\underline 1;\underline 1,\underline 1)
+16(\underline 1,\underline 1;\underline 1,\underline 1)
.\eqn\reptc
$$
These states have modular weights
$$\vec n_{T_2}=(0,-{1\over 2},-{1\over 2}).\eqn\wtc
$$
In addition, this sector possesses states with positive chirality
oscillators:
$$\phi_{T_2}=6
(\underline 1,\underline 2;\underline 1,\underline 1)^{p_2=1}+
6(\underline 1,\underline 2;\underline 1,\underline 1)^{p_3=1}.\eqn\reptd
$$
Here the modular weights are
$$\vec n_{T_2}^{p_2=1}=(0,-{3\over 2},-{1\over 2}),\qquad
\vec n_{T_2}^{p_3=1}=(0,-{1\over 2},-{3\over 2}).\eqn\wtd
$$
Finally there are 20 representations of this kind with negative chirality
oscillators with respect to either the second or the third complex plane
($q_{2,3}=1$):
$$\phi_{T_2}=10
(\underline 1,\underline 2;\underline 1,\underline 1)^{q_2=1}+
10
(\underline 1,\underline 2;\underline 1,\underline 1)^{q_3=1}.\eqn\repte
$$
This implies the following set of modular weights:
$$\vec n_{T_2}^{q_2=1}=(0,{1\over 2},-{1\over 2}),\qquad
\vec n_{T_2}^{q_3=1}=(0,-{1\over 2},{1\over 2}).\eqn\wte
$$

Now, using eq.\bprimes, we are ready to compute the anomaly coefficients
for the four non-Abelian gauge group factors
($C(E_6)=12$, $C(E_7)=18$, $C(SU(2))=2$, $T({\underline{27}})=3$,
$T({\underline{56}})=6$, $T(\underline 2)=1/2$):
$$\eqalign{{b'}_{E_6}^{1}&=-12+3\lbrack 2(-1)+4(1)+16(1)\rbrack =42,\cr
{b'}_{E_6}^{2,3}&=-12+3\lbrack 2(-1)+4(1)+16(0)\rbrack =-6,\cr
{b'}_{SU(2)_1}^{1}&=-2+{1\over 2}\lbrack 56(1)+32(0)+32(1)\rbrack =42,\cr
{b'}_{SU(2)_1}^{2,3}&
=-2+{1\over 2}\lbrack 28(1)+28(-1)+32(-{1\over 2})\cr
&+6(-2)+6(0)+10(0)+10(2)\rbrack =-6,\cr
{b'}_{E_7}^{1}&=-18+6\lbrack 2(-1)\rbrack =-30,\cr
{b'}_{E_7}^{2,3}&=-18+6\lbrack 2(1)\rbrack =-6,\cr
{b'}_{SU(2)_2}^{1}&=-2+{1\over 2}\lbrack 56(-1)
+32(0)+32(0)\rbrack =-30,\cr
{b'}_{SU(2)_2}^{2,3}&
=-2+{1\over 2}\lbrack 56(1)+32(-{1\over 2})+16(-{5\over 2})
+16(-{1\over 2})\rbrack =-6.\cr}\eqn\bprimesmo
$$
For the gravitational anomaly eq.\gravi\ we obtain the following
coefficients ($\delta_M^1=2$, $\delta_M^{2,3}=-2$):
$$\eqalign{{b'}_{\rm grav}^1&=22+2-218+166(-1)+112(1)+128(0)+512(1)=264
,\cr
{b'}_{\rm grav}^{2,3}&=22-2-218+166(1)+56(1)+56(-1)+64(-{1\over 2})\cr
&+32(-{5\over 2})+32(-{1\over 2})+448(0)+12(-2)+12(0)+20(2)+20(0)=-144.
\cr}\eqn\gravbmo
$$
Since the second and third
complex plane are both rotated by $\vec\theta_1$
and $\vec\theta_2$, the anomaly coefficients with respect to these two
planes agree: ${b'}_a^{2,3}={b'}_{\rm grav}^{2,3}/24=-6$.

The first plane provides a non-vanishing contribution to the threshold
corrections. The relative unification masses $M_X$
among the coupling constants
of the four gauge group factors are given by
$${M_X\over M_{\rm string}}=
\lbrack
(T_1+\bar T_1)|\eta(T_1)|^4\rbrack^{{{b'}_b^{1}-{b'}_a^{1}
\over 2(b_a-b_b)}}
.\eqn\unificex
$$
The relevant $\beta$-function coefficients are given as
$b_{E_6}=30$, $b_{E_7}=-42$, $b_{SU(2)_1}=b_{SU(2)_2}=54$.
Then we obtain the following unification masses $M_X$:
$$\eqalign{&a=E_6,b=SU(2)_1:\qquad {M_X\over M_{\rm string}}=1,\cr
&a=E_7,b=SU(2)_2:\qquad {M_X\over M_{\rm string}}=1,\cr
&a=E_6,b=E_7:\qquad {M_X\over M_{\rm string}}=
\lbrack
(T_1+\bar T_1)|\eta(T_1)|^4\rbrack^{-1/2}>1,\cr
&a=E_6,b=SU(2)_2:\qquad {M_X\over M_{\rm string}}=
\lbrack
(T_1+\bar T_1)|\eta(T_1)|^4\rbrack^{3/2}<1,\cr
&a=E_7,b=SU(2)_1:\qquad {M_X\over M_{\rm string}}=
\lbrack
(T_1+\bar T_1)|\eta(T_1)|^4\rbrack^{-3/8}>1.\cr}\eqn\unifmasse
$$

\vfill
\endpage

\item{Table   1:} List of all Abelian orbifolds with, in the second
column, the corresponding
twist vectors. The
results for each case from the requirements of
modular anomaly cancellation and
coupling constant unification assuming
the spectrum of the minimal supersymmetric standard model ($3/5k_1=k_2
=k_3$) are shown in colums 3--6.
$+$ stands for an allowed model, whereas $-$ rules out the
particular case. A model with
viable minimal unification must have a $+$ in the
last column.

\bigskip
\bigskip

\begintable   \|   |         |
Unification | Unification | Unification
    \crnorule
 Orbifold
                        \|
 $(\theta_1,\theta_2,\theta_3 )$  |
        Anom. canc.
               |  $T=T_1=T_2=T_3$|
                  $T=T_1=T_2=T_3$|
                  $T_1>>T_2,T_3$\crnorule
  \|  |  |  no anom. canc. |
   plus anom. canc. |
   plus anom. canc. \crthick
${\bf Z}_3$ \| $(1/3,1/3,1/3)$ | -- | -- | -- | -- \cr
${\bf Z}_4$ \| $(1/2,1/4,1/4)$ | + | -- | -- | -- \cr
${\bf Z}_6$ \| $(1/2,1/3,1/6)$ | + | -- | -- | + \cr
${\bf Z}_6'$ \| $(2/3,1/6,1/6)$ | + | -- | -- | -- \cr
${\bf Z}_7$ \| $(4/7,2/7,1/7)$ | -- | -- | -- | -- \cr
${\bf Z}_8$ \| $(1/2,1/8,3/8)$ | + | -- | -- | -- \cr
${\bf Z}_8'$ \| $(1/4,1/8,5/8)$ | + | + | + | + \cr
${\bf Z}_{12}$ \| $(1/3,1/12,7/12)$ | + | + | -- | -- \cr
${\bf Z}_{12}'$ \| $(1/2,1/12,5/12)$ | + | + | -- | --
\endtable

\vfill
\endpage

\item{Table   1:} (continued)

\bigskip
\bigskip

\begintable   \|   |         |
Unification | Unification | Unification
    \crnorule
 Orbifold
                        \|
 $(\theta_1,\theta_2,\theta_3 )$  |
        Anom. canc.
               |  $T=T_1=T_2=T_3$|
                  $T=T_1=T_2=T_3$|
                  $T_1>>T_2,T_3$\crnorule
  \|  |  |  no anom. canc. |
   plus anom. canc. |
   plus anom. canc. \crthick
${\bf Z}_{2}\times {\bf Z}_2$ \| $(1/2,1/2,0)$ | + | -- | -- | + \crnorule
                              \| $(0,1/2,1/2)$ |   |   |   |   \cr
${\bf Z}_{2}\times {\bf Z}_4$ \| $(1/2,1/2,0)$ | + | -- | -- | + \crnorule
                              \| $(0,1/4,3/4)$ |   |   |   |   \cr
${\bf Z}_{2}\times {\bf Z}_6$ \| $(1/2,1/2,0)$ | + | + | + | + \crnorule
                              \| $(0,1/6,5/6)$ |   |   |   |   \cr
${\bf Z}_{2}\times {\bf Z}_6'$ \| $(1/2,1/2,0)$ | + | -- | -- | + \crnorule
                              \| $(1/6,2/3,1/6)$ |   |   |   |   \cr
${\bf Z}_{3}\times {\bf Z}_3$ \| $(1/3,2/3,0)$ | + | -- | -- | + \crnorule
                              \| $(0,1/3,2/3)$ |   |   |   |   \cr
${\bf Z}_{3}\times {\bf Z}_6$ \| $(1/3,2/3,0)$ | + | + | + | + \crnorule
                              \| $(0,1/6,5/6)$ |   |   |   |   \cr
${\bf Z}_{4}\times {\bf Z}_4$ \| $(1/4,3/4,0)$ | + | -- | -- | + \crnorule
                              \| $(0,1/4,3/4)$ |   |   |   |   \cr
${\bf Z}_{6}\times {\bf Z}_6$ \| $(1/6,5/6,0)$ | + | -- | -- | + \crnorule
                              \| $(0,1/6,5/6)$ |   |   |   |
\endtable

\vfill
\endpage

\item{Table  2:} Kac--Moody conformal dimension of the
standard model particles. The two rightmost columns  give
the value of the total conformal dimension for two
choices of the levels consistent with GUT-like gauge coupling
boundary conditions at the string scale.

\bigskip
\bigskip

\begintable
\ \| $SU(3)$ | $SU(2)$ | $U(1)$ | ${\rm Total}  (3/5k_1=k_2=k_3=1)$ |
${\rm Total}  (3/5k_1=k_2=k_3=2)$  \crthick
$Q$ \| ${4\over {9+3k_3}}$ | ${3\over {8+4k_2}}$ | ${1\over {36k_1}}$
 | ${3/     5}$ | $  37/       80  $ \cr
$U$ \| ${4\over {9+3k_3}}$ | $0$ | ${4\over {9k_1}}$ |
$3/5     $ | $2/5     $ \cr
$D$ \| ${4\over {9+3k_3}}$ | $0$ | $1\over {9k_1}$ |
$2/5     $ | $ 3/10        $ \cr
$L,H$ \| $0$ | ${3\over {8+4k_2}}$ | $1\over {4k_1}$ |
$2/5     $ | $ 21 /      80 $ \cr
$E$ \| $0$ | $0$ | $1\over {k_1}$ | $3/5       $ |
$3/10       $
\endtable

\vfill
\endpage

\item{Table   3:} Maximum allowed number of oscillators and
allowed values of overall modulus modular weights for all
possible twisted sectors of Abelian orbifolds.

\bigskip
\bigskip

\begintable
 $(\vert v_1\vert ,\vert v_2\vert ,\vert v_3\vert )$
                        \| $ p_{\rm max}$  | $q_{\rm max}$|$h_{KM}=
\sum \vert v_i\vert ^2/2$
 | $h_{KM}=3/5\ (Q,U,E)$ | $h_{KM}=2/5\ (L,H,D)$  \crthick
$(0,0,0)$  \| $0$   | $0$ |$n=-1$ | $n=-1$ | $n=-1$ \cr
$(1/3,1/3,2/3)$\| $1$ | $0$ |  $-3\leq n\leq -2$ | $n=-2$| $n=-2$ \cr
$(1/2,1/4,1/4)$\| $2$ | $1$ | $-4\leq n\leq -1$ | $n=-2$ |$-3\leq n\leq
-2$ \cr
$(1/3,1/6,1/6)$ \| $4$ | $2$ | $-6\leq n\leq 0$| $n=-2$ |$-4\leq n\leq
-1  $ \cr
$(1/2,1/3,1/6)$\| $3$ | $1$ | $-5\leq n\leq -1$ | $n=-2$ | $-3\leq n
\leq -2$ \cr
$(3/7,2/7,1/7)$ \| $4$ | $1$ | $-6\leq n\leq -1$ | $n=-2$ | $-4\leq n
\leq -2$     \cr
$(1/2,1/8,3/8)$ \| $4$ | $1$ | $-6\leq n\leq  -1$ | $n=-2$ | $-4\leq n
\leq -2$      \cr
$(1/4,1/8,3/8)$ \| $5$ | $1$ | $-7\leq n\leq -1$ | $-3\leq  n\leq -2$ |
$-4\leq n\leq -2$     \cr
$(1/3,1/12, 5/12)$ \| $7$ | $1$ | $-9\leq n\leq -1$ | $-3\leq n\leq -2$ |
$-5\leq n\leq -2$      \cr
$(1/2,1/12,5/12)$ \| $6$ | $1$ | $-8\leq n\leq -1$ | $-3\leq n\leq -2$ |
$-5\leq n\leq -2$       \cr
$(0,1/2,1/2)$ \|  $1$ | $1$ | $-2\leq n\leq 0$ | $n=-1$ | $n=-1$   \cr
$(0,1/3,1/3)$ \| $2$ | $2$ | $-3\leq n\leq 1$ | $n=-1$ | $-2\leq n\leq
0$   \cr
$(0,1/4,1/4)$ \| $3$ | $3$ | $-4\leq n\leq 2$ | $n=-1$ | $-2\leq n\leq
0$   \cr
$(0,1/6,1/6)$ \| $5$ | $5$ | $-6\leq n\leq 4$ | $-2\leq n\leq 0$ |
$-3\leq n\leq 1$
\endtable

\vfill
\endpage

\item{Table  4:} Values of the weak mixing angle and $\alpha _3$
in non-minimal models including extra massless chiral fields.
The notation  for each model is given in the text. The values of
the parameters $\gamma $ and $A,B$ are also given.

\bigskip
\bigskip

\begintable
   \
 \| MODEL |$\sin^2\theta^0_W$| $\alpha _3^0$| $\gamma^0 $ | A|
B | B/A \crthick
  \         \| Minimal | 0.218 | 0.20 | 2.7-3.7 | 28/5 | 20 | 25/7 \cr
a)   \|$ O+T$ | 0.274 | 0.056 | 2.7-3.0 | 18/5 | 14 | 35/9 \cr
b) \| $O+T+E+{\bar E}$ | 0.24 | 0.074 | 7.8-9.8 | 24/5 | 16 | 10/3 \cr
c) \| $O+T+2(E+{\bar E})$ | 0.207 | 0.109 | $\sim 0$ | 6 | 18 | 3 \cr
d) \| $O+T+H+{\bar H}+E+{\bar E}$ | 0.251 | 0.109 | $\sim 0$ | 22/5 |
18 | 45/11 \cr
e) \| $D+{\bar D}$ | 0.207 | 0.109 | $\sim 0$ | 6 | 18 | 3 \cr
f) \| $U+{\bar U}$ | 0.173 | 0.205 | 0.7-0.9 | 36/5 | 20 | 25/9 \cr
g) \| $Q+{\bar Q}$ | 0.296 | 0.109 | $\sim 0$ | 14/5 | 18 | 45/7 \cr
h) \| $D+{\bar D}+E+{\bar E}$ | 0.205 | 0.173 | 1.1-1.6 | 36/5|20|25/9\cr
i) \| $U+{\bar U}+Q+{\bar Q}$ | 0.251 | 0.109 | $\sim 0$ | 22/5|18|45/11
\cr
j) \| $D+{\bar D}+Q+{\bar Q}$ | 0.285 | 0.075 | 1.0-1.3 |  16/5 | 16 |
5
\endtable

\vfill
\endpage

\refout

\vfill\eject\bye